\documentclass[a4paper,12pt]{article}
\usepackage[utf8]{inputenc}
\usepackage[T1]{fontenc}
\usepackage{epsfig,graphicx,xcolor,amsbsy,amssymb,latexsym,amsfonts,amsmath,tcolorbox,setspace,mathrsfs}
\usepackage{pstricks}
\usepackage{color}
\usepackage{soul}
\usepackage{accents}
\usepackage{dsfont}
\usepackage{pifont}

\usepackage{simpler-wick}
\usepackage{placeins}
\usepackage{wrapfig,framed,caption}
\usepackage[makeroom]{cancel}
\usepackage{amsfonts}
\usepackage{dsfont}
\usepackage{mathrsfs}

\usepackage[style=numeric,sorting=none,style=numeric-comp,maxnames=10]{biblatex}

\definecolor{shadecolor}{gray}{0.925}

\usepackage{eurosym}
\usepackage{enumitem}
\usepackage[a4paper]{geometry}
\usepackage{hyperref}
\usepackage{graphicx}
\usepackage{tikz}
\usetikzlibrary{cd}
\usetikzlibrary{decorations.pathreplacing,decorations.markings,snakes}
\usetikzlibrary{decorations.pathmorphing}
\usepackage{xcolor}
\usepackage{amsmath,amssymb,amsfonts,pstricks,setspace}
\usepackage[enableskew]{youngtab}
\usepackage{ytableau}

\numberwithin{equation}{section}

\newcommand{\bea}{\begin{eqnarray}\displaystyle}
\newcommand{\eea}{\end{eqnarray}}

\newcommand{\figref}[1]{Fig.~\protect\ref{#1}}

\newcommand{\quivbox}[3]{\draw[ thick,fill=white] (#1-#3,#2-#3) -- (#1-#3,#2+#3) -- (#1+#3,#2+#3) -- (#1+#3,#2-#3) -- (#1-#3,#2-#3);}

\newcommand{\sqbox}[2]{\draw[ thick,fill=white] (#1-0.5,#2-0.5) -- (#1-0.5,#2+0.5) -- (#1+0.5,#2+0.5) -- (#1+0.5,#2-0.5) -- (#1-0.5,#2-0.5);}

\newcommand{\trbox}[2]{\draw[ thick,fill=gray] (#1-0.5,#2-0.5) -- (#1-0.5,#2+0.5) -- (#1+0.5,#2+0.5) -- (#1-0.5,#2-0.5);}
\newcommand{\gsqbox}[2]{\draw[ thick,fill=green] (#1-0.5,#2-0.5) -- (#1-0.5,#2+0.5) -- (#1+0.5,#2+0.5) -- (#1+0.5,#2-0.5) -- (#1-0.5,#2-0.5);}
\newcommand{\rsqbox}[2]{\draw[ thick,fill=red] (#1-0.5,#2-0.5) -- (#1-0.5,#2+0.5) -- (#1+0.5,#2+0.5) -- (#1+0.5,#2-0.5) -- (#1-0.5,#2-0.5);}

\newcommand{\wtrbox}[2]{\draw[ thick] (#1-0.5,#2-0.5) -- (#1-0.5,#2+0.5) -- (#1+0.5,#2+0.5) -- (#1-0.5,#2-0.5);}

\newcommand{\btrbox}[2]{\draw[ thick,fill=black] (#1+0.5,#2+0.5) -- (#1+0.5,#2-0.5) -- (#1-0.5,#2-0.5) -- (#1+0.5,#2+0.5);}

\newcommand{\stab}[2]{\draw[green,ultra thick,fill] (#1,#2) circle (0.08);}

\newcommand{\sstab}[2]{\draw[orange,ultra thick,fill] (#1,#2) circle (0.08);}

\newcommand{\istab}[2]{\draw[red,ultra thick,fill] (#1,#2) circle (0.08);}

\newcommand{\ydi}[1]{\scalebox{.5}{\ydiagram{#1}}}

\newcommand{\Qt}{Q_{\tau}}

\newcommand{\dd} {\mathrm{d}}

\usepackage{float}
\newtcolorbox{summary}[2][]{colbacktitle=blue!10!white, colback=yellow!10!white,coltitle=blue!70!black, title={#2},fonttitle=\bfseries,#1}

\newcommand{\modif}[1]{#1}

\title{
{\bf Wall-crossing of Instantons on the Blow-up}\\[40pt]}

\author{\large \textsc{Baptiste~Filoche\footnote{\tt bfiloche@mail.tsinghua.edu.cn}}~~,~~\textsc{Stefan~Hohenegger\footnote{\tt s.hohenegger@ipnl.in2p3.fr}}
~,~\,and\,~\textsc{Taro~Kimura\footnote{\tt taro.kimura@ube.fr}}
}

\begin{document}

\maketitle
\thispagestyle{empty}
\begin{center}
\renewcommand{\thefootnote}{\fnsymbol{footnote}}\vspace{-0.5cm}
${}^{\footnotemark[1]}$ Beijing Institute of Mathematical Sciences and Applications, Beijing 101408, China\\[0.1cm]
Yau Mathematical Sciences Center, Tsinghua University, Beijing 100084, China
\\[0.2cm]
${}^{\footnotemark[2]}$ Université Claude Bernard Lyon 1, CNRS/IN2P3, IP2I Lyon, UMR 5822, Villeurbanne, F-69100, France\\[0.2cm]
${}^{\footnotemark[3]}$ Universit\'e Bourgogne Europe, CNRS, IMB UMR 5584, 21000 Dijon, France\\[2.5cm]
\end{center}

\begin{abstract}

We study the instanton counting in four dimensional $\mathcal{N}=2$ supersymmetric gauge theories on the blow-up of $\mathbb{C}^2$: we start by formulating the instanton moduli space as a quiver variety, which we regularise by introducing two stability parameters, thus endowing it with a structure of infinitely many chambers separated by walls. Within a given chamber, we formulate the instanton partition function as a contour integral, which can be evaluated using the Jeffrey-Kirwan residue prescription. We characterise the physically relevant contributions in terms of bipartite oriented graphs and show that they can more efficiently be classified in terms of combinatorial objects called super-partitions. Within a given chamber, only certain types of super-partitions contribute and we show that the corresponding selection criteria are equivalent to stability conditions that have previously been proposed in the literature. We use this formalism to compare how the instanton counting changes when moving across walls between neighbouring chambers and provide explicit expressions for the corresponding partition functions. In a limiting chamber and using our approach, we show how to reproduce the Nakajima-Yoshioka blow-up formula.
\end{abstract}

\newpage

\tableofcontents

\newpage

\section{Introduction and Overview}

\paragraph{}
The introduction of supersymmetric localization~\cite{Duistermaat:1982vw,berline1982classes,Atiyah:1984px,Pestun:2016zxk} has launched a large program to compute exact non-perturbative observables in gauge theories with extended supersymmetry. Techniques associated with localization have for instance been used to compute instanton corrections to the pre-potential of Seiberg-Witten theories through the computation of instanton partition functions \modif{as conjectured by Nekrasov~\cite{Nekrasov:2002qd}.
This conjecture was independently proven by Nakajima and Yoshioka~\cite{Nakajima:2003pg} and Nekrasov and Okounkov~\cite{Nekrasov:2003rj}.
Nakajima-Yoshioka's approach to this problem is to consider instantons on the blow-up of the Euclidean space-time~\cite{Nakajima:2003pg,Nakajima:2005fg}.} The blow-up is obtained by considering the manifold obtained from the replacement of the origin by a $2$-sphere.
While interesting in its own right, studying supersymmetric gauge theories on the blow-up, is also fruitful for understanding the theories on the original Euclidean space. For example, as demonstrated in \cite{Nakajima:2003pg}, it provides a way to iteratively compute instanton corrections (\emph{e.g.} for the partition function) through the so-called {\it blow-up formula}. Indeed, the latter establishes a (bilinear) relation between the partition function on the blow-up and on the original geometry. It therefore encodes crucial information on the supersymmetric gauge theory (on the Euclidean space-time), for example:
\begin{itemize}
\item[\emph{(i)}] 
    The blow-up formula can be used to construct the non-perturbative sector of a supersymmetric gauge theory purely from its perturbative sector. This idea has for example extensively been used to compute the non-perturbative partition function of theories which cannot be directly accessed through localization~\cite{Huang:2017mis,Gu:2018gmy,Gu:2019dan,Gu:2019pqj,Gu:2020fem,Kim:2019uqw,Kim:2023glm,Kim:2021gyj,Bonelli:2012ny,Kuhn:2021tul,Kuhn:2022xyw}.
\item[\emph{(ii)}] For gauge theories on the $\Omega$-background~\cite{Nekrasov:2002qd}, the blow-up formula allows to explicitly link the partition function in two limits of the  corresponding deformation parameters~\cite{Sun:2016obh}, namely the unrefined limit and the Nekrasov-Shatashvili limit~\cite{Nekrasov:2009rc}. This relation provides interesting insights about the resurgent structure of topological string partition functions~\cite{Huang:2017mis,Grassi:2016nnt,Gu:2023wum,Francois:2023trm}.
    \item[\emph{(iii)}] In the context of the Bethe/Gauge correspondence~\cite{Nekrasov:2009rc}, non-perturbative partition functions can be interpreted as wave-functions and $\tau$-functions of known integrable systems. In particular, a specific version of the blow-up formula can be formulated as the action of a bilinear derivative operator, the so-called Hirota differential~\cite{Nakajima:2003pg}, on the partition function. Furthermore, it is known \cite{Gamayun:2012ma,Gamayun:2013auu} that the discrete Fourier transform of the non-perturbative partition function on the $\Omega$-background (known as the dual partition function), can be interpreted as the $\tau$-function of the quantum Painlevé equation. In the unrefined limit, the latter is equivalent to the blow-up formula, written in terms of the Hirota differential~\cite{Bershtein:2014yia,Nekrasov:2020qcq,Jeong:2020uxz,Bonelli:2024wha,Bonelli:2025bmt,Bonelli:2025mfw,Bonelli:2025owb}.
 
\end{itemize}

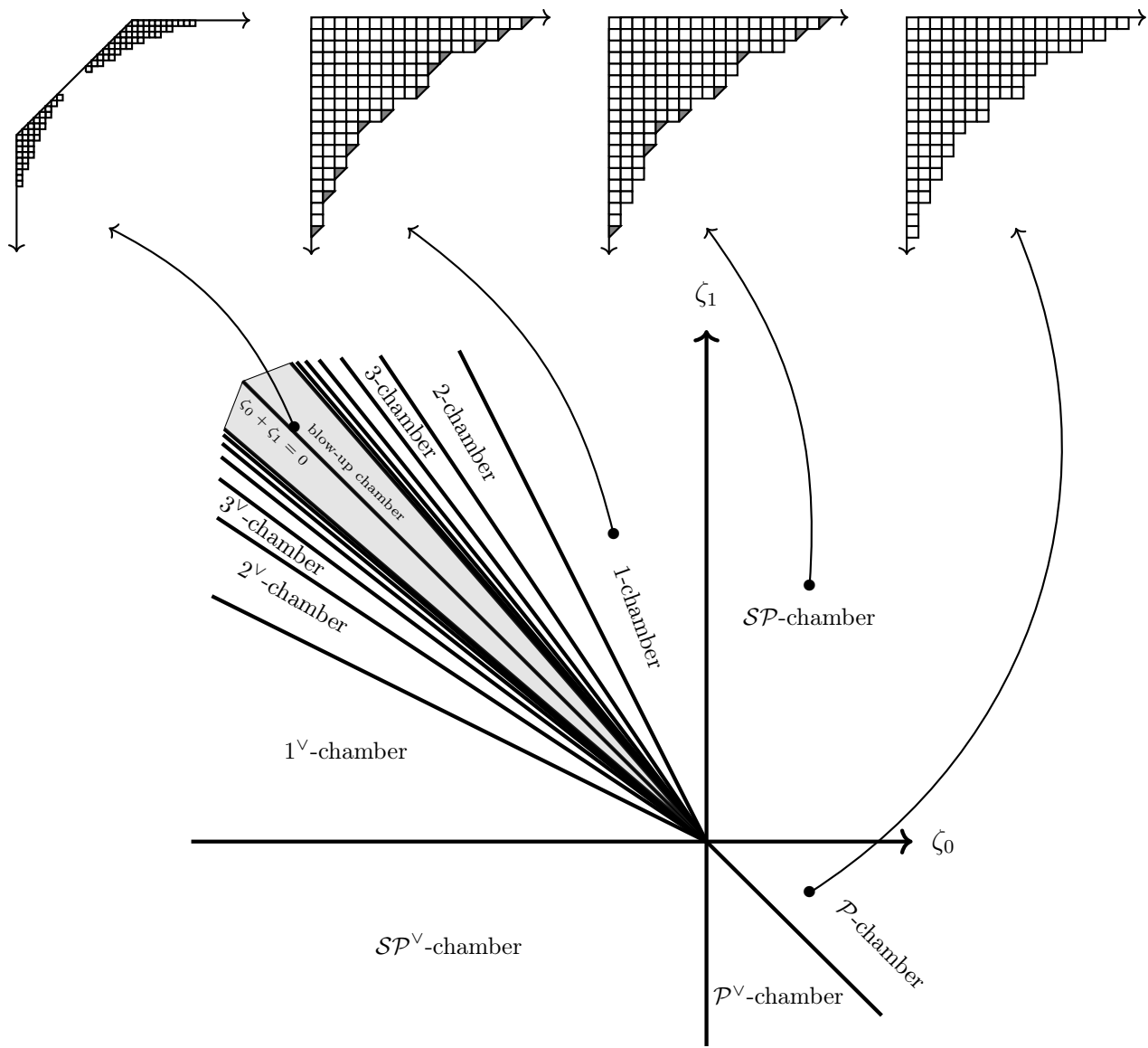
\begin{figure}[tbp]
    \centering
    \begin{tikzpicture}
        \begin{scope}[shift={(13,0)},scale=0.17]
        \draw[thick,->] (0.5,0.5) -- (20,0.5);
        \draw[thick,->] (-0.5,-0.5) -- (-0.5,-20);
        \foreach \i in {0,...,17}{
            \sqbox{\i}{0};}
        \foreach \i in {0,...,15}{
            \sqbox{\i}{-1};}
        \foreach \i in {0,...,14}{
            \sqbox{\i}{-2};}
        \foreach \i in {0,...,10}{
            \sqbox{\i}{-3};}
        \foreach \i in {0,...,10}{
            \sqbox{\i}{-4};}
        \foreach \i in {0,...,9}{
            \sqbox{\i}{-5};}
        \foreach \i in {0,...,8}{
            \sqbox{\i}{-6};}
        \foreach \i in {0,...,6}{
            \sqbox{\i}{-7};}
        \foreach \i in {0,...,5}{
            \sqbox{\i}{-8};}
        \foreach \i in {0,...,3}{
            \sqbox{\i}{-9};}
        \foreach \i in {0,...,3}{
            \sqbox{\i}{-10};}
        \foreach \i in {0,...,2}{
            \sqbox{\i}{-11};}
        \foreach \i in {0,...,2}{
            \sqbox{\i}{-12};}
        \foreach \i in {0,...,2}{
            \sqbox{\i}{-13};}
        \foreach \i in {0,1}{
            \sqbox{\i}{-14};}
        \foreach \i in {0,1}{
            \sqbox{\i}{-15};}
        \sqbox{0}{-16}
        \sqbox{0}{-17}
        \sqbox{0}{-18}\sqbox{3}{-11}\sqbox{4}{-9}\sqbox{6}{-8}
        \sqbox{9}{-6}\sqbox{11}{-3}\sqbox{16}{-1}\sqbox{18}{0}
        \end{scope}
        \begin{scope}[shift={(8.666,0)},scale=0.17]
        \draw[thick,->] (0.5,0.5) -- (20,0.5);
        \draw[thick,->] (-0.5,-0.5) -- (-0.5,-20);
        \foreach \i in {0,...,17}{
            \sqbox{\i}{0};}
        \foreach \i in {0,...,15}{
            \sqbox{\i}{-1};}
        \foreach \i in {0,...,14}{
            \sqbox{\i}{-2};}
        \foreach \i in {0,...,10}{
            \sqbox{\i}{-3};}
        \foreach \i in {0,...,10}{
            \sqbox{\i}{-4};}
        \foreach \i in {0,...,9}{
            \sqbox{\i}{-5};}
        \foreach \i in {0,...,8}{
            \sqbox{\i}{-6};}
        \foreach \i in {0,...,6}{
            \sqbox{\i}{-7};}
        \foreach \i in {0,...,5}{
            \sqbox{\i}{-8};}
        \foreach \i in {0,...,3}{
            \sqbox{\i}{-9};}
        \foreach \i in {0,...,3}{
            \sqbox{\i}{-10};}
        \foreach \i in {0,...,2}{
            \sqbox{\i}{-11};}
        \foreach \i in {0,...,2}{
            \sqbox{\i}{-12};}
        \foreach \i in {0,...,2}{
            \sqbox{\i}{-13};}
        \foreach \i in {0,1}{
            \sqbox{\i}{-14};}
        \foreach \i in {0,1}{
            \sqbox{\i}{-15};}
        \sqbox{0}{-16}
        \sqbox{0}{-17}
        \trbox{0}{-18}\trbox{3}{-11}\trbox{4}{-9}\trbox{6}{-8}
        \trbox{9}{-6}\trbox{11}{-3}\trbox{16}{-1}\trbox{18}{0}
        \end{scope}
        \begin{scope}[shift={(4.333,0)},scale=0.17]
        \draw[thick,->] (0.5,0.5) -- (20,0.5);
        \draw[thick,->] (-0.5,-0.5) -- (-0.5,-20);
        \foreach \i in {0,...,17}{
            \sqbox{\i}{0};}
        \foreach \i in {0,...,15}{
            \sqbox{\i}{-1};}
        \foreach \i in {0,...,13}{
            \sqbox{\i}{-2};}
        \foreach \i in {0,...,10}{
            \sqbox{\i}{-3};}
        \foreach \i in {0,...,9}{
            \sqbox{\i}{-4};}
        \foreach \i in {0,...,9}{
            \sqbox{\i}{-5};}
        \foreach \i in {0,...,8}{
            \sqbox{\i}{-6};}
        \foreach \i in {0,...,6}{
            \sqbox{\i}{-7};}
        \foreach \i in {0,...,5}{
            \sqbox{\i}{-8};}
        \foreach \i in {0,...,3}{
            \sqbox{\i}{-9};}
        \foreach \i in {0,...,3}{
            \sqbox{\i}{-10};}
        \foreach \i in {0,...,2}{
            \sqbox{\i}{-11};}
        \foreach \i in {0,...,2}{
            \sqbox{\i}{-12};}
        \foreach \i in {0,...,1}{
            \sqbox{\i}{-13};}
        \foreach \i in {0,1}{
            \sqbox{\i}{-14};}
        \foreach \i in {0}{
            \sqbox{\i}{-15};}
        \sqbox{0}{-16}
        \sqbox{0}{-17}
        \trbox{0}{-18}\trbox{3}{-11}\trbox{4}{-9}\trbox{6}{-8}
        \trbox{9}{-6}\trbox{11}{-3}\trbox{16}{-1}\trbox{18}{0}
        \trbox{14}{-2}
        \trbox{10}{-4}
        \trbox{2}{-13}
        \trbox{1}{-15}
        \end{scope}
        \begin{scope}[shift={(0,0)},scale=0.084]
        \draw[thick,->] (19.5,0.5) -- (40,0.5);
        \draw[thick,->] (-0.5,-19.5) -- (-0.5,-40);
        \draw[thick] (19.5,0.5) -- (-0.5,-19.5);
        \sqbox{7}{-13}
        \sqbox{6}{-14}
        \sqbox{5}{-15}
        \foreach \i in {0,...,1}{
            \sqbox{5}{-15-\i};}
        \sqbox{4}{-16}
        \foreach \i in {0,...,2}{
            \sqbox{4}{-16-\i};}
        \sqbox{3}{-17}
        \foreach \i in {0,...,3}{
            \sqbox{3}{-17-\i};}
        \sqbox{2}{-18}
        \foreach \i in {0,...,5}{
            \sqbox{2}{-18-\i};}
        \sqbox{1}{-19}
        \foreach \i in {0,...,6}{
            \sqbox{1}{-19-\i};}
        \foreach \i in {0,...,8}{
            \sqbox{0}{-20-\i};}

        \foreach \i in {0,...,10}{
            \sqbox{20+\i}{0};}
        \sqbox{20}{0}
        \foreach \i in {0,...,7}{
            \sqbox{19+\i}{-1};}
        \sqbox{19}{-1}
        \foreach \i in {0,...,6}{
            \sqbox{18+\i}{-2};}
        \sqbox{18}{-2}
        \foreach \i in {0,...,4}{
            \sqbox{17+\i}{-3};}
        \sqbox{17}{-3}
        \foreach \i in {0,...,3}{
            \sqbox{16+\i}{-4};}
        \sqbox{16}{-4}
        \foreach \i in {0,...,2}{
            \sqbox{15+\i}{-5};}
        \sqbox{15}{-5}
        \foreach \i in {0,...,2}{
            \sqbox{14+\i}{-6};}
        \sqbox{14}{-6}
        \foreach \i in {0,...,1}{
            \sqbox{13+\i}{-7};}
        \sqbox{13}{-7}
        \foreach \i in {0}{
            \sqbox{12+\i}{-8};}
        \sqbox{12}{-8}
        \end{scope}
        \begin{scope}[shift={(10,-12)},scale=1.5]
            \node at (0,5.35) {$\zeta_1$};
            \node at (2.3,0) {$\zeta_0$};
            \draw[ultra thick,->] (-5,0) -- (2,0);
            \draw[ultra thick,->] (0,-2) -- (0,5);
            \draw[ultra thick] (-4.5,4.5) -- (1.7,-1.7);
            \draw[ultra thick] (-4.8,2.4) -- (0,0);
            \draw[ultra thick] (-4.75,3.1666) -- (0,0);
            \draw[ultra thick] (-4.73,3.5475) -- (0,0);
            \draw[ultra thick] (-4.71,3.76) -- (0,0);
            \draw[ultra thick] (-4.70,3.8916) -- (0,0);
            \draw[ultra thick] (-4.69,3.977) -- (0,0);
            \draw[ultra thick] (-4.68,4.03375) -- (0,0);
            \draw[fill=gray,fill opacity=0.2] (-4.68,4.03375) -- (0,0) --  (-4.5,4.5) -- (-4.68,4.03375);
            \draw[ultra thick] (-2.4,4.8) -- (0,0);
            \draw[ultra thick] (-3.1666,4.75) -- (0,0);
            \draw[ultra thick] (-3.5475,4.73) -- (0,0);
            \draw[ultra thick] (-3.76,4.71) -- (0,0);
            \draw[ultra thick] (-3.8916,4.70) -- (0,0);
            \draw[ultra thick] (-3.977,4.69) -- (0,0);
            \draw[ultra thick] (-4.03375,4.68) -- (0,0);
            \draw[fill=gray,fill opacity=0.2]
              (-4.03375,4.68) -- (0,0) -- (-4.5,4.5) -- (-4.03375,4.68);
            \node at (1,-0.5) {$\bullet$};
            \draw[->, thick, bend right=40] (1,-0.5) to (3,6);
            \node at (1,2.5) {$\bullet$};
            \draw[->, thick, bend right=20] (1,2.5) to (0,6);
            \node at (-0.9,3) {$\bullet$};
            \draw[->, thick, bend right=20] (-0.9,3) to (-2.9,6);
            \node at (-4,4.05) {$\bullet$};
            \draw[->, thick, bend right=20] (-4,4.05) to (-5.8,6);
            \node at (-2.5,-1) {\footnotesize $\mathcal{SP}^\vee$-chamber};
            \node at (0.7,-1.5) {\footnotesize $\mathcal{P}^\vee$-chamber};
            \node at (-3.5,0.9) {\footnotesize $1^\vee$-chamber};
            \node[rotate=-30] at (-4,2.4) {\footnotesize $2^\vee$-chamber};
            \node[rotate=-90+30] at (-2.33,4) {\footnotesize $2$-chamber};
\node at (1,2.2) {\footnotesize $\mathcal{SP}$-chamber};   
\node[rotate=-70] at (-0.65,2.2) {\footnotesize $1$-chamber};   
\node[rotate=-54] at (-2.95,4.2) {\footnotesize $3$-chamber};            
%
\node[rotate=-35] at (-4.2,3) {\footnotesize $3^\vee$-chamber};   
%
\node[rotate=-45] at (1.7,-1) {\footnotesize $\mathcal{P}$-chamber};   
\node[rotate=-45] at (-4.2,4) {\tiny $\zeta_0+\zeta_1=0$};
\node[rotate=-45] at (-3.4,3.6) {\tiny blow-up chamber};
        \end{scope}
    \end{tikzpicture}
    \caption{The lines correspond to walls separating the different stability chambers, in the grey region the walls accumulate towards the line $\zeta_0+\zeta_1=0$. The diagram exhibits a duality structure upon mirroring along this line. The blow-up formula is given by the configurations in the limit chamber next to the accumulation line. Above, we present typical super-partitions stable in some chambers of interest. }
    \label{fig:pertdm}
\end{figure}

\noindent
In order to study supersymmetric gauge theories on blow-up geometries, various techniques have been explored. A fruitful approach to the problem of instanton counting in such situations has emerged by considering the moduli space of instantons as a quiver variety~\cite{King,Nakajima:2008eq,Nakajima:2008ss,Nakajima:2009qjc}. In order to extract meaningful information (\emph{e.g.} to calculate the non-perturbative partition function of the gauge theory), this moduli space needs to be regularised, which can be achieved by introducing so-called \emph{stability conditions}. The latter represent additional geometric information, which subdivides the moduli space into `chambers' separated by 'walls': In this paper we are exclusively interested in the blow-up $\widehat{\mathbb{C}}^2$ of $\mathbb{C}^2$, for which the chamber structure is schematically shown in the bottom part of Figure~\ref{fig:pertdm}: $\widehat{\mathbb{C}}^2$ is obtained by replacing the Euclidean space-time origin by a complex curve $\mathbb{P}^1$. The moduli space of instantons on $\widehat{\mathbb{C}}^2$ is regularised by two parameters, which are called $\zeta_0$ and $\zeta_1$ in Figure~\ref{fig:pertdm}. This subdivision into chambers shown in the latter and more importantly the condition for instantons to be stable within any such chamber is a priori not evident. This problem has been studied in a different context in ~\cite{Nakajima:2008eq}. In this paper we propose a condition motivated by the physical BPS counting: in the blow-up geometry, the complex curve $\mathcal{C}$ can support magnetic flux and the partition function therefore actually counts instanton-monopole configurations. Using the ADHM construction, we formulate the (equivariant) volume of the moduli space of such configurations as a contour integral. In order to extract the contribution corresponding to a given instanton- and flux number, is equivalent to selecting the contour to include a specific set of poles. This choice is made using the Jeffrey-Kirwan residue prescription~\cite{Jeffrey:1993cun} and depends on the values of $(\zeta_0,\zeta_1)$, \emph{i.e.} the chamber as indicated in the bottom of Figure~\ref{fig:pertdm}.

\paragraph{}
As we shall explain, the choice of poles can graphically be represented in the form of oriented bi-partite graphs, which unambiguously determine the physical contributions to the instanton partition function. While this formulation allows to explicitly compute the partition function term by term, it is less practical for studying the physical states or comparing the contents of two neighbouring chambers across a wall. Indeed, a single partite graph only provides a choice of poles but does not guarantee that the corresponding residues are non-vanishing. We therefore provide a more efficient description of the non-vanishing contributions to the partition function in terms of the combinatorics of super-partitions. The latter are a generalisation of integer partitions, which appear in the counting of instantons of supersymmetric gauge theories. Super-partitions have recently also appeared in problems related to BPS state counting and super-Macdonald polynomials~\cite{Galakhov:2023mak,Galakhov:2024foa,Galakhov:2024cry,Galakhov:2025phf,Kanno:2025ifd}. Graphically, they can be visualised by super-Young diagrams that involve new types of so-called boundary boxes. The latter are represented by triangles in the top part of Figure~\ref{fig:pertdm}: while the instanton counting of the gauge theory on $\mathbb{C}^2$ ($\mathcal{P}$-chamber) is organised in terms of integer partitions (represented in terms of regular Young diagrams), the counting in the $\mathcal{SP}$-chamber is organised in terms of super-partitions, which allow us to give a compact form for the partition function in this chamber. We furthermore show, how the counting in a general $n$-chamber can equally be organised in terms of super-partitions satisfying certain conditions regarding their shape. We demonstrate that these are equivalent to stability conditions imposed in previous works~\cite{Nakajima:2008eq}. In this way, we can compare the instanton counting in neighbouring chambers, across a dividing wall. In the limit $n\to\infty$ (called the blow-up chamber in Figure~\ref{fig:pertdm}), we show that only super-partitions can contribute that can be split into two integer partitions (in a well-defined manner, as indicated in the top-left part of Figure~\ref{fig:pertdm}), allowing us not only to compute the partition function in a compact form, but also to deduce the blow-up formula.

\paragraph{}
This paper is organised as follows: In Section~\ref{sec:reviewcount}, we review the construction of the moduli space and set-up the counting problem. In Section~\ref{sec:classification}, using the Jeffrey-Kirwan residue prescription~\cite{Jeffrey:1993cun} we show that the poles selected by this procedure are captured by bipartite directed graphs, which thus encode the contributions of instanton-monopole configurations to the partition function of a gauge theory on the $\widehat{\mathbb{C}}^2$ blow-up. We furthermore show that the counting of such objects is captured by the combinatorics of super-partitions. In Section~\ref{sec:pfandblowup}, we provide expressions for the instanton partition functions in the different chambers and use our description to provide an alternative derivation of the blow-up formula. Finally, Section~\ref{Sect:Conclusion} contains our Conclusion as well as an outlook for applications of our results for future work. 

\section{Instanton counting on the blow-up of $\mathbb{C}^2$}\label{sec:reviewcount}

\paragraph{}
The problem of counting instantons of $\mathcal{N}=2$ $SU(N)$ pure super-Yang-Mills theory on the blow-up of $\mathbb{C}^2$ (or equivalently rank $N$ perverse coherent sheaves on the blow-up of $\mathbb{P}^2$) has been studied in detail in~\cite{King,Nakajima:2003pg,Nakajima:2003uh,Nakajima:2005fg,Nakajima:2008eq,Nakajima:2008ss,Nakajima:2009qjc}. In this Section, we review the definition of the problem and introduce the relevant instanton moduli spaces. In addition, we provide a dual description of the problem in terms of free-field correlators.

\subsection{Introduction and definitions}\label{subsec:introdef}

\begin{figure}[tbp]
    \centering
    \begin{tikzpicture}[scale=1.4]
        \draw[thick,->] (0,0) -- (2,0); 
        \draw[thick,->] (0,0) -- (0,2);
        \node at (0,0) {\small $\bullet$};
        \node at (-0.2,-0.3) {\small $(0,0)$};
        \node at (2.4,0) {$\mathbb{C}_1$};
        \node at (0.05,2.3) {$\mathbb{C}_2$};
        \node at (4,1) {$\longrightarrow$};
        \begin{scope}[shift={(2,-0.375)}]
        \draw[thick,->] (4.75,0) -- (6.75,0);
        \draw[thick,->] (4,0.75) -- (4,2.75);
        \draw[thick] (4,0)++(-15:1) arc[start angle=-15, end angle=105, radius=1];
        \node at (5,0) {\small $\bullet$};
        \node at (4,1) {\small $\bullet$};
        \node at (7.15,0) {$\mathbb{C}_1$};
        \node at (4.05,3) {$\mathbb{C}_2$};
        \node at (4.9,0.9) {$\mathcal{C}$};
        \node at (3.8,0.8) {\small $p_-$};
        \node at (4.8,-0.2) {\small $p_+$};
        \end{scope}
    \end{tikzpicture}
    \caption{Graphical depiction of $\mathbb{C}^2=\mathbb{C}_1\times \mathbb{C}_2$ and $\widehat{\mathbb{C}}^2$ with the marked fixed points of the $T$-action and $\mathcal{C}\cong \mathbb{P}^1$ is an exceptional divisor of $\widehat{\mathbb{C}}^2$ given by $z_1=z_2=0$.}
    \label{fig:blowupOmega}
\end{figure}
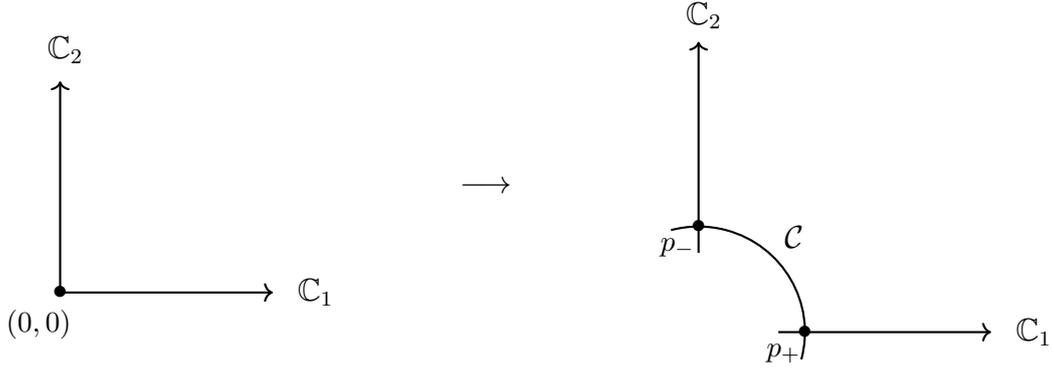

\paragraph{Geometric background}
In the original problem associated with the counting of $\mathcal{N}=2$ $SU(N)$ instantons~\cite{Nekrasov:2002qd}, the space-time is considered to be Euclidean $\mathbb{R}^4\cong \mathbb{C}^2$. In order to regularise the integration over the space-time appearing throughout the computations, one needs to introduce the so-called $\Omega$-background~\cite{Nekrasov:2002qd} which we define as the following $T$-action on the space-time coordinates:
\begin{align}
    T:= U(1)_{\varepsilon_1} \times U(1)_{\varepsilon_2}: (z_1,z_2) \longrightarrow (q_1 z_1,q_2 z_2)\,,
\end{align}
for $(z_1,z_2)\in \mathbb{C}^2$ and with $q_{j}=e^{2i\pi \varepsilon_{j}}$ for $j = 1,2$. By performing space-time integration in a $T$-equivariant fashion, we extract regularised quantities. For instance, the $T$-equivariant volume of $\mathbb{C}^2$ is $(\varepsilon_1\varepsilon_2)^{-1}$. We move from this background and consider the blow-up of $\mathbb{C}^2$ at the point $(0,0)$ which we denote as $\widehat{\mathbb{C}}^2$. The blow-up $\widehat{\mathbb{C}}^2$ is defined as the following embedding in $\mathbb{C}^2\times \mathbb{P}^1$:
\begin{align}\label{eq:coordBC2}
    \widehat{\mathbb{C}}^2 := \left\{ (z_1,z_2) \in \mathbb{C}^2 \,, \,\, [z:w] \in \mathbb{P}^1 \, \, | \,\, z_1w = zz_2 \right\}.
\end{align}
The $\Omega$-background can be naturally extended to the blow-up~\cite{Nakajima:2003pg} by defining the $T$-action on $\widehat{\mathbb{C}}^2$ to be:
\begin{align}
    T : (z_1,z_2),[z:w] \longrightarrow (q_1z_1,q_2z_2),[q_1z:q_2w]\,.
\end{align}
While the original $\mathbb{C}^2$ has only the fixed point $(0,0)$ under the $T$-action, the blow-up admits two fixed points: $p_+ = ((0,0),[1:0])$ and $p_-= ((0,0),[0:1])$. The blow-up procedure and the fixed points of the $\Omega$-background are depicted by~\figref{fig:blowupOmega}.
\newline
Since the map $\mathbb{C}^2 \to \widehat{\mathbb{C}}^2$ is bi-rational, the $T$-equivariant volume of $\mathbb{C}^2$ is unchanged by the blow-up procedure. Indeed, we parametrise the chart $U_{p_+}$ around $p_+$ by $(z_1/z,z_2)$ and the chart $U_{p_-}$ around $p_-$ by $(z_1,z_2/w)$. $p_+$ has therefore tangent weight $(\varepsilon_1-\varepsilon_2,\varepsilon_2)$ under the $T$-action, and $p_-$ has weight $(\varepsilon_1,\varepsilon_2-\varepsilon_1)$. The $T$-equivariant volume of $\widehat{\mathbb{C}}^2$ is therefore given by:
\begin{align}
    {\rm vol}_T(\widehat{\mathbb{C}}^2) = \frac{1}{\varepsilon_1(\varepsilon_2-\varepsilon_1)} + \frac{1}{\varepsilon_2(\varepsilon_1-\varepsilon_2)} = \frac{1}{\varepsilon_1 \varepsilon_2}\,.
\end{align}

\paragraph{Physical problem}
We consider instantons in $\mathcal{N}=2$ super-Yang-Mills theory with gauge group $G$ on $\widehat{\mathbb{C}}^2$. Instantons are defined as solutions of the anti-self-dual Yang-Mills equation:
\begin{align}
    F = - \star F\,,
\end{align}
with $F$ the ${\rm Lie}(G)$-valued 2-form field strength. The blow-up $\widehat{\mathbb{C}}^2$ admits an exceptional divisor $\mathcal{C}\cong \mathbb{P}^1$ given by $z_1=z_2=0$ depicted in~\figref{fig:blowupOmega}. 
Importantly, topologically non-trivial configurations on the blow-up $\widehat{\mathbb{C}}^2$ are now characterized by two quantum numbers corresponding to:
\begin{align}\label{eq:twoqnum}
    \frac{1}{2\pi}\int_{\mathcal{C}} {\rm tr}\, F = p \in \mathbb{Z}\,, && \frac{1}{4\pi^2} \int_{\mathcal{S}} {\rm tr}\, F \wedge F  = k \in \mathbb{N}\,.  
\end{align}
with $\mathcal{S}$ a compactification of the Euclidean space-time. Comparing this situation with its blow-down version, the problem of instanton counting on the blow-up corresponds to moving to a background with magnetic fluxes turned-on at the origin, or equivalently in presence of magnetic monopole of charge $p$ at the origin. For $G=SU(N)$, the first Chern class over $\mathcal{C}$ is vanishing, however at a generic point of the Coulomb branch the gauge group breaks down to $U(1)^{N-1}$, each $U(1)$ gauge group can have a non-vanishing associated first Chern class over $\mathcal{C}$ and a configuration is therefore indexed by an element of $\mathbb{Z}^{N-1}$. The problem of counting instanton on the blow-up can be understood as the counting of mixed instanton-monopoles configurations.

\subsection{The moduli space of instantons on $\widehat{\mathbb{C}}^2$}

\begin{figure}[t]
    \centering
    \begin{tikzpicture}[scale=1.75]
        \draw[thick,->] (0,0) .. controls (-0.5,0.5) and (-1,0.5) .. (-1,0) ..controls (-1,-0.5) and (-0.5,-0.5) .. (0,0);
        \draw[thick,->] (0,0) .. controls (-0.5,0.5) and (-1,0.5) .. (-1,0);
        \draw[thick,->] (0,0) .. controls (-0.5,-0.75) and (-1.25,-0.75) .. (-1.25,0) ..controls (-1.25,0.75) and (-0.5,0.75) .. (0,0);
        \draw[thick,->] (0,0) .. controls (-0.5,0.75) and (-1.25,0.75) .. (-1.25,0);
        \draw[thick] (0,0.06) -- (2,0.06);
        \draw[thick] (0,-0.06) -- (2,-0.06);
        \draw[thick,->] (0,0.06) -- (1,0.06);
        \draw[thick,->] (2,-0.06) -- (1,-0.06);
        \quivbox{2}{0}{0.35}
        \draw[thick,fill=white] (0,0) circle (0.35);
        \node[above] at (1,0.06) {$J$};
        \node[below] at (1,-0.06) {$I$};
        \node at (-0.65,0) {$B_2$};
        \node at (-1.55,0) {$B_1$};
        \node at (0,0) {\Large $\mathfrak{K}$};
        \node at (2,0) {\Large $\mathfrak{N}$};
        \draw[thick] (5,0.6) -- (7,0);
        \draw[thick] (5,-0.6) -- (7,0);
        \draw[thick,->] (5,0.6) -- (6,0.3);
        \draw[thick,->] (7,0) -- (6,-0.3);
        \node[above] at (6,0.3) {$J$};
        \node[below] at (6,-0.3) {$I$};
        \node at (4.3,0) {$B_1$};
        \node at (4.95,0) {$B_2$};
        \node at (5.65,0) {$d$};
        \draw[thick,->] (4.75,-0.6) .. controls (4.5,0) and (4.5,0) .. (4.75,0.6);
        \draw[thick,->] (4.565,0) -- (4.565,-0.001);
        \draw[thick,->] (4.9,-0.6) .. controls (4.65,0) and (4.65,0) .. (4.9,0.6);
        \draw[thick,->] (4.715,0) -- (4.715,-0.001);
        \draw[thick,->] (5.25,-0.6) .. controls (5.5,0) and (5.5,0) .. (5.25,0.6);
        \draw[thick,->] (5.435,0) -- (5.435,0.001);
        \draw[thick,fill=white] (5,0.6) circle (0.35);
        \draw[thick,fill=white] (5,-0.6) circle (0.35);
        \quivbox{7}{0}{0.35}
        \node at (5,0.6) {\Large $\mathfrak{K}_0$};
        \node at (5,-0.6) {\Large $\mathfrak{K}_1$};
        \node at (7,0) {\Large $\mathfrak{N}$};
    \end{tikzpicture}
    \caption{On the left, the ADHM quiver for $k$ $U(N)$ instantons on $\mathbb{C}^2$, with $\mathfrak{K}\cong \mathbb{C}^k$ and $\mathfrak{N}\cong \mathbb{C}^N$. On the right, the ADHM quiver for $k_0+k_1$ $U(N)$ instantons on $\widehat{\mathbb{C}}^2$, with $\mathfrak{K}_{0,1} \cong \mathbb{C}^{k_{0,1}}$.}
    \label{fig:ADHMquiver}
\end{figure}
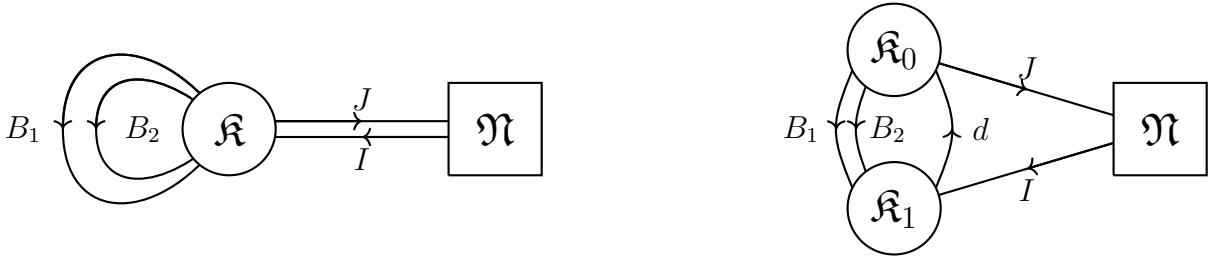

\paragraph{Instantons on $\mathbb{C}^2$}
We start by recalling the construction of the moduli space of $k$ instantons of $U(N)$ on $\mathbb{C}^2$ denoted by $\mathcal{M}_{N,k}$.
The ADHM construction allows to construct this moduli space as a quiver variety~\cite{Atiyah:1978ri,Nakajima:1994nid} with quiver depicted on the l.h.s. of~\figref{fig:ADHMquiver} and variety defined as:
\begin{align}\label{eq:MNk}
    \mathcal{M}_{N,k}(\zeta) := \left\{(B_1,B_2,I,J)\,\, | \, \mu_\mathbb{C} = 0 \,, \mu_{\mathbb{R}}=\zeta \mathds{1}  \right\} \big/U(k)\,,
\end{align}
with $\mathds{1}$ the $k$-dimensional identity matrix and $\mu_{\mathbb{C}}$, $\mu_{\mathbb{R}}$ the complex and real moment maps defined as:
\begin{align}
    \mu_{\mathbb{C}}=[B_1,B_2] + IJ\,, && \mu_{\mathbb{R}} =[B_1,B_1^\dagger] + [B_2,B_2^\dagger] + I I^\dagger -JJ^\dagger\,.
\end{align}
Furthermore, the $U(k)$ action defined as:
\begin{align}
    (B_1,B_2,I,J) \xrightarrow{g \in U(k)} (gB_1 g^{-1},gB_2 g^{-1},gI,Jg^{-1}) \,.
\end{align}
The real parameter $\zeta$ is introduced in \eqref{eq:MNk} as a regularisation parameter. This definition does not hinge on the exact value of $\zeta$, but is only sensitive to its sign. This leads to two distinct realisations, namely $\zeta >0 $ and $\zeta<0$. An equivalent realisation of this regularisation is so-called {\it (co-)stability} condition on the maps $(B_1,B_2,I,J)$~\cite{Nakajima:2003pg} which leads to the relations:
\begin{align}\label{eq:stabbd}
    &\mathfrak{K} = \mathbb{C}[B_1,B_2] I(\mathfrak{N)}\,, && \text{for} && \zeta>0\,,\nonumber\\
    &\mathfrak{K} = \mathbb{C}[B_1^\dagger,B_2^\dagger] J^\dagger (\mathfrak{N}), && \text{for} && \zeta<0\,,
\end{align}
where $\mathfrak{N}\cong \mathbb{C}^N$ is the framing vector space which encodes the constraints from the $U(N)$ gauge group and $\mathfrak{K}\cong \mathbb{C}^k$ is the instanton vector space modelling instantons of charge $k$.

\paragraph{Instantons on $\widehat{\mathbb{C}}^2$}
The moduli space of instantons on the blow-up can be similarly defined as a quiver variety~\cite{King,Nakajima:2008eq}.
This can be understood by considering an alternative realisation of $\widehat{\mathbb{C}}^2$ as a toric variety. The blow-up $\widehat{\mathbb{C}}^2$ corresponds to the total space ${\rm Tot}(\mathcal{O}(-1)\to \mathbb{P}^1)$ and can be realised as the following K\"ahler quotient:
\begin{align}
    \widehat{\mathbb{C}}^2 = \left\{(u,v,w)\in \mathbb{C}^3 \,|\, |u|^2 + |v|^2 - |w|^2 >0 \right\}/\left((u,v,w)\sim (e^{i\alpha} u,e^{i\alpha} v,e^{-i\alpha}w)\, ,\, e^{i\alpha} \in U(1)\right)\,,\label{DefKaehlerQuot}
\end{align}
where the exceptional $\mathbb{P}^1$ corresponds to $w=0$. In the ADHM construction, the space-time coordinates are promoted to maps on vector spaces; we promote the coordinates of the K\"ahler quotient following:
\begin{align}\label{eq:coorversmap}
    (u,v,w) \longrightarrow(B_1,B_2,d)\,.
\end{align}
As a consequence of the quotient $U(1)$, the maps $B_1$ and $B_2$ are raising the $U(1)$-charge by $+1$ while the $d$ map lowers the charge by $-1$.
In order to account for the K\"ahler quotient $U(1)$ charge, the instanton vector space must split into a pair $\mathfrak{K}_0\oplus\mathfrak{K}_1$ such that $\mathfrak{K}_0\cong \mathbb{C}^{k_0}$ and $\mathfrak{K}_1\cong \mathbb{C}^{k_1}$ with respective charge $0$ and $1$. The $B_1,B_2$ and $d$ maps are therefore homomorphism between these spaces:
\begin{align}
    B_1,B_2 \in {\rm Hom}(\mathfrak{K}_0,\mathfrak{K}_1)\,, && d\in {\rm Hom}(\mathfrak{K}_1,\mathfrak{K}_0)\,.
\end{align}
The coordinates on the K\"ahler quotient~\eqref{DefKaehlerQuot} are related to the coordinates of~\eqref{eq:coordBC2} by the reparametrisation:
\begin{align}\label{eq:reparam}
    (u,v,w) \longrightarrow (z_1=uw,z_2=vw),[u,v] \in \mathbb{C}^2 \times \mathbb{P}^1\,,
\end{align}
which verifies the relation $z_1v=z_2u$. In absence of a framing vector space, this relation is promoted, in the ADHM construction, to the complex moment map:
\begin{align}
    \mu_{\mathbb{C}}^{(0)}:= B_1 d B_2 - B_2 d B_1 \in \rm{Hom}(\mathfrak{K}_0,\mathfrak{K}_1)\,.
\end{align}
In addition to the instanton vector spaces $\mathfrak{K}_0$ and $\mathfrak{K}_1$, one should introduce the framing vector space $\mathfrak{N}$ which encodes the choice of gauge group for the instantons. It manifests itself as a choice of boundary conditions. The blow-up manifold is asymptotically $\mathbb{C}^2$ and the framing space is therefore not charged under the quotient $U(1)$ and the framing vector space $\mathfrak{N}\cong \mathbb{C}^N$ is in the trivial representation of the K\"ahler quotient $U(1)$. Similarly to the standard ADHM quiver, the framing is realised by a pair of maps $(I,J)$ which deforms the unframed complex moment map, the only possible choice which is compatible with the $U(1)$ charge assignment is:
\begin{align}
    I\in \rm{Hom}(\mathfrak{N},\mathfrak{K}_1)\,, && J \in \rm{Hom}(\mathfrak{K}_0,\mathfrak{N})\,,
\end{align}
with the complex moment map~\footnote{The moduli space of instantons on blow-up is not hyper-K\"ahler. Therefore the complex moment map does not take its origin from a quaternionic structure and should be interpreted instead as a relation on the quiver. We still denote this relation as the complex moment map by analogy with the blow-down case.}:
\begin{align}
    \mu_{\mathbb{C}}= B_1dB_2 - B_2 d B_1 + IJ\,.
\end{align}
The associated quiver is given by the r.h.s. of~\figref{fig:ADHMquiver} and the instanton moduli space on the blow-up is defined as:
\begin{align}
    \widehat{\mathcal{M}}_{N,k_0,k_1}(\zeta_0,\zeta_1) := \left\{ (B_1,B_2,d,I,J)\,\, | \, \mu_{\mathbb{C}}=0, \, \mu_{\mathbb{R},0} = \zeta_0 \mathds{1}, \, \mu_{\mathbb{R},1} = \zeta_1 \mathds{1} \right\} \big/ U(k_0) \times U(k_1)\,,\label{InstModSpace}
\end{align}
with the real regularisation parameters $\zeta_0$ and $\zeta_1$ and the real moment maps defined as:
\begin{subequations}
\begin{align}
    &\mu_{\mathbb{R},0} := -B_1^\dagger B_1 - B_2^\dagger B_2 + d d^\dagger -J^\dagger J\,,\\
    &\mu_{\mathbb{R},1} := B_1 B_1^\dagger + B_2 B_2^\dagger - d^\dagger d +I I^\dagger\,.
\end{align}
\end{subequations}
The dual gauge group $U(k_0)\times U(k_1)$ action is canonically defined as:
\begin{align}
    (B_1,B_2,d,I,J) \xrightarrow[]{(g_0,g_1)\in U(k_0)\times U(k_1)} (g_1 B_1 g_0^{-1},g_1 B_2 g_0^{-1},g_0 d g_1^{-1},g_1I,Jg_0^{-1})\,,
\end{align}
and the framing gauge group $U(N)$ action is:
\begin{align}
    (B_1,B_2,d,I,J) \xrightarrow{h\in U(N)} (B_1,B_2,d,Ih^{-1},hJ)\,.
\end{align}
So far, we have not introduced the $\Omega$-background action on the instanton moduli space and we should stress that the two instanton vector spaces do not correspond to configurations near the two fixed points of $\widehat{\mathbb{C}}^2$ under the $\Omega$-background. The $T$-action on the quiver maps can be read off the reparametrisation~\eqref{eq:reparam} and the map assignment~\eqref{eq:coorversmap} and is given by:
\begin{align}
    (B_1,B_2,d,I,J) \xrightarrow{(q_1,q_2)\in T} (q_1^{-1}B_1,q_2^{-1}B_2,d,q_{12}^{-1}I,J)\,,
\end{align}
with the convention $q_{12}:=q_1q_2$.

\subsection{Contour integral formula}\label{subsec:coutourintegral}

\paragraph{}
We are now interested in computing the equivariant volume of the instanton moduli space~\eqref{InstModSpace} at fixed $k_0,k_1$ and $N$,
\begin{align}
    Z_{N,k_0,k_1}^{\rm inst.} = \int_{\widehat{\mathcal{M}}_{N,k_0,k_1}} 1 \, .
\end{align}
This quantity corresponds physically to a particular sector of the instanton partition function of a $U(N)$ gauge theory on the blow-up. 
We formulate the problem of computing the partition function on $\widehat{\mathbb{C}}^2$ in terms of a contour integral. We start by going to a basis of the quiver representation in which the overall $U(N)\times U(k_0) \times U(k_1)$ group action is diagonal:
\begin{subequations}
\begin{align}
    B_1 = \bigoplus_{a=1}^{k_0} \bigoplus_{b=1}^{k_1} B_{1,ab}\,, &&  B_2 = \bigoplus_{a=1}^{k_0} \bigoplus_{b=1}^{k_1} B_{2,ab}\,, &&  d = \bigoplus_{a=1}^{k_0} \bigoplus_{b=1}^{k_1} d_{ab}\,,
\end{align}
\begin{align}
    J = \bigoplus_{a=1}^{k_0} \bigoplus_{\alpha=1}^N J_{\alpha,a}\,, && I = \bigoplus_{b=1}^{k_1} \bigoplus_{\alpha=1}^N I_{\alpha,b}\,, && \mu_{\mathbb{C}} = \bigoplus_{a=1}^{k_0} \bigoplus_{b=1}^{k_1} \mu_{\mathbb{C},ab}\,.
\end{align}
\end{subequations}
In this basis the $U(N)\times U(k_0) \times U(k_1)$-action reduces to:
\begin{align}
    &g_0 = \exp (2i\pi\phi^{(0)})\,, && \phi^{(0)} = {\rm diag}(\phi_1^{(0)},\ldots, \phi_{k_0}^{(0)})\,,\nonumber\\
    &g_1 = \exp (2i\pi\phi^{(1)})\,, && \phi^{(0)} = {\rm diag}(\phi_1^{(1)},\ldots, \phi_{k_1}^{(1)})\,,\\
    &h = \exp (2i\pi{\sf a})\,, && {\sf a}={\rm diag}({\sf a}_1,\ldots, {\sf a}_N)\,,\nonumber
\end{align}
which generates together with the $T$-action the $\widetilde{T}$-action:
\begin{align}\label{eq:ttilde}
    \widetilde{T} = U(1)^{k_0} \times U(1)^{k_1} \times U(1)^N \times T\,.
\end{align}
The weights of the ADHM variables and equations under the $\widetilde{T}$-action are given by:
\begin{align}
    w_{B_{1,ab}}^{\widetilde{T}} = \phi_b^{(1)}-\phi_a^{(0)} -\varepsilon_1\,, && w_{B_{2,ab}}^{\widetilde{T}} = \phi_b^{(1)}-\phi_a^{(0)} -\varepsilon_2\,, && w_{d_{ab}}^{\widetilde{T}} = \phi_{a}^{(0)} - \phi_b^{(1)}\,,
\end{align}
\begin{align}
    w_{J_{\alpha,a}}^{\widetilde{T}} = {\sf a}_{\alpha} - \phi^{(0)}_a\,, && w^{\widetilde{T}}_{I_{\alpha,b}} = -\phi^{(1)}_b + {\sf a}_\alpha -\varepsilon_{12}\,, && w^{\widetilde{T}}_{\mu_{\mathbb{C},ab}} = \phi^{(1)}_b - \phi^{(0)}_a - \varepsilon_{12}\,.
\end{align}
We obtain the following contour integral:
\begin{align}\label{eq:contour}
    Z_{N,k_0,k_1}^{\rm inst.} = \frac{1}{k_0! k_1!} \oint_{\mathrm{JK}_{\zeta_0,\zeta_1}} &\prod_{a=1}^{k_0} \frac{d \phi_a^{(0)}}{2i\pi}\prod_{b=1}^{k_1} \frac{d \phi_b^{(1)}}{2i\pi}  \frac{\prod_{a\neq a^\prime =1}^{k_0} (\phi_a^{(0)} - \phi_{a^\prime}^{(0)})\prod_{b\neq b^\prime =1}^{k_1} (\phi_b^{(1)} - \phi_{b^\prime}^{(1)})}{\prod_{a=1}^{k_0}P(\phi^{(1)}_a,{\sf a}) \prod_{b=1}^{k_1}\widetilde{P}(\phi^{(0)}_b,{\sf a})}\nonumber \\
    &\times\prod_{a=1}^{k_0} \prod_{b=1}^{k_1} \frac{(\phi^{(1)}_b - \phi^{(0)}_a - \varepsilon_{12})}{(\phi_b^{(1)}-\phi_a^{(0)} -\varepsilon_1)(\phi_b^{(1)}-\phi_a^{(0)} -\varepsilon_2)(\phi_{a}^{(0)} - \phi_b^{(1)})}\,,
\end{align}
where we defined:\modif{
\begin{align}
    P(\phi^{(0)}_a,{\sf a}) := \prod_{\alpha=1}^N (\phi^{(0)}_a -{\sf a}_{\alpha}) \,, && \widetilde{P}(\phi^{(1)}_b,{\sf a}) :=  \prod_{\alpha=1}^N (-\phi^{(1)}_b + {\sf a}_\alpha -\varepsilon_{12})\,,
\end{align}}
and we used the formula for the Haar measure of $U(k)$:
\begin{align}
    \frac{d \phi}{{\rm vol}\, U(k)} = \frac{1}{k!} \prod_{a=1}^k\frac{d \phi_a}{2i\pi} \prod_{a\neq b=1}^k (\phi_a - \phi_b)\,.
\end{align}
The contour integral formula~\eqref{eq:contour} is obtained by integrating over the dual gauge group $U(k_0)\times U(k_1)$ and it is schematically given by three types of contributions: the weights of the variables appearing in the denominator, the weights of the constraints and the integration group Haar measure both appearing in the numerator. The result of the integral depends on the choice of contour. We fix this choice using the Jeffrey-Kirwan (JK) residue prescription~\cite{Jeffrey:1993cun}. The choice of real regularisation parameters $(\zeta_0,\zeta_1)$ corresponds to a choice of stability vector $\vec \eta$ for the JK prescription. We will detail the prescription and the classification of contours in Section~\ref{sec:classification}.

\subsection{Vertex operator formalism}\label{subsec:freefield}

\paragraph{}
Before analysing the resulting partition functions in more details, we quickly discuss an alternative description of the instanton counting problem in terms of vertex operators based on the free field realisation, which allows to obtain the contour integral~\eqref{eq:contour} from a different point of view.  

\paragraph{}
The free-field algebra is expressed most naturally by considering a five dimensional uplift of our original setup by considering an interacting system of $D0$- and $D4$-branes on $\widehat{\mathbb{C}}^2 \times \mathbb{S}^1$. On the blow-down $\mathbb{C}^2\times \mathbb{S}^1$, the $D0$- and $D4$-branes are represented by the operators:
\begin{align}
    {\sf A}(z) = \,: \exp\left(\sum_{m\neq0} a_m z^m  \right)\!:\,, && {\sf X}(z) = \, : \exp \left( \sum_{m\neq0} x_m z^m \right):\,,
\end{align}
where $::$ denotes the normal ordering  and $a_m$, $x_m$ satisfy the free-field algebra:
\begin{align}
    [a_n,a_m] = \frac{1}{n}(1-q_1^n)(1-q_2^n)(1+q_{12}^n) \delta_{n+m,0}\,, && [x_n,a_m] = \frac{1}{n} \delta_{n+m,0}\,, &&[a_n,x_m] = \frac{q_{12}^{-n}}{n} \delta_{n+m,0}\,.
\end{align}
For the moment, we do not discuss the zero modes of the vertex operators for simplicity.
Using this representation, the blow-down partition function for $k$ $D0$-branes interacting with $N$ $D4$-branes is obtained by considering the correlation function~\cite{Kimura:2019hnw,Kimura:2022zsm,Kimura:2023bxy}:%
\footnote{We may similarly consider the partition function of the six dimensional gauge theory compactified on a torus based on the doubled free field formalism~\cite{Kimura:2016dys}.}
\begin{align}
    \mathcal{Z}_{k,N}^{\mathbb{C}^2\times\mathbb{S}^1} = \frac{1}{k!} \oint \prod_{I=1}^k\frac{\mathrm{d}\varphi_I}{2i\pi \varphi_I} \left\langle  {\sf A}^{-1}(\varphi_I) :\prod_{\alpha=1}^N {\sf X}({\sf a}^{\rm 5d}_\alpha): \right\rangle\,,
\end{align}
where the coordinate at which the $D4$-branes are inserted correspond to the five dimensional Coulomb branch moduli $\underline{\sf a}^{\rm 5d}$.

\paragraph{} On the blow-up the situation is slightly different since the exceptional curve $\mathcal{C}$ allows to include non-trivial $D2$-branes wrapping $\mathcal{C}\times \mathbb{S}^1 \subset \mathbb{C}^2\times \mathbb{S}^1$. The $D0$-branes can then melt on the $D2$-branes to form $D0$-$D2$ bound states. At the level of the free-field algebra, the bound states admit modified commutation relation compared to the free $D0$-branes, we therefore include two types of $D0$-like vertex operators:
\begin{align}
    {\sf A}_0(x) =\, :\exp \left(\sum_{m\neq0} a^{(0)}_n x^n\right)\!:\,, && {\sf A}_1(x) =\, :\exp \left(\sum_{m\neq0} a^{(1)}_n x^n\right)\!:\,, && {\sf X}(z) = \, : \exp \left( \sum_{m\neq0} x_m z^m \right):\,.
\end{align}
In order to be compatible with the original algebra, the generators must satisfy the three constraints:
\begin{subequations}\label{Constr}
\begin{align}&[a_n^{(0)},a_m^{(0)}]+[a_n^{(0)},a_m^{(1)}]+[a_n^{(1)},a_m^{(0)}]+[a_n^{(1)},a_m^{(1)}]=[a_n,a_m]\,,\label{Constr1}\\
& [a_n^{(0)},x_m]+[a_n^{(1)},x_m]=[a_n,x_m]\,,\\
& [x_n,a_m^{(0)}]+[x_n,a_m^{(1)}] = [x_n,a_m]\,,\label{Constr3}
\end{align}
\end{subequations}
which lead to the following set of commutation relations:
\begin{align}\label{eq:kappamatrix}
    [a^{(i)}_n,a^{(j)}_m]=\kappa^{ij}_{mn}\,, && \kappa_{mn} =  \frac{\delta_{n+m,0}}{n} (1+q_{12}^n) \begin{bmatrix}
        1 & -q_1^n -q_2^n +q_{12}^n \\ -1 & 1
    \end{bmatrix}\,.
\end{align}
This matrix $\kappa$ is interpreted as a deformation of the Cartan matrix of type $\widehat{A}_1$~\cite{Kimura:2015rgi}, which comes from the quiver structure as shown in Figure~\ref{fig:ADHMquiver}.
The constraints \eqref{Constr} imply that $\det \kappa_{mn} =[a_n,a_m]$ and: 
\begin{align}
    &[x_n,a^{(0)}_m] = [a_n^{(1)},x_m] = 0\,, &&[x_n,a_m^{(1)}] = \frac{1}{n} \delta_{n+m,0}\,, &&[a_n^{(0)},x_m] = \frac{q_{12}^{-n}}{n} \delta_{n+m,0}\,.
\end{align}
The five dimensional version of the contour integral~\eqref{eq:contour} can then be obtained by considering the correlation function:
\begin{align}
    \mathcal{Z}^{\widehat{\mathbb C}^2\times \mathbb{S}^1}_{k_0,k_1,N} = \frac{1}{k_0!k_1!} \oint_{\mathrm{JK}_{\zeta_0,\zeta_1}} \prod_{a=1}^{k_0}\frac{\dd \varphi_a^{(0)}}{2i\pi \varphi_a^{(0)}} \prod_{b=1}^{k_1} \frac{\dd \varphi_b^{(1)}}{2i\pi \varphi_b^{(1)}} \left\langle {\sf A}^{-1}_0(\varphi_a^{(0)}) {\sf A}_1^{-1} (\varphi_b^{(1)}) :\prod_{\alpha=1}^N {\sf X}({\sf a}^{\rm 5d}_\alpha): \right\rangle\,.
\end{align}
In four dimensions, the problem of counting $D0$-$D2$ bound states becomes a problem of counting instanton-monopole configurations. 

\section{Classification of chamber configurations}\label{sec:classification}

\paragraph{}
In this Section, we discuss the chamber structure associated with the instanton counting problem on $\widehat{\mathbb{C}}^2$. Using the JK residue prescription, we detail the combinatorics encoding the instanton-monopole configurations in each chambers and detail the general pattern of evolution as we cross the walls in the stability plane.

\subsection{Stability and chamber structure}\label{subsec:chambers}

\begin{figure}[tbp]
    \centering
    \begin{tikzpicture}
    \begin{scope}[shift={(10,-10)},scale=2.2]
            \node at (0,5.35) {$\zeta_1$};
            \node at (2.3,0) {$\zeta_0$};
            \draw[ultra thick,->] (-5,0) -- (2,0);
            \draw[ultra thick,->] (0,-2) -- (0,5);
            \draw[ultra thick] (-4.5,4.5) -- (1.7,-1.7);
            \node at (-4.8,4.8) {\small $\zeta_0+\zeta_1=0$};
            \draw[ultra thick] (-4.8,2.4) -- (0,0);
            \draw[ultra thick] (-4.75,3.1666) -- (0,0);
            \draw[ultra thick] (-4.73,3.5475) -- (0,0);
            \draw[ultra thick] (-4.71,3.76) -- (0,0);
            \draw[ultra thick] (-4.70,3.8916) -- (0,0);
            \draw[ultra thick] (-4.69,3.977) -- (0,0);
            \draw[ultra thick] (-4.68,4.03375) -- (0,0);
            \draw[fill=gray,fill opacity=0.2] (-4.68,4.03375) -- (0,0) --  (-4.5,4.5) -- (-4.68,4.03375);
            \draw[ultra thick] (-2.4,4.8) -- (0,0);
            \draw[ultra thick] (-3.1666,4.75) -- (0,0);
            \draw[ultra thick] (-3.5475,4.73) -- (0,0);
            \draw[ultra thick] (-3.76,4.71) -- (0,0);
            \draw[ultra thick] (-3.8916,4.70) -- (0,0);
            \draw[ultra thick] (-3.977,4.69) -- (0,0);
            \draw[ultra thick] (-4.03375,4.68) -- (0,0);
            \draw[fill=gray,fill opacity=0.2]
              (-4.03375,4.68) -- (0,0) -- (-4.5,4.5) -- (-4.03375,4.68);
            \node at (-2.5,-1) {\footnotesize $\mathcal{SP}^\vee$-chamber};
            \node at (1,2.5) {\footnotesize $\mathcal{SP}$-chamber};
            \node at (0.7,-1.5) {\footnotesize $\mathcal{P}^\vee$-chamber};
            \node at (1.3,-0.5) {\footnotesize $\mathcal{P}$-chamber};
            \node at (-3.5,0.9) {\footnotesize $1^\vee$-chamber};
            \node at (-0.9,3.5) {\footnotesize $1$-chamber};
            \node[rotate=-28] at (-4,2.4) {\footnotesize $2^\vee$-chamber};
            \node[rotate=-90+32] at (-2.33,4) {\footnotesize $2$-chamber};
        \end{scope}
    \end{tikzpicture}
    \caption{The space of stability conditions with its chamber structure, the gray area is a region where walls accumulate near the $\zeta_0+\zeta_1=0$ line.}
    \label{fig:chambers}
\end{figure}
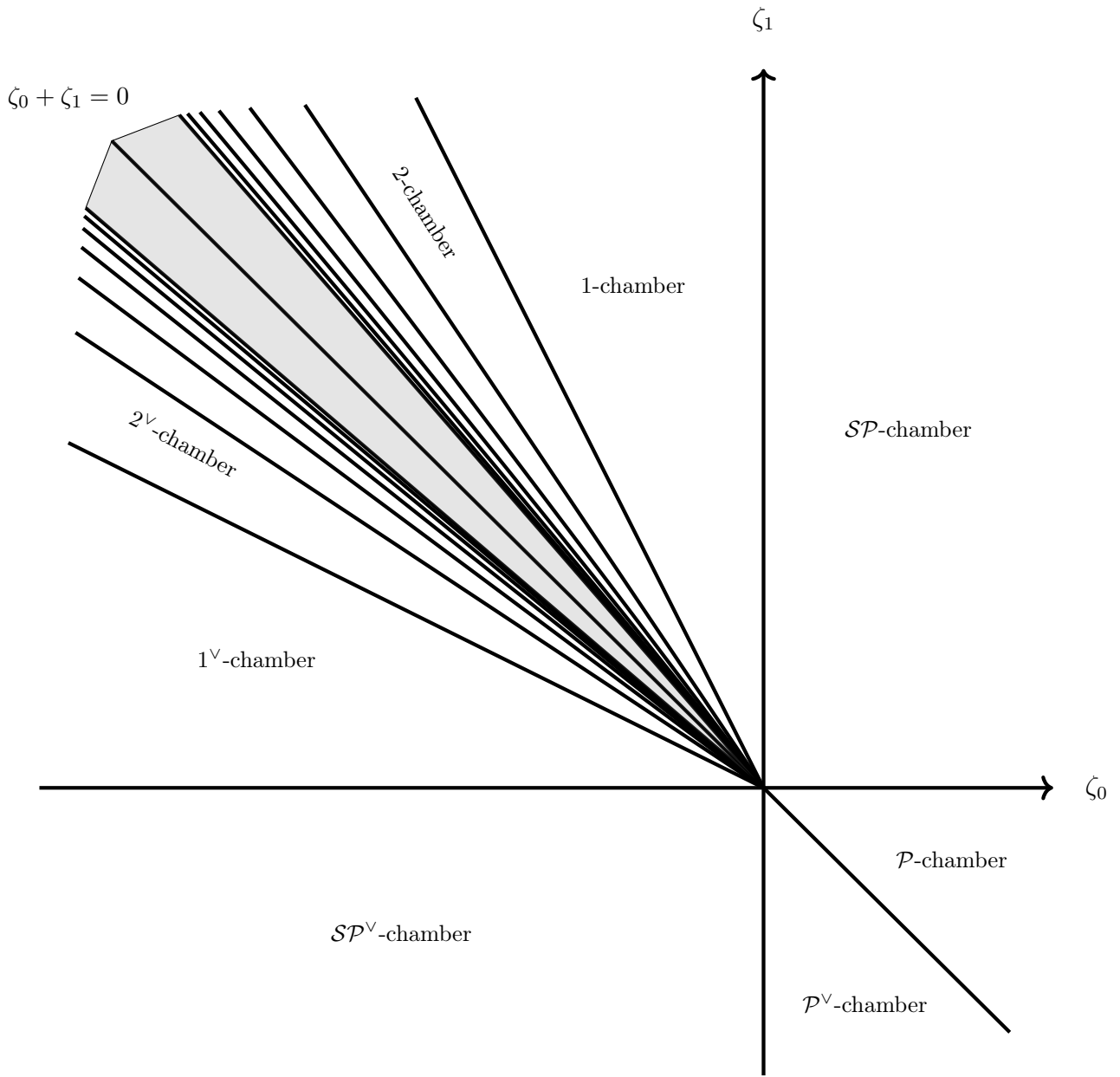

\paragraph{}
The stability conditions~\eqref{eq:stabbd} associated with the standard ADHM construction of the moduli space of instantons on $\mathbb{C}^2$ only depend on the sign of the real regularisation parameter $\zeta$ and lead to physically equivalent instanton partition functions. On the blow-up however these conditions generalises in the presence of two regularisation parameters as in (\ref{InstModSpace}). Indeed, these conditions exhibit a \emph{chamber} structure. To be precise, in the following we shall call a chamber as a connected region of the $(\zeta_0,\zeta_1)$-plane which admits the same instanton partition function. We now simply state the result which we obtained using the JK prescription, the chamber structure is given by~\figref{fig:chambers}.
We classify the different chambers in following way:
\begin{itemize}
    \item the $\mathcal{P}$-chamber: $\{(\zeta_0,\zeta_1) \in \mathbb{R}^2 \,|\, \zeta_0>\zeta_0+\zeta_1>0\}$, we show in Section~\ref{subsec:pchamber} that in this chamber the configurations organise into standard integer partitions, and the partition function reduces to the instanton partition function on the blow-down $\mathbb{C}^2$;
    \item the $\mathcal{SP}$-chamber: $\{(\zeta_0,\zeta_1) \in \mathbb{R}^2 \, |\, \zeta_1>0\,, \, \zeta_0>0\}$, we show in Section~\ref{subsec:spchamber} that configurations in this chamber are indexed by super-partitions which are partitions of half-integers;
    \item the $n$-chamber: in the region where $\zeta_1 > \zeta_1+\zeta_0 >0$ walls with slope $-(m+1)/m$ for $m\in \mathbb{N}^*$ passing through the $(0,0)$ point accumulate towards the $\zeta_0 + \zeta_1=0$ line. The $n$-chamber is the region between the wall with slope $-(n+1)/n$ and $-n/(n-1)$ for $n \geq 2$, the $1$-chamber lying between the slope $-2$ wall and the wall given by $\zeta_0=0$. The configurations in the $n$-chamber are indexed by $n$-stable super-partitions which we define in Section~\ref{subsec:nchamber};
    \item the Blow-up chamber or $\infty$-chamber: in the limit $n \to \infty$ the configurations restrict to a special class of super-partitions which we detail in Section~\ref{subsec:buchamber}. The partition function in this chamber factorises into two contributions associated with the fixed points $p_+$ and $p_-$ as given by~\figref{fig:blowupOmega} as expected from the blow-up formula~\cite{Nakajima:2003pg}. We detail this process in Section~\ref{subsec:blowupformula};
    \item the dual chambers denoted by $^\vee$ which lead to the same contributions as the original chambers up to $\varepsilon_{1,2}\to-\varepsilon_{1,2}$ and ${\sf a}_{\alpha} \to {\sf a}_\alpha - \varepsilon_{12}$.
\end{itemize} 

\subsection{Jeffrey-Kirwan residues on the blow-up}

\paragraph{}
We now discuss the process of computing the integral~\eqref{eq:contour} using the JK residues prescription~\cite{Jeffrey:1993cun}.
The integral~\eqref{eq:contour} admits poles of the form:
\begin{align}\label{eq:poles}
    \phi_a^{(0)}-\phi_b^{(1)}\,, && \phi_b^{(1)} - \phi_a^{(0)} - \varepsilon_{1,2}\,, && {\sf a}_\alpha-\phi_b^{(1)}-\varepsilon_{12}\,, && \phi_a^{(0)}-{\sf a}_\alpha\,, 
\end{align}
with $a\in \{1,\ldots,k_0\}$, $b\in \{1,\ldots,k_1\}$ and $\alpha \in \{1,\ldots,N\}$ arising from the weights of respectively the $d,B_{1,2},I$ and $J$ maps. A choice of contour corresponds to picking a set of $k_0+k_1$ poles. We encode such a choice by representing a pole by a charge vector with labels the charge in the integration variables of each pole of~\eqref{eq:poles}:
\begin{align}\label{eq:chargevector}
    e_a^{(0)}-e_b^{(1)}\,, && -e_a^{(0)} + e_b^{(1)}\,, && -e_b^{(1)}\,, && e_a^{(0)}\,,
\end{align}
with the convention
\begin{align}
    e_a^{(0)}:=(\overbrace{0,\ldots,0,1,0,\ldots,0}^{k_0 \text{ entries}}\,|\,\overbrace{0,\ldots,0}^{k_1 \text{ entries}}) \in \{0,1\}^{k_0+k_1}\,,
\end{align}
with a one in the $a$-th position of the first block and a second block with $k_1$ zeros and similarly
\begin{align}
    e_b^{(1)}:=(\overbrace{0,\ldots,0}^{k_0 \text{ entries}}\,|\,\overbrace{0,\ldots,0,1,0,\ldots,0}^{k_1 \text{ entries}}) \in \{0,1\}^{k_0+k_1}\,,
\end{align}
with a one in the $b$-th position of the second block and a first block with $k_0$ zeros. Using this notation a choice of $k_0+k_1$ poles picked corresponds to a square matrix of size $k_0+k_1$ which is a choice of $k_0+k_1$ charge vectors among~\eqref{eq:chargevector}:
\begin{align}
    V:=(v_1,\ldots,v_{k_0+k_1})\,.
\end{align}
A contour contributes to the partition function if $V$ is invertible. Denoting the cone generated by $V$ as:
\begin{align}
    {\rm Cone}(V) := \left\{\sum_{i=1}^{k_0+k_1} c_i v_i\quad \text{for} \quad c_i \geq 0 \right\}\,.
\end{align}
and the stability vector $\vec{\eta}$ as the following function of the regularisation parameters $\zeta_0$ and $\zeta_1$:
\begin{align}
    \vec \eta = (\zeta_0,\ldots,\zeta_0\,|\,\zeta_1,\ldots,\zeta_1) \in \mathbb{R}^{k_0+k_1}\,,
\end{align}
a contour contributes only when the stability vector $\vec{\eta}$ sits inside Cone$(V)$.
Furthermore, some contours are admissible but lead to vanishing residues due to the factors appearing in the integrand numerator of~\eqref{eq:contour} of the form:
\begin{align}\label{eq:haarandmom}
    \phi_a^{(0)}-\phi^{(0)}_{a^\prime}\,, && \phi_b^{(1)}-\phi^{(1)}_{b^\prime} \,, && \phi^{(1)}_b - \phi^{(0)}_a - \varepsilon_{12}\,.
\end{align}
The two first conditions arise from the Haar measures on the dual gauge groups while the third condition arise from the moment map. Actually, a necessary but not sufficient condition to have a non-vanishing contour is that it can pick only once any one of the poles of the form ${\sf a}_\alpha - \phi_b^{(1)}-\varepsilon_{12}$ or $\phi_a^{(0)}-{\sf a}_\alpha$ for $\alpha\in\{1,\ldots,N\}$. 

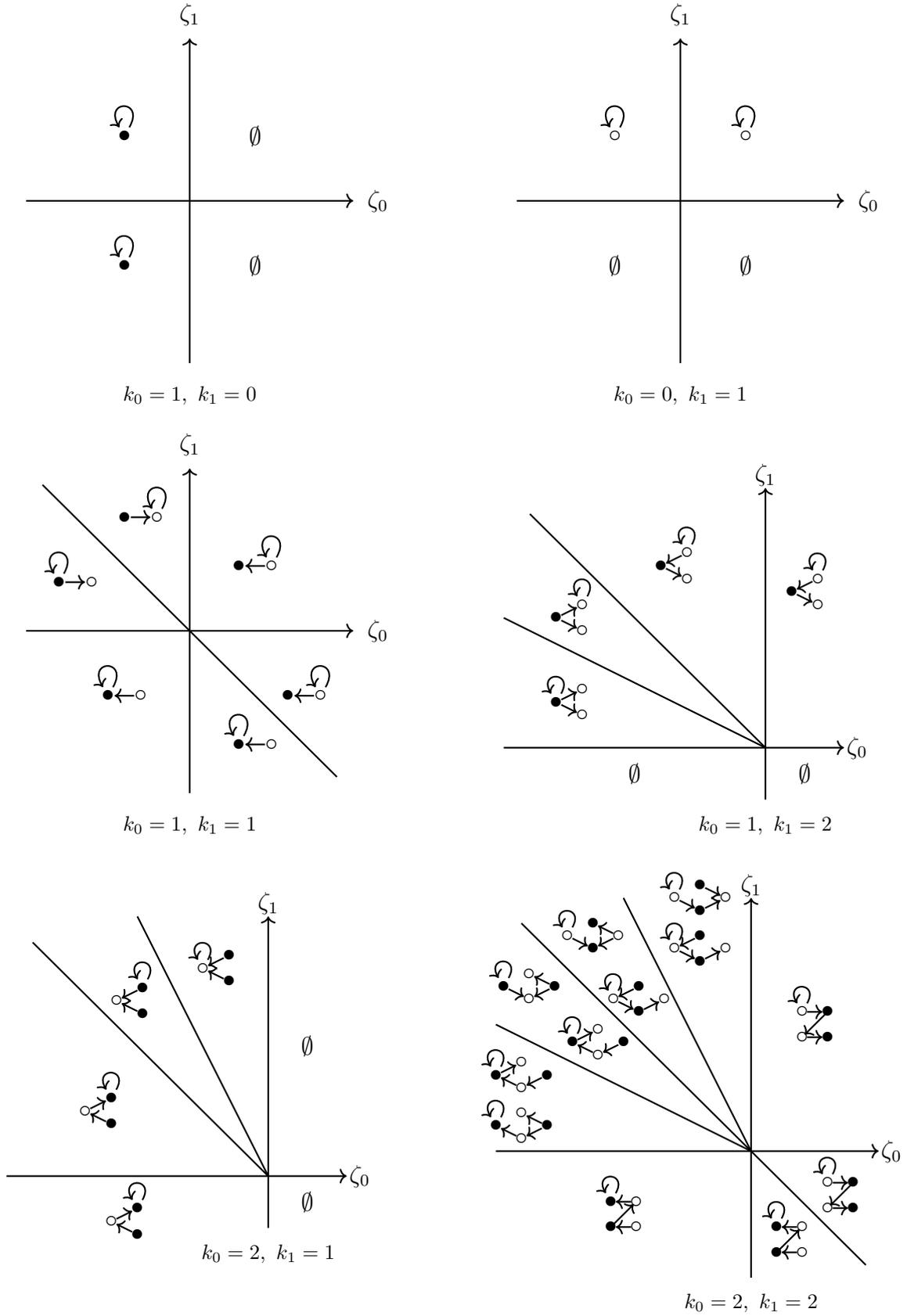
\begin{figure}[tbp]
    \centering
    \begin{tikzpicture}[scale=1.1]
        \begin{scope}[scale=0.25,shift={(0,-1.5)}]
            \draw[thick,->] (-10,0) -- (10,0);
            \draw[thick,->] (0,-10) -- (0,10);
            \node (v1) at (-4,-4) {$ $};
            \node at (-4,-4) {$\bullet$};
            \draw[->,thick] (v1) edge[out=60, in=120, looseness=8,loop,distance=20mm] (v1);
            \node (v2) at (-4,4) {$ $};
            \node at (-4,4) {$\bullet$};
            \draw[->,thick] (v2) edge[out=60, in=120, looseness=8,loop,distance=20mm] (v2);
            \node at (4,4) {$\emptyset$};
            \node at (4,-4) {$\emptyset$};
            \node at (0,-12) {\footnotesize $k_0=1,\,\,k_1=0$};
            \node at (11.5,0) {$\zeta_0$};
            \node at (0,11.5) {$\zeta_1$};
        \end{scope}
        \begin{scope}[scale=0.25,shift={(30,-1.5)}]
            \draw[thick,->] (-10,0) -- (10,0);
            \draw[thick,->] (0,-10) -- (0,10);
            \node (v1) at (4,4) {$ $};
            \node at (-4,-4) {$\emptyset$};
            \draw[->,thick] (v1) edge[out=60, in=120, looseness=8,loop,distance=20mm] (v1);
            \node (v2) at (-4,4) {$ $};
            \node at (-4,4) {$\circ$};
            \draw[->,thick] (v2) edge[out=60, in=120, looseness=8,loop,distance=20mm] (v2);
            \node at (4,4) {$\circ$};
            \node at (4,-4) {$\emptyset$};
            \node at (0,-12) {\footnotesize $k_0=0,\,\,k_1=1$};
            \node at (11.5,0) {$\zeta_0$};
            \node at (0,11.5) {$\zeta_1$};
        \end{scope}
        \begin{scope}[scale=0.25,shift={(0,-28)}]
            \draw[thick,->] (-10,0) -- (10,0);
            \draw[thick,->] (0,-10) -- (0,10);
            \draw[thick] (-9,9) -- (9,-9);
            \node at (0,-12) {\footnotesize $k_0=1,\,\,k_1=1$};
            \begin{scope}[shift={(-3,7)}]
                \node (v1) at (-5,-4) {$ $};
                \node at (-5,-4) {$\bullet$};
                \node (v2) at (-3,-4) {$\circ$};
                \draw[->,thick] (v1) edge[out=60, in=120, looseness=8,loop,distance=20mm] (v1);
                \draw[<-,thick] (-3.4,-4) -- (-4.6,-4);
            \end{scope}
            \begin{scope}[shift={(8,-3)}]
                \node (v1) at (-5,-4) {$ $};
                \node at (-5,-4) {$\bullet$};
                \node (v2) at (-3,-4) {$\circ$};
                \draw[->,thick] (v1) edge[out=60, in=120, looseness=8,loop,distance=20mm] (v1);
                \draw[->,thick] (-3.4,-4) -- (-4.6,-4);
            \end{scope}
            \node (v1) at (-5,-4) {$ $};
            \node at (-5,-4) {$\bullet$};
            \node (v2) at (-3,-4) {$\circ$};
            \draw[->,thick] (v1) edge[out=60, in=120, looseness=8,loop,distance=20mm] (v1);
            \draw[->,thick] (-3.4,-4) -- (-4.6,-4);
            
            \node (v1) at (5,4) {$ $};
            \node at (5,4) {$\circ$};
            \node (v2) at (3,4) {$\bullet$};
            \draw[->,thick] (v1) edge[out=60, in=120, looseness=8,loop,distance=20mm] (v1);
            \draw[<-,thick] (3.4,4) -- (4.6,4);
            \begin{scope}[shift={(-7,3)}]
                \node (v1) at (5,4) {$ $};
                \node at (5,4) {$\circ$};
                \node (v2) at (3,4) {$\bullet$};
                \draw[->,thick] (v1) edge[out=60, in=120, looseness=8,loop,distance=20mm] (v1);
                \draw[->,thick] (3.4,4) -- (4.6,4);
            \end{scope}
            \begin{scope}[shift={(3,-8)}]
                \node (v1) at (5,4) {$ $};
                \node at (5,4) {$\circ$};
                \node (v2) at (3,4) {$\bullet$};
                \draw[->,thick] (v1) edge[out=60, in=120, looseness=8,loop,distance=20mm] (v1);
                \draw[<-,thick] (3.4,4) -- (4.6,4);
            \end{scope}
            \node at (11.5,0) {$\zeta_0$};
            \node at (0,11.5) {$\zeta_1$};
        \end{scope}
        \begin{scope}[scale=0.40,shift={(22,-22)}]
            \draw[thick,->] (-10,0) -- (3,0);
            \draw[thick,->] (0,-2) -- (0,10);
            \draw[thick] (0,0) -- (-9,9);
            \draw[thick] (0,0) -- (-10,5);
            \node at (3.5,0) {$\zeta_0$};
            \node at (0,10.5) {$\zeta_1$};
            \node at (-5,-1) {$\emptyset$};
            \node at (1.5,-1) {$\emptyset$};
            \node at (0,-3) {\footnotesize $k_0=1,\,\,k_1=2$};
            \begin{scope}[shift={(0,0.75)}]
                \node (v1) at (-8,1) {$ $};
                \node (v2) at (-7,0.5) {$\circ$};
                \node (v3) at (-7,1.5) {$\circ$};
                \node at (-8,1) {$\bullet$};
                \draw[thick,->] (-7.8,1.1) -- (-7.2,1.4);
                \draw[thick,->] (-7.8,0.9) -- (-7.2,0.6);
                \draw[->,thick] (v1) edge[out=60, in=120, looseness=8,loop,distance=10mm] (v1);
            \end{scope}
            \begin{scope}[shift={(0,4)}]
                \node (v3) at (-7,1.5) {$ $};
                \node (v2) at (-7,0.5) {$\circ$};
                \node at (-7,1.5) {$\circ$};
                \node at (-8,1) {$\bullet$};
                \draw[thick,->] (-7.8,1.1) -- (-7.2,1.4);
                \draw[thick,->] (-7.8,0.9) -- (-7.2,0.6);
                \draw[->,thick] (v3) edge[out=60, in=120, looseness=8,loop,distance=10mm] (v3);
            \end{scope}
            \begin{scope}[shift={(4,6)}]
                \node (v3) at (-7,1.5) {$ $};
                \node (v2) at (-7,0.5) {$\circ$};
                \node at (-7,1.5) {$\circ$};
                \node at (-8,1) {$\bullet$};
                \draw[thick,<-] (-7.8,1.1) -- (-7.2,1.4);
                \draw[thick,->] (-7.8,0.9) -- (-7.2,0.6);
                \draw[->,thick] (v3) edge[out=60, in=120, looseness=8,loop,distance=10mm] (v3);
            \end{scope}
            \begin{scope}[shift={(9,5)}]
                \node (v3) at (-7,1.5) {$ $};
                \node (v2) at (-7,0.5) {$\circ$};
                \node at (-7,1.5) {$\circ$};
                \node at (-8,1) {$\bullet$};
                \draw[thick,<-] (-7.8,1.1) -- (-7.2,1.4);
                \draw[thick,->] (-7.8,0.9) -- (-7.2,0.6);
                \draw[->,thick] (v3) edge[out=60, in=120, looseness=8,loop,distance=10mm] (v3);
            \end{scope}
        \end{scope}
        \begin{scope}[scale=0.4,shift={(3,-38.5)}]
            \draw[thick,->] (-10,0) -- (3,0);
            \draw[thick,->] (0,-2) -- (0,10);
            \draw[thick] (0,0) -- (-9,9);
            \draw[thick] (0,0) -- (-5,10);
            \begin{scope}[shift={(1,1.5)}]
                \node (v3) at (-7,1.5) {$ $};
                \node (v2) at (-7,0.5) {$\bullet$};
                \node at (-7,1.5) {$\bullet$};
                \node at (-8,1) {$\circ$};
                \draw[thick,->] (-7.8,1.1) -- (-7.2,1.4);
                \draw[thick,<-] (-7.8,0.9) -- (-7.2,0.6);
                \draw[->,thick] (v3) edge[out=60, in=120, looseness=8,loop,distance=10mm] (v3);
            \end{scope}
            \begin{scope}[shift={(2,-2.75)}]
                \node (v3) at (-7,1.5) {$ $};
                \node (v2) at (-7,0.5) {$\bullet$};
                \node at (-7,1.5) {$\bullet$};
                \node at (-8,1) {$\circ$};
                \draw[thick,->] (-7.8,1.1) -- (-7.2,1.4);
                \draw[thick,<-] (-7.8,0.9) -- (-7.2,0.6);
                \draw[->,thick] (v3) edge[out=60, in=120, looseness=8,loop,distance=10mm] (v3);
            \end{scope}
            \begin{scope}[shift={(2.2,5.75)}]
                \node (v3) at (-7,1.5) {$ $};
                \node (v2) at (-7,0.5) {$\bullet$};
                \node at (-7,1.5) {$\bullet$};
                \node at (-8,1) {$\circ$};
                \draw[thick,<-] (-7.8,1.1) -- (-7.2,1.4);
                \draw[thick,<-] (-7.8,0.9) -- (-7.2,0.6);
                \draw[->,thick] (v3) edge[out=60, in=120, looseness=8,loop,distance=10mm] (v3);
            \end{scope}
            \begin{scope}[shift={(5.5,7)}]
                \node (v3) at (-8,1) {$ $};
                \node (v2) at (-7,0.5) {$\bullet$};
                \node at (-7,1.5) {$\bullet$};
                \node at (-8,1) {$\circ$};
                \draw[thick,<-] (-7.8,1.1) -- (-7.2,1.4);
                \draw[thick,<-] (-7.8,0.9) -- (-7.2,0.6);
                \draw[->,thick] (v3) edge[out=60, in=120, looseness=8,loop,distance=10mm] (v3);
            \end{scope}
            \node at (1.5,5) {$\emptyset$};
            \node at (1.5,-1) {$\emptyset$};
            \node at (0,-3) {\footnotesize $k_0=2,\,\,k_1=1$};
            \node at (3.5,0) {$\zeta_0$};
            \node at (0,10.5) {$\zeta_1$};
        \end{scope}
        \begin{scope}[scale=0.39,shift={(22,-38.5)}]
            \draw[thick,->] (-10,0) -- (5,0);
            \draw[thick,->] (0,-5) -- (0,10);
            \draw[thick] (0,0) -- (-10,5);
            \draw[thick] (4.5,-4.5) -- (-9,9);
            \draw[thick] (0,0) -- (-5,10);
            \node at (0,-6) {\footnotesize $k_0=2,\,\,k_1=2$};
            \begin{scope}[shift={(-5.5,-2)}]
                \node (v1) at (0,0) {$ $};
                \node at (0,0) {$\bullet$};
                \draw[->,thick] (v1) edge[out=60, in=120, looseness=8,loop,distance=10mm] (v1);
                \node at (1,0) {$\circ$};
                \node at (0,-1) {$\bullet$};
                \node at (1,-1) {$\circ$};
                \draw[thick, <-] (0.2,-1) -- (0.8,-1);
                \draw[thick, ->] (0.1,-0.9) -- (0.9,-0.1);
                \draw[thick, <-] (0.2,0) -- (0.8,0);
            \end{scope}
            \begin{scope}[shift={(1,-3)}]
                \node (v1) at (0,0) {$ $};
                \node at (0,0) {$\bullet$};
                \draw[->,thick] (v1) edge[out=60, in=120, looseness=8,loop,distance=10mm] (v1);
                \node at (1,0) {$\circ$};
                \node at (0,-1) {$\bullet$};
                \node at (1,-1) {$\circ$};
                \draw[thick, <-] (0.2,-1) -- (0.8,-1);
                \draw[thick, ->] (0.1,-0.9) -- (0.9,-0.1);
                \draw[thick, <-] (0.2,0) -- (0.8,0);
            \end{scope}
            \begin{scope}[shift={(3,-1.25)}]
                \node (v1) at (0,0) {$ $};
                \node at (0,0) {$\circ$};
                \draw[->,thick] (v1) edge[out=60, in=120, looseness=8,loop,distance=10mm] (v1);
                \node at (1,0) {$\bullet$};
                \node at (0,-1) {$\circ$};
                \node at (1,-1) {$\bullet$};
                \draw[thick, ->] (0.2,-1) -- (0.8,-1);
                \draw[thick, <-] (0.1,-0.9) -- (0.9,-0.1);
                \draw[thick, ->] (0.2,0) -- (0.8,0);
            \end{scope}
            \begin{scope}[shift={(2,5.5)}]
                \node (v1) at (0,0) {$ $};
                \node at (0,0) {$\circ$};
                \draw[->,thick] (v1) edge[out=60, in=120, looseness=8,loop,distance=10mm] (v1);
                \node at (1,0) {$\bullet$};
                \node at (0,-1) {$\circ$};
                \node at (1,-1) {$\bullet$};
                \draw[thick, ->] (0.2,-1) -- (0.8,-1);
                \draw[thick, <-] (0.1,-0.9) -- (0.9,-0.1);
                \draw[thick, ->] (0.2,0) -- (0.8,0);
            \end{scope}
            \begin{scope}[shift={(-10,3)}]
                \node (v1) at (0,0) {$ $};
                \draw[->,thick] (v1) edge[out=60, in=120, looseness=8,loop,distance=10mm] (v1);
                \node at (0,0) {$\bullet$};
                \node at (1,0.5) {$\circ$};
                \node at (1,-0.5) {$\circ$};
                \node at (2,0) {$\bullet$};
                \draw[thick,->] (0.2,0.1) -- (0.8,0.4);
                \draw[thick,<-] (0.2,-0.1) -- (0.8,-0.4);
                \draw[thick,<-] (1.2,-0.4) -- (1.8,-0.1);
            \end{scope}
            \begin{scope}[shift={(-10,1)}]
                \node (v1) at (0,0) {$ $};
                \draw[->,thick] (v1) edge[out=60, in=120, looseness=8,loop,distance=10mm] (v1);
                \node at (0,0) {$\bullet$};
                \node at (1,0.5) {$\circ$};
                \node at (1,-0.5) {$\circ$};
                \node at (2,0) {$\bullet$};
                \draw[thick,<-] (1.2,0.4) -- (1.8,0.1);
                \draw[thick,<-] (0.2,-0.1) -- (0.8,-0.4);
                \draw[thick,<-] (1.2,-0.4) -- (1.8,-0.1);
            \end{scope}
            \begin{scope}[shift={(-9.7,6.5)}]
                \node (v1) at (0,0) {$ $};
                \draw[->,thick] (v1) edge[out=60, in=120, looseness=8,loop,distance=10mm] (v1);
                \node at (0,0) {$\bullet$};
                \node at (1,0.5) {$\circ$};
                \node at (1,-0.5) {$\circ$};
                \node at (2,0) {$\bullet$};
                \draw[thick,<-] (1.2,0.4) -- (1.8,0.1);
                \draw[thick,->] (0.2,-0.1) -- (0.8,-0.4);
                \draw[thick,<-] (1.2,-0.4) -- (1.8,-0.1);
            \end{scope}
            \begin{scope}[shift={(-7,4.3)}]
                \node (v1) at (0,0) {$ $};
                \draw[->,thick] (v1) edge[out=60, in=120, looseness=8,loop,distance=10mm] (v1);
                \node at (0,0) {$\bullet$};
                \node at (1,0.5) {$\circ$};
                \node at (1,-0.5) {$\circ$};
                \node at (2,0) {$\bullet$};
                \draw[thick,->] (0.2,0.1) -- (0.8,0.4);
                \draw[thick,<-] (0.2,-0.1) -- (0.8,-0.4);
                \draw[thick,<-] (1.2,-0.4) -- (1.8,-0.1);
            \end{scope}
            \begin{scope}[shift={(-3,8)}]
                \node (v1) at (0,0) {$ $};
                \draw[->,thick] (v1) edge[out=60, in=120, looseness=8,loop,distance=10mm] (v1);
                \node at (0,0) {$\circ$};
                \node at (1,0.5) {$\bullet$};
                \node at (1,-0.5) {$\bullet$};
                \node at (2,0) {$\circ$};
                \draw[thick,<-] (0.2,0.1) -- (0.8,0.4);
                \draw[thick,->] (0.2,-0.1) -- (0.8,-0.4);
                \draw[thick,->] (1.2,-0.4) -- (1.8,-0.1);
            \end{scope}
            \begin{scope}[shift={(-3,10)}]
                \node (v1) at (0,0) {$ $};
                \draw[->,thick] (v1) edge[out=60, in=120, looseness=8,loop,distance=10mm] (v1);
                \node at (0,0) {$\circ$};
                \node at (1,0.5) {$\bullet$};
                \node at (1,-0.5) {$\bullet$};
                \node at (2,0) {$\circ$};
                \draw[thick,->] (1.2,0.4) -- (1.8,0.1);
                \draw[thick,->] (0.2,-0.1) -- (0.8,-0.4);
                \draw[thick,->] (1.2,-0.4) -- (1.8,-0.1);
            \end{scope}
            \begin{scope}[shift={(-7.2,8.5)}]
                \node (v1) at (0,0) {$ $};
                \draw[->,thick] (v1) edge[out=60, in=120, looseness=8,loop,distance=10mm] (v1);
                \node at (0,0) {$\circ$};
                \node at (1,0.5) {$\bullet$};
                \node at (1,-0.5) {$\bullet$};
                \node at (2,0) {$\circ$};
                \draw[thick,<-] (1.2,0.4) -- (1.8,0.1);
                \draw[thick,->] (0.2,-0.1) -- (0.8,-0.4);
                \draw[thick,<-] (1.2,-0.4) -- (1.8,-0.1);
            \end{scope}
            \begin{scope}[shift={(-5.4,6)}]
                \node (v1) at (0,0) {$ $};
                \draw[->,thick] (v1) edge[out=60, in=120, looseness=8,loop,distance=10mm] (v1);
                \node at (0,0) {$\circ$};
                \node at (1,0.5) {$\bullet$};
                \node at (1,-0.5) {$\bullet$};
                \node at (2,0) {$\circ$};
                \draw[thick,<-] (0.2,0.1) -- (0.8,0.4);
                \draw[thick,->] (0.2,-0.1) -- (0.8,-0.4);
                \draw[thick,->] (1.2,-0.4) -- (1.8,-0.1);
            \end{scope}
            \node at (5.5,0) {$\zeta_0$};
            \node at (0,10.5) {$\zeta_1$};
        \end{scope}
    \end{tikzpicture}
    \caption{Poles picked by the JK prescription for $k_0,k_1=0,1,2$ as functions of the stability parameters $\zeta_{0}$ and $\zeta_1$.}
    \label{fig:JKblow1}
\end{figure}

\paragraph{}
A convenient way to represent the sequence of poles picked by the JK prescription on the blow-up is to use an oriented bipartite graph consisting of black and white nodes connected by arrows. Concretely, we have
\begin{itemize}
\item $k_0$ black nodes labelled by $a\in \{1,\ldots ,k_0 \}$ 
\item $k_1$ white nodes labelled by $b\in \{1,\ldots ,k_1 \}$
\item the charge vector $e_a^{(0)}-e_b^{(1)}$ is represented by an arrow leaving the $b$-th white node and joining the $a$-th black node
\item the charge vector $-e_a^{(0)}+e_b^{(1)}$ is associated with an arrow leaving the $a$-th black node and joining the $b$-th white node 
\item charge vectors $e_a^{(0)}$ are represented by arrows starting and ending on the $a$-th black, while $-e_b^{(1)}$ are represented by arrows starting and ending on the $b$-th white node (`self-loops')
\end{itemize}
A simple example for such a bi-partite graph for $k_0=k_1=2$ is shown below:
\begin{align}
    V= \big(e_1^{(0)},-e_1^{(0)}+e_1^{(1)},-e_1^{(0)}+e_2^{(1)},-e_2^{(0)}+e_2^{(1)}\big) && \longrightarrow && 
    \begin{tikzpicture}[scale=0.8,baseline=(current bounding box.center)]
    \begin{scope}
        \node (v1) at (0,0) {$ $};
        \draw[->,thick] (v1) edge[out=60, in=120, looseness=8,loop,distance=10mm] (v1);
        \node at (0,0) {$\bullet$};
        \node at (1,0.5) {$\circ$};
        \node at (1,-0.5) {$\circ$};
        \node at (2,0) {$\bullet$};
        \draw[thick,->] (0.2,0.1) -- (0.8,0.4);
        \draw[thick,->] (0.2,-0.1) -- (0.8,-0.4);
        \draw[thick,<-] (1.2,-0.4) -- (1.8,-0.1);
        \node at (-0.2,-0.2) {\tiny $1$};
        \node at (1,0.8) {\tiny $1$};
        \node at (1,-0.8) {\tiny $2$};
        \node at (2.2,-0.2) {\tiny $2$};
    \end{scope}
    \end{tikzpicture}
\end{align}
For better readability, we have explicitly labelled the nodes of the graph, which we shall refrain from doing in the following. Imposing the Haar measure condition, one obtains a very specific type of graphs which admit $N$ connected components labelled by the Coulomb moduli ${\sf a}_\alpha$ and each component is a tree-like graph. In order to completely characterise these trees it is sufficient to consider the case $N=1$ since in higher rank cases the same contributions distributed within the different components appear. All non-vanishing admissible bipartite graphs appearing for $k_0,k_1 \leq 2$ are given by~\figref{fig:JKblow1}. There is no contribution with $(k_0,k_1)=(2,0)$ or $(k_0,k_1)=(0,2)$ since it is not possible to draw a connected graph for these values. 
Using this representation, the configurations in neighbouring chambers can be characterised and compared across walls.

\paragraph{}
Importantly, all admissible contours from the perspective of the JK prescription are not stable contributions given the quiver~\figref{fig:ADHMquiver}. Indeed, one should also impose that all configurations appearing in a chamber are holomorphic (or anti-holomorphic) in the $B_1$ and $B_2$ maps. More concretely, to each choice of set of fixed points one can associate a chain of maps which construct a sub-vector space of the instanton vector space. If the arrows appearing in the bipartite tree are not pointing in the same direction in each branch of the tree, then the chain of maps mixes $B_1,B_2,d$ and $B_1^\dagger,B_2^\dagger,d^\dagger$, and the constructed sub-vector space is non-holomorphic in the maps. Such configurations are therefore not stable.

\paragraph{}
In the region where $\zeta_0<0$ and $\zeta_1>0$, all the different graphs can be related by a sequence of moves: flipping arrows or moving self-loops. Given the set of graphs in a chamber, the set of graphs obtained in the dual chamber (as defined in Section~\ref{subsec:chambers}) can be obtained by flipping all arrows, exchanging white nodes and black nodes and exchanging $k_0$ and $k_1$. 

\subsection{Fixed points as super-partitions}

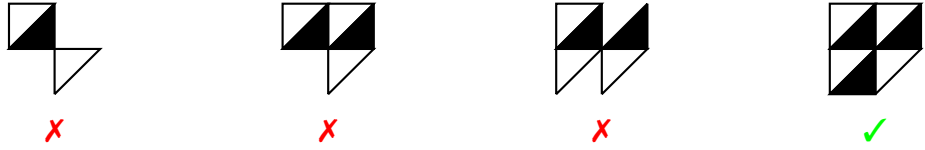
\begin{figure}[tbp]
    \centering
    \begin{tikzpicture}
        \begin{scope}[scale=0.6]
            \wtrbox{0}{0}
            \btrbox{0}{0}\wtrbox{1}{-1}
            \node at (0.5,-2.3) {{\color{red} \ding{55}}};
        \end{scope}
        \begin{scope}[scale=0.6,shift={(6,0)}]
            \btrbox{0}{0}\wtrbox{0}{0}\wtrbox{1}{-1}
            \btrbox{1}{0}\wtrbox{1}{0}
            \node at (0.5,-2.3) {{\color{red} \ding{55}}};
        \end{scope}
        \begin{scope}[scale=0.6,shift={(12,0)}]
            \btrbox{0}{0}\wtrbox{0}{0}\wtrbox{1}{-1}
            \btrbox{1}{0}\wtrbox{0}{-1}
            \node at (0.5,-2.3) {{\color{red} \ding{55}}};
        \end{scope}
        \begin{scope}[scale=0.6,shift={(18,0)}]
            \btrbox{0}{0}\wtrbox{0}{0}\wtrbox{1}{-1}
            \btrbox{1}{0}\wtrbox{1}{0}
            \btrbox{0}{-1}\wtrbox{0}{-1}
            \node at (0.5,-2.3) {{\color{green} \ding{51}}};
        \end{scope}
    \end{tikzpicture}
    \caption{Configurations cancelled by the complex moment map constraint are indicated by {\color{red} \ding{55}} while preserved configurations are indicated by {\color{green} \ding{51}}.}
    \label{fig:mucpart}
\end{figure}

\paragraph{From graphs to super-partitions}
While the bipartite graph are convenient to describe the choice of poles and chamber dependence, they do not allow to directly compute the instanton partition function because of the ambiguity between $B_1$ and $B_2$. Indeed, a single graph can lead to several distinct choices of poles. In addition, we did not include the moment map condition arising from~\eqref{eq:haarandmom} which forces certain valid (as previously defined) graphs to yield vanishing residues in the contour integral. A more suitable description is given by the characters of the instanton spaces $\mathfrak{K}_0$ and $\mathfrak{K}_1$ near the fixed points of the $\widetilde{T}$-action~\eqref{eq:ttilde}. The characters are defined as:
\begin{align}\label{eq:defk01}
    {\bf K}_0 :={\rm ch}\, \mathfrak{K}_0=\sum_{a=1}^{k_0} e^{2i\pi \phi_a^{(0)}}\,, && {\bf K}_1:={\rm ch} \, \mathfrak{K}_1 = \sum_{b=1}^{k_1} e^{2i\pi \phi_b^{(1)}}\,.
\end{align}
Each fixed point is associated with a choice of pole picking. The resulting explicit expression for $\phi_a^{(0)}$ and $\phi_b^{(1)}$ in terms of the variables ${\sf a}_\alpha$, $\varepsilon_1$ and $\varepsilon_2$ gives the character near the associated fixed point. As a consequence of the splitting in $N$ components of a typical expression takes the form:
\begin{align}
    \mathbf{K}_{0}\big|_{\rm fix. } = \sum_{\alpha=1}^N \sum_{(n,m)\in \lambda_\alpha^{(0)}} e^{2i\pi({\sf a}_\alpha + n\varepsilon_1+m\varepsilon_2)}\,,&& \mathbf{K}_{1}\big|_{\rm fix. } = \sum_{\alpha=1}^N \sum_{(n,m)\in \lambda_\alpha^{(1)}} e^{2i\pi({\sf a}_\alpha + n\varepsilon_1+m\varepsilon_2)}\,,
\end{align}
where $\lambda_\alpha^{(0)}$ and $\lambda_\alpha^{(1)}$ are partitions, represented as finite subsets of $\mathbb{Z}^2$ such that $\sum_{\alpha=1}^N |\lambda_\alpha^{(i)}|=k_{i}$ for $i = 1,2$. We use a graphical depiction for these partitions by associating to each element  $(n,m)\in \lambda^{(0)}_\alpha$ a black triangle \begin{tikzpicture}[scale=0.4]\btrbox{0}{0}
\end{tikzpicture} with length-1 smaller edge and bottom-right corner at position $(n+1,m+1)$. Similarly to each $(n,m)\in \lambda_\alpha^{(1)}$ we associate a white triangle of same size \begin{tikzpicture}[scale=0.4]\wtrbox{0}{0}
\end{tikzpicture} with top-left corner at coordinate $(n,m)$. If we have both $(n,m)\in \lambda^{(0)}_\alpha$ and $(n,m)\in \lambda^{(1)}_\alpha$ we obtain a full box \begin{tikzpicture}[scale=0.4]\wtrbox{0}{0}\btrbox{0}{0}
\end{tikzpicture} which can be thought as the usual boxes appearing in a Young diagram. As an example we consider a graph appearing in the $\mathcal{SP}$-chamber with $k_0=1$ and $k_1=3$:
\begin{align}
\begin{tikzpicture}[scale=1,baseline=(current bounding box.center)]
    \begin{scope}
        \node (v1) at (0,0) {$ $};
        \draw[->,thick] (v1) edge[out=60, in=120, looseness=8,loop,distance=10mm] (v1);
        \node at (0,0) {$\circ$};
        \node at (1,0) {$\bullet$};
        \node at (2,0.6) {$\circ$};
        \node at (2,-0.6) {$\circ$};
        \draw[thick,->] (0.15,0) -- (0.85,0);
        \draw[thick,->] (1.15,0.1) -- (1.9,0.55);
        \draw[thick,->] (1.15,-0.1) -- (1.9,-0.55);
        \node at (0.45,-0.25) {\tiny $d$};
        \node at (1.35,0.48) {\tiny $B_{1,2}$};
        \node at (1.35,-0.45) {\tiny $B_{1,2}$};
    \end{scope}
\end{tikzpicture}
&&\longrightarrow&&
    \begin{tikzpicture}[scale=0.5,baseline=(current bounding box.center)]
        \wtrbox{0}{0}\btrbox{0}{0}\wtrbox{1}{0}\wtrbox{0}{-1}
        \node at (-0.5,0.5) {{\color{red}{$\bullet$}}};
    \end{tikzpicture} && \longrightarrow && \begin{cases}
        \lambda^{(0)}= \ydi{1}\\
        \lambda^{(1)}= \ydi{2,1}
    \end{cases},
\end{align}
with the red dot indicating the point with coordinates $(0,0)$ and to match the convention of Young diagrams we use a frame with the vertical axis pointing downward. In that situation, we obtain the partitions $\lambda^{(0)}=\{1\}$ and $\lambda^{(1)}=\{2,1\}$. We should stress at this point that there exists only one possibility as a consequence of the constraints imposed by the Haar measure which restricts that the choice of maps $(d,B_1,B_1)$ and $(d,B_2,B_2)$ both lead to a vanishing contribution. Using this graphical depiction, one can also read off the constraint arising from the complex moment map~\eqref{eq:haarandmom} which cancels configurations with a black and a white triangle in a diagonal position with missing triangles in between. Examples of such configurations are given by \figref{fig:mucpart}.
All these constraints force the graphical depiction of the partitions $\lambda^{(0)}_\alpha$ and $\lambda^{(1)}_\alpha$ to collectively take the form of {\it super-partitions}. 

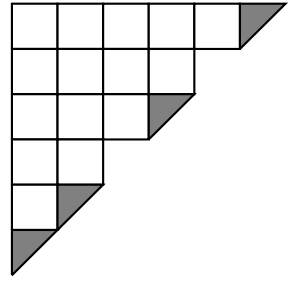
\begin{figure}[tbp]
    \centering
    \begin{tikzpicture}
        \node at (0,0) {$\Lambda = \{11/2,4,7/2,2,3/2,1/2\}$};
        \begin{scope}[shift={(5.5,1.8)},scale=0.6]
            \sqbox{0}{0}\sqbox{1}{0}\sqbox{2}{0}\sqbox{3}{0}\sqbox{4}{0}\trbox{5}{0}
            \sqbox{0}{-1}\sqbox{1}{-1}\sqbox{2}{-1}\sqbox{3}{-1}
            \sqbox{0}{-2}\sqbox{1}{-2}\sqbox{2}{-2}\trbox{3}{-2}
            \sqbox{0}{-3}\sqbox{1}{-3}
            \sqbox{0}{-4}\trbox{1}{-4}
            \trbox{0}{-5}
        \end{scope}
    \end{tikzpicture}
    \caption{An example of a super-partition and its graphical representation as a super-Young diagram.}
    \label{fig:superYoungdiagram}
\end{figure}

\paragraph{Super-partitions}
A super-partition $\Lambda\in\mathcal{SP}$ ($\mathcal{SP}$ denoting the set of all super-partitions) is defined as a set of positive integers and half-integers $\Lambda=(\Lambda_1,\ldots,\Lambda_l )$ such that $\Lambda_1 \geq \Lambda_2\geq \ldots \geq \Lambda_l$ and such that the decreasing sub-sequence formed only by half-integers is strictly decreasing. A super-partition can represented graphically by a super-Young diagram in which integers are depicted by lines of boxes while half-integers are depicted by line of boxes ending on a triangle. An example of a super-Young diagram is given by~\figref{fig:superYoungdiagram}.
From a combinatorial point of view, we can also describe a super-partition $\Lambda$ as a pair of partitions $(\lambda^{(0)},\lambda^{(1)})$ such that $\lambda^{(0)} \subseteq \lambda^{(1)} \subseteq \lambda^{(0)} \cup \partial_+ \lambda^{(0)}$ where $\partial_+ \lambda^{(0)}$ denotes the positive boundary of the partition $\lambda^{(0)}$, i.e. the set of positions where boxes can be added to the partition $\lambda^{(0)}$. In terms of super-Young diagram $\lambda^{(0)}$ is the partition formed only by the boxes while $\lambda^{(1)}$ is formed by both triangles and boxes. Given a super-partition $\Lambda$, we denote $\partial \Lambda$ the set of triangle boxes appearing in its super-Young diagram. 

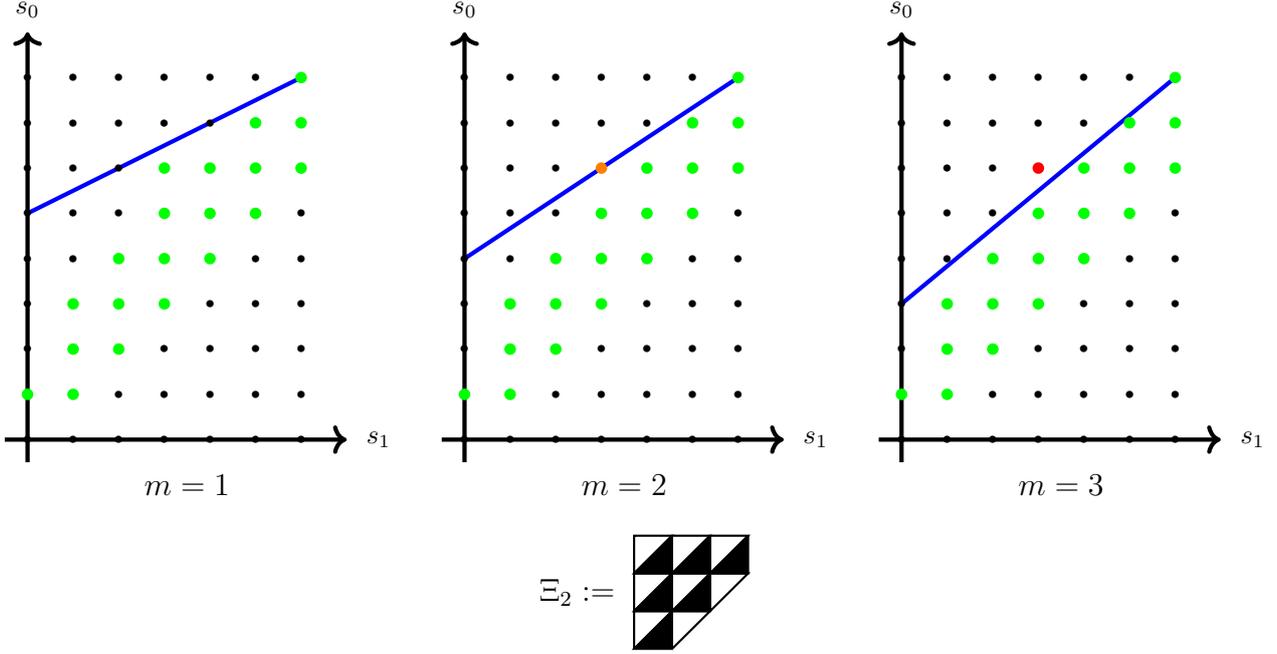
\begin{figure}[tbp]
    \centering
    \begin{tikzpicture}
    \begin{scope}[scale=0.6]
        \draw[ultra thick,blue] (0,5) -- (6,8);
        \draw[ultra thick,->] (-0.5,0) -- (7,0);
        \draw[ultra thick,->] (0,-0.5) -- (0,9);
        \node at (7.7,0) {\footnotesize $s_1$};
        \node at (0,9.5) {\footnotesize $s_0$};
        \node at (3.5,-1) {$m=1$};
        \foreach \i in {0,...,6}{
        \foreach \j in {0,...,8}{
            \node at (\i,\j) {\tiny $\bullet$};
            }
        }
        \stab{0}{1}\stab{1}{1}\stab{1}{2}\stab{2}{2}\stab{2}{3}\stab{3}{3}\stab{1}{3}\stab{2}{4}\stab{3}{4}\stab{4}{4}\stab{3}{5}\stab{4}{5}\stab{5}{5}\stab{5}{6}\stab{3}{6}\stab{4}{6}\stab{5}{7}\stab{6}{6}\stab{6}{7}\stab{6}{8}
    \end{scope}
    \begin{scope}[shift={(5.75,0)},scale=0.6]
        \draw[ultra thick,blue] (0,4) -- (6,8);
        \draw[ultra thick,->] (-0.5,0) -- (7,0);
        \draw[ultra thick,->] (0,-0.5) -- (0,9);
        \node at (7.7,0) {\footnotesize $s_1$};
        \node at (0,9.5) {\footnotesize $s_0$};
        \node at (3.5,-1) {$m=2$};
        \foreach \i in {0,...,6}{
        \foreach \j in {0,...,8}{
            \node at (\i,\j) {\tiny $\bullet$};
            }
        }
        \stab{0}{1}\stab{1}{1}\stab{1}{2}\stab{2}{2}\stab{2}{3}\stab{3}{3}\stab{1}{3}\stab{2}{4}\stab{3}{4}\stab{4}{4}\stab{3}{5}\stab{4}{5}\stab{5}{5}\stab{5}{6}\sstab{3}{6}\stab{4}{6}\stab{5}{7}\stab{6}{6}\stab{6}{7}\stab{6}{8}    
    \end{scope}
    \begin{scope}[shift={(11.5,0)},scale=0.6]
        \draw[ultra thick,blue] (0,3) -- (6,8);
        \draw[ultra thick,->] (-0.5,0) -- (7,0);
        \draw[ultra thick,->] (0,-0.5) -- (0,9);
        \node at (7.7,0) {\footnotesize $s_1$};
        \node at (0,9.5) {\footnotesize $s_0$};
        \node at (3.5,-1) {$m=3$};
        \foreach \i in {0,...,6}{
        \foreach \j in {0,...,8}{
            \node at (\i,\j) {\tiny $\bullet$};
            }
        }
        \stab{0}{1}\stab{1}{1}\stab{1}{2}\stab{2}{2}\stab{2}{3}\stab{3}{3}\stab{1}{3}\stab{2}{4}\stab{3}{4}\stab{4}{4}\stab{3}{5}\stab{4}{5}\stab{5}{5}\stab{5}{6}\istab{3}{6}\stab{4}{6}\stab{5}{7}\stab{6}{6}\stab{6}{7}\stab{6}{8}   
    \end{scope}
    \end{tikzpicture}
    \begin{center}
        \begin{tikzpicture}[scale=0.5]
            \wtrbox{0}{0}\btrbox{0}{0}\wtrbox{1}{0}\wtrbox{2}{0}\btrbox{1}{0}\btrbox{2}{0}
            \wtrbox{0}{-1}\btrbox{0}{-1}
            \wtrbox{0}{-2}\btrbox{0}{-2}
            \wtrbox{1}{-1}\btrbox{1}{-1}
            \wtrbox{1}{-2}
            \wtrbox{2}{-1}
            \node at (-2,-1) {$\Xi_2:=$};
        \end{tikzpicture}
    \end{center}
    \caption{The green dots denote a pair $(s_0,s_1)$ of dimensions associated with a sub-super-partition of $\Xi_2$ which satisfies~\eqref{eq:stabcond} strictly, the orange dots denote a similar pair which is an equality case of~\eqref{eq:stabcond}, while the red dots violate the inequality; the blue line corresponds to the condition~\eqref{eq:stabcond} for different $m$.}
    \label{fig:recursivestab}
\end{figure}

\paragraph{Stability for super-partitions}
We have imposed the stability conditions the real moments maps~\eqref{InstModSpace} and the JK residue prescription, one can translate such condition directly at the level of super-partitions.
The main idea is to consider a super-partition as the result of a recursive process of adding black and white triangle during which two different process are in competition:
\begin{enumerate}
    \item Adding (when possible) a black triangle \begin{tikzpicture}[scale=0.4]\btrbox{0}{0}
    \end{tikzpicture}
    to form a full box \begin{tikzpicture}[scale=0.4]\wtrbox{0}{0}\btrbox{0}{0}
    \end{tikzpicture}, which reduces $\dim\mathfrak{S}_1/\mathfrak{S}_0$ for $(\mathfrak{S}_{0} ,\mathfrak{S}_1)\subseteq (\mathfrak{K}_{0},\mathfrak{K}_1)$ some sub-vector spaces of the instanton vector spaces since one gets rid of a white triangle \begin{tikzpicture}[scale=0.4]\wtrbox{0}{0}
    \end{tikzpicture} in the process.
    \item Adding (when possible) a white triangle \begin{tikzpicture}[scale=0.4]\wtrbox{0}{0}
    \end{tikzpicture} to the super-partitions, which increases $\dim \mathfrak{S}_1/\mathfrak{S}_0$.
\end{enumerate}
The balance between these two process is encoded in the stability condition. As detailed in ~\cite{Nakajima:2008eq}, any non-trivial sub-vector spaces $(\mathfrak{S}_{0} ,\mathfrak{S}_1)$ of $(\mathfrak{K}_{0},\mathfrak{K}_1)$  which can be described combinatorially by a super-partition, must satisfy the condition:
\begin{align}\label{eq:stabcond}
    \zeta_0 \dim \mathfrak{S}_0 + \zeta_1 \dim \mathfrak{S}_1 \leq \zeta_0 \dim \mathfrak{K}_0 + \zeta_1\dim \mathfrak{K}_1\,.
\end{align}
More concretely, this is a condition on the pair of dimensions $(s_0,s_1):=(\dim\mathfrak{S}_0,\dim \mathfrak{S}_1)$ which are admissible when recursively constructing the super-partition. Such pairs of integers are depicted in \figref{fig:recursivestab} for a concrete example.

\subsection{Specific chambers}

\subsubsection{$\mathcal{P}$-chamber}\label{subsec:pchamber}

\begin{figure}[tbp]
    \centering
    \begin{tikzpicture}
        \node (v1) at (0,0) {$ $};
        \node at (0,0) {$\circ$};
        \node (v2) at (1,0) {$ $};
        \node (v3) at (2,0.6) {$ $};
        \node (v4) at (2,-0.6) {$ $};
        \node (v5) at (3,0.6) {$ $};
        \node (v6) at (3,-0.6) {$ $};
        \node (v7) at (4,1.2) {$ $};
        \node (v8) at (4,0) {$ $};
        \node (v9) at (5,1.2) {$ $};
        \node (v10) at (5,0) {$ $};
        \node at (1,0) {$\bullet$};
        \node at (2,0.6) {$\circ$};
        \node at (2,-0.6) {$\circ$};
        \node at (3,-0.6) {$\bullet$};
        \node at (3,0.6) {$\bullet$};
        \node at (4,1.2) {$\circ$};
        \node at (4,0) {$\circ$};
        \node at (5,0) {$\bullet$};
        \node at (5,1.2) {$\bullet$};
        \draw[->,thick] (v1) edge[out=60, in=120, looseness=8,loop,distance=10mm] (v1);
        \draw[->,thick] (v1) -- (v2);
        \draw[->,thick] (v2) -- (v3);
        \draw[->,thick] (v3) -- (v5);
        \draw[->,thick] (v5) -- (v7);
        \draw[->,thick] (v7) -- (v9);
        \draw[->,thick] (v2) -- (v4);
        \draw[->,thick] (v4) -- (v6);
        \draw[->,thick] (v5) -- (v8);
        \draw[->,thick] (v8) -- (v10);
        \draw[rounded corners=0.3cm, thick,dashed] 
        (-0.2, -0.25) rectangle (1.2, 0.25);
        \draw[rounded corners=0.3cm, thick,dashed] 
        (-0.2+2, -0.25+0.6) rectangle (1.2+2, 0.25+0.6);
        \draw[rounded corners=0.3cm, thick,dashed] 
        (-0.2+2, -0.25-0.6) rectangle (1.2+2, 0.25-0.6);
        \draw[rounded corners=0.3cm, thick,dashed] 
        (-0.2+4, -0.25+1.2) rectangle (1.2+4, 0.25+1.2);
        \draw[rounded corners=0.3cm, thick,dashed] 
        (-0.2+4, -0.25+0) rectangle (1.2+4, 0.25+0);
    \end{tikzpicture}
    \caption{A tree graph in the $\mathcal{P}$-chamber, all nodes can be grouped in pair of one black node and one white node.}
    \label{fig:treeP}
\end{figure}
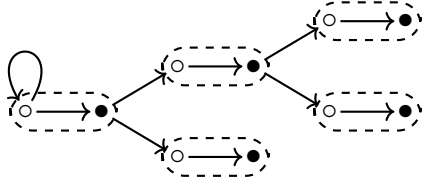

\paragraph{}
We start by discussing the $\mathcal{P}$-chamber, given by the domain $\{(\zeta_0,\zeta_1) \in \mathbb{R}^2 \,|\, \zeta_0>\zeta_0+\zeta_1>0\}$. In that situation, the bipartite trees restrict to a very specific type which satisfies the property that any branch of the tree will have an even number of arrows going in the same direction. This property implies that $k_0=k_1$ and we can indeed verify that all graphs with $k_0\neq k_1$ are unstable in the $\mathcal{P}$-chamber.
Indeed, each branch of such tree can be associated with a sequence of maps of the form $(d,B_{1,2},d,B_{1,2},\ldots , d,B_{1,2})$. Graphically this guarantees that each white triangle is paired with a black triangle at the same position. An example of such graph is given in~\figref{fig:treeP}.
The moment map constraint previously described enforces that the boxes arrange following a Young diagram (we denote the set of all Young diagram $\mathcal{P}$). In this chamber, the two instanton spaces are identical $\mathfrak{K}_0=\mathfrak{K}_1=\mathfrak{K}$ and the fixed points are indexed by partitions. Therefore, the instanton counting problem reduces to the standard counting problem on $\mathbb{C}^2$ given by the quiver on the l.h.s. of~\figref{fig:ADHMquiver}. The stability condition in the $\mathcal{P}$-chamber matches the stability condition for the $\mathbb{C}^2$ case~\eqref{eq:stabbd}:
\begin{align}
    \mathfrak{K}=\mathbb{C}[\widetilde{B}_1:=B_1d,\widetilde{B}_2:=B_2d]I(\mathfrak{N})\,.
\end{align}
The instanton vector space can be constructed using the maps $\widetilde{B}_1$ and $\widetilde{B}_2$ which corresponds as the geometric coordinate to $z_1$ and $z_2$~\eqref{eq:reparam} the coordinates on the original $\mathbb{C}^2$. We recall that the generating function of Young diagrams, interpreted as the partition function of 4d $U(1)$ $\mathcal{N}=4$ theory on $\mathbb{C}^2$, is given by:
\begin{align}\label{eq:ZP}
    Z_{\mathcal{P}}(\Qt) = \sum_{\lambda \in \mathcal{P}} \Qt^{|\lambda|} =(\Qt;\Qt)_\infty^{-1}\,,
\end{align}
where $\Qt$ is a formal counting parameter with $|Q_\tau| < 1$ and $(a;q)_n$ the Pochhammer symbol is defined as:
\begin{align}
    (a;q)_n := \prod_{k=0}^{n-1}(1-a q^k)\,.
\end{align}

\subsubsection{$\mathcal{SP}$-chamber}\label{subsec:spchamber}

\begin{figure}[tbp]
    \centering
    \begin{tikzpicture}
        \node (v1) at (0,0) {$ $};
        \node at (0,0) {$\circ$};
        \node (v2) at (1,0) {$ $};
        \node (v3) at (2,0.6) {$ $};
        \node (v4) at (2,-0.6) {$ $};
        \node (v5) at (3,0.6) {$ $};
        \node (v6) at (3,-0.6) {$ $};
        \node (v7) at (4,1.2) {$ $};
        \node (v8) at (4,0) {$ $};
        \node (v9) at (5,1.2) {$ $};
        \node (v10) at (5,0) {$ $};
        \node (v11) at (6,1.2) {$ $};
        \node (v12) at (6,0.6) {$ $};
        \node (v13) at (6,-0.6) {$ $};
        \node at (1,0) {$\bullet$};
        \node at (2,0.6) {$\circ$};
        \node at (2,-0.6) {$\circ$};
        \node at (3,-0.6) {$\bullet$};
        \node at (3,0.6) {$\bullet$};
        \node at (4,1.2) {$\circ$};
        \node at (4,0) {$\circ$};
        \node at (5,0) {$\bullet$};
        \node at (5,1.2) {$\bullet$};
        \node at (6,1.2) {$\circ$};
        \node at (6,0.6) {$\circ$};
        \node at (6,-0.6) {$\circ$};
        \draw[->,thick] (v1) edge[out=60, in=120, looseness=8,loop,distance=10mm] (v1);
        \draw[->,thick] (v1) -- (v2);
        \draw[->,thick] (v2) -- (v3);
        \draw[->,thick] (v3) -- (v5);
        \draw[->,thick] (v5) -- (v7);
        \draw[->,thick] (v7) -- (v9);
        \draw[->,thick] (v2) -- (v4);
        \draw[->,thick] (v4) -- (v6);
        \draw[->,thick] (v5) -- (v8);
        \draw[->,thick] (v8) -- (v10);
        \draw[->,thick] (v9) -- (v11);
        \draw[->,thick] (v10) -- (v12);
        \draw[->,thick] (v10) -- (v13);
        \draw[rounded corners=0.3cm, thick,dashed] 
        (-0.2, -0.25) rectangle (1.2, 0.25);
        \draw[rounded corners=0.3cm, thick,dashed] 
        (-0.2+2, -0.25+0.6) rectangle (1.2+2, 0.25+0.6);
        \draw[rounded corners=0.3cm, thick,dashed] 
        (-0.2+2, -0.25-0.6) rectangle (1.2+2, 0.25-0.6);
        \draw[rounded corners=0.3cm, thick,dashed] 
        (-0.2+4, -0.25+1.2) rectangle (1.2+4, 0.25+1.2);
        \draw[rounded corners=0.3cm, thick,dashed] 
        (-0.2+4, -0.25+0) rectangle (1.2+4, 0.25+0);
    \end{tikzpicture}
    \caption{A bipartite graph in the $\mathcal{SP}$-chamber differing from a graph in the $\mathcal{P}$-chamber at the branches edges.}
    \label{fig:treeSP}
\end{figure}
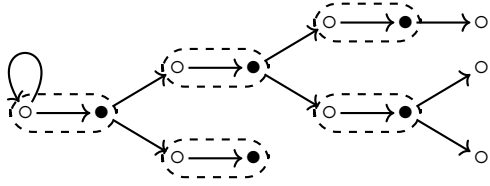

\paragraph{}
The $\mathcal{SP}$-chamber is given by the stability domain $\{(\zeta_0,\zeta_1) \in \mathbb{R}^2 \, |\, \zeta_1>0\,, \, \zeta_0>0\}$. Similarly to the $\mathcal{P}$-chamber, the stable bipartite trees appearing in this chamber all have arrows pointing in the same direction, however the number of maps on each branch is not necessarily even and this chamber will admit stable configurations for $k_0\neq k_1$ but no configurations for $k_0>k_1$. In practice all branches will be either of the form $(d,B_{1,2},d,B_{1,2},\ldots,d,B_{1,2})$ or $(d,B_{1,2},d,B_{1,2},\ldots,d)$.
The bipartite tree will differ from the one appearing in the $\mathcal{P}$-chamber only at the extremity of each branch. Graphically the super-Young diagram differs from a Young diagram on its boundary by $k_1-k_0$ white triangles.
An example of such graph is given in~\figref{fig:treeSP}.
These configurations correspond to super-Young diagrams and we give all such configurations for $k_0,k_1\leq4$ in Table~\ref{tab:superchamber}.
\begin{table}[tbp]
    \centering
    \begin{tabular}{c|c|c|c}
        $(k_0,k_1)$ &  & $(k_0,k_1)$ & \\
        \hline \hline
        $(0,1)$ & \begin{tikzpicture}[scale=0.4,baseline=(current bounding box.center)]
            \wtrbox{0}{0}
        \end{tikzpicture} & $(2,3)$ & \begin{tikzpicture}[scale=0.4,baseline=(current bounding box.center)]
            \node at (0,0.5) {$ $};
            \wtrbox{0}{0}\btrbox{0}{0} \wtrbox{0}{-2} \wtrbox{0}{-1}\btrbox{0}{-1}
            \node at (0,-2.5) {$ $};
        \end{tikzpicture}\quad\begin{tikzpicture}[scale=0.4,baseline=(current bounding box.center)]
            \wtrbox{0}{0}\btrbox{0}{0} \wtrbox{1}{0} \wtrbox{0}{-1}\btrbox{0}{-1}
        \end{tikzpicture}\quad\begin{tikzpicture}[scale=0.4,baseline=(current bounding box.center)]
            \wtrbox{0}{0}\btrbox{0}{0} \wtrbox{1}{0} \wtrbox{0}{-1}\btrbox{1}{0}
        \end{tikzpicture}\quad\begin{tikzpicture}[scale=0.4,baseline=(current bounding box.center)]
            \wtrbox{0}{0}\btrbox{0}{0} \wtrbox{1}{0} \wtrbox{2}{0}\btrbox{1}{0}
        \end{tikzpicture} \\
        \hline
        $(1,1)$ & \begin{tikzpicture}[scale=0.4,baseline=(current bounding box.center)]
            \wtrbox{0}{0}\btrbox{0}{0}
        \end{tikzpicture} & $(3,3)$ & \begin{tikzpicture}[scale=0.4,baseline=(current bounding box.center)]
            \wtrbox{0}{0}\btrbox{0}{0} \wtrbox{1}{0} \btrbox{1}{0} \wtrbox{2}{0} \btrbox{2}{0}
        \end{tikzpicture}\quad\begin{tikzpicture}[scale=0.4,baseline=(current bounding box.center)]
            \node at (0,0.5) {$ $};
            \wtrbox{0}{0}\btrbox{0}{0} \wtrbox{0}{-1} \btrbox{0}{-1} \wtrbox{0}{-2} \btrbox{0}{-2}
            \node at (0,-2.5) {$ $};
        \end{tikzpicture}\quad\begin{tikzpicture}[scale=0.4,baseline=(current bounding box.center)]
            \wtrbox{0}{0}\btrbox{0}{0} \wtrbox{0}{-1} \btrbox{0}{-1} \btrbox{1}{0} \wtrbox{1}{0}
        \end{tikzpicture} \\
        \hline
        $(1,2)$ & \begin{tikzpicture}[scale=0.4,baseline=(current bounding box.center)]
            \wtrbox{0}{0}\btrbox{0}{0} \wtrbox{1}{0} 
        \end{tikzpicture}\quad\begin{tikzpicture}[scale=0.4,baseline=(current bounding box.center)]
            \wtrbox{0}{0}\btrbox{0}{0} \wtrbox{0}{-1}
        \end{tikzpicture} & $(2,4)$ & \begin{tikzpicture}[scale=0.4,baseline=(current bounding box.center)]
            \node at (0,0.5) {$ $};
            \wtrbox{0}{0}\btrbox{0}{0} \wtrbox{1}{0} \wtrbox{0}{-1}\btrbox{0}{-1} \wtrbox{0}{-2}
            \node at (0,-2.5) {$ $};
        \end{tikzpicture}\quad\begin{tikzpicture}[scale=0.4,baseline=(current bounding box.center)]
            \node at (0,0.5) {$ $};
            \wtrbox{0}{0}\btrbox{0}{0} \wtrbox{1}{0} \btrbox{1}{0} \wtrbox{2}{0}\wtrbox{0}{-1}
        \end{tikzpicture} \\
        \hline
        $(2,2)$ & \begin{tikzpicture}[scale=0.4,baseline=(current bounding box.center)]
            \node at (0,0.5) {$ $};
            \wtrbox{0}{0}\btrbox{0}{0} \wtrbox{1}{0} \btrbox{1}{0}
        \end{tikzpicture}\quad\begin{tikzpicture}[scale=0.4,baseline=(current bounding box.center)]
            \node at (0,0.5) {$ $};
            \wtrbox{0}{0}\btrbox{0}{0} \wtrbox{0}{-1}\btrbox{0}{-1}
        \end{tikzpicture} & $(3,4)$ & \begin{tikzpicture}[scale=0.4,baseline=(current bounding box.center)]
            \node at (0,0.5) {$ $};
            \wtrbox{0}{0}\btrbox{0}{0} \wtrbox{1}{0} \btrbox{1}{0} \wtrbox{0}{-1}\btrbox{0}{-1} \wtrbox{0}{-2}
            \node at (0,-2.5) {$ $};
        \end{tikzpicture}\quad \begin{tikzpicture}[scale=0.4,baseline=(current bounding box.center)]
            \node at (0,0.5) {$ $};
            \wtrbox{0}{0}\btrbox{0}{0} \wtrbox{1}{0} \btrbox{1}{0} \wtrbox{0}{-1}\btrbox{0}{-1} \wtrbox{1}{-1}
            \node at (0,-1.5) {$ $};
        \end{tikzpicture}\quad\begin{tikzpicture}[scale=0.4,baseline=(current bounding box.center)]
            \node at (0,0.5) {$ $};
            \wtrbox{0}{0}\btrbox{0}{0} \wtrbox{1}{0} \btrbox{1}{0} \wtrbox{2}{0} \btrbox{2}{0} \wtrbox{3}{0}
        \end{tikzpicture} + tr.  \\
        \hline
        $(1,3)$ & \begin{tikzpicture}[scale=0.4,baseline=(current bounding box.center)]
            \node at (0,0.5) {$ $};
            \wtrbox{0}{0}\btrbox{0}{0} \wtrbox{1}{0} \wtrbox{0}{-1}
        \end{tikzpicture} & $(4,4)$ & \begin{tikzpicture}[scale=0.4,baseline=(current bounding box.center)]
            \node at (0,0.5) {$ $};
            \wtrbox{0}{0}\btrbox{0}{0} \wtrbox{1}{0} \btrbox{1}{0} \wtrbox{2}{0} \btrbox{2}{0} \wtrbox{0}{-1} \btrbox{0}{-1}
        \end{tikzpicture}\quad\begin{tikzpicture}[scale=0.4,baseline=(current bounding box.center)]
            \node at (0,0.5) {$ $};
            \wtrbox{0}{0}\btrbox{0}{0} \wtrbox{1}{0} \btrbox{1}{0} \wtrbox{1}{-1} \btrbox{1}{-1} \wtrbox{0}{-1} \btrbox{0}{-1}
            \node at (0,-1.5) {$ $};
        \end{tikzpicture}\quad\begin{tikzpicture}[scale=0.4,baseline=(current bounding box.center)]
            \node at (0,0.5) {$ $};
            \wtrbox{0}{0}\btrbox{0}{0} \wtrbox{1}{0} \btrbox{1}{0} \wtrbox{2}{0} \btrbox{2}{0} \wtrbox{3}{0} \btrbox{3}{0}
        \end{tikzpicture} + tr. \\
    \end{tabular}
    \caption{Configurations of super-partitions in the $\mathcal{SP}$-chamber with the + tr. indicating additional transposed partitions, all other cases for $k_{0,1}\leq4$ admit no configurations.}
    \label{tab:superchamber}
\end{table}
In this chamber, the stability condition can be written in the following way:
\begin{align}
    \mathfrak{K}_0 = \mathbb{C}[B_1d,B_2d]I(\mathfrak{N})\,, && \mathfrak{K}_1 = \mathbb{C}[dB_1,dB_2]dI(\mathfrak{N})\,,
\end{align}
with $k_1 \geq k_0$ and the partitions associated to $\mathfrak{K}_0$ and $\mathfrak{K}_1$ differ by $k_1-k_0$ boxes freely positioned on the boundary.

\paragraph{}
It is possible to associate to the set of all super-partitions $\mathcal{SP}$ a refined partition function for the number of super-partitions $\Lambda$ with $|\Lambda|$ boxes and $|\partial \Lambda|$ triangle boxes~\cite{Desrosiers:2004pm}: 
\begin{align}\label{eq:ZSPp}
    Z^{\prime}_{\mathcal{SP}}(\Qt,z) := \sum_{\Lambda \in \mathcal{SP}} \Qt^{|\Lambda|} z^{|\partial\Lambda|} = \frac{(-z;\Qt)_\infty}{(\Qt;\Qt)_\infty}\,,
\end{align}
where, in addition to $Q_\tau$, we introduce another formal counting parameter $z$ for the number of triangle boxes.
More physically we are interested in constructing a partition function for the instanton number and magnetic flux given by~\eqref{eq:twoqnum}. In that framework, the instanton number is given by the Chern character and the magnetic flux by the first Chern class on the exceptional $\mathbb{P}^1$. Since the number of boxes in the super-partition is associated with the second Chern class evaluated on the (compactified) space-time, we obtain the physical partition function of 4d $U(1)$ $\mathcal{N}=4$ theory on the blow-up: 
\begin{align}
    Z_{\mathcal{SP}}(\Qt,z) := \sum_{\Lambda\in \mathcal{SP}} \Qt^{|\Lambda| - \binom{|\partial\Lambda|}{2}} z^{|\partial \Lambda|} = (\Qt;\Qt)_\infty^{-1} (z;\Qt)_\infty^{-1}\,,
\end{align}
which can be derived from~\eqref{eq:ZSPp} using the $q$-binomial formulas:
\begin{align}
    (-z;\Qt)_\infty = \sum_{n=0}^\infty \frac{\Qt^{\frac{n(n-1)}{2}}}{(\Qt;\Qt)_n} z^n\,, && (z;\Qt)_\infty^{-1} = \sum_{n=0}^\infty \frac{z^n}{(\Qt;\Qt)_n} \,.
\end{align}
Physically, this partition function corresponds to the Vafa-Witten partition function~\cite{Vafa:1994tf} in presence of a refinement parameter associated with the counting of magnetic flux (as introduced in \cite[Section~2]{Dijkgraaf:2007sw} in the context of Vafa-Witten partition function of $U(N)$ theories on ALE spaces). In the context of the blow-up, such a partition function has been analysed for higher rank gauge theories in the blow-up chamber in~\cite{Kuhn:2022xyw}.

\subsubsection{Blow-up chamber}\label{subsec:buchamber}

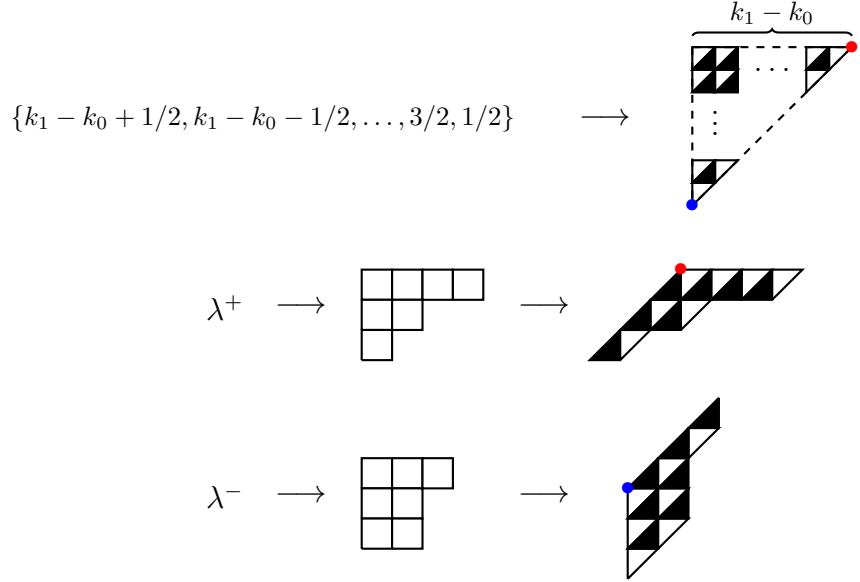
\begin{figure}[tbp]
    \centering
    \begin{tikzpicture}
        \begin{scope}
            \node at (1.5,0) {\footnotesize $\{k_1-k_0+1/2,k_1-k_0-1/2,\ldots,3/2,1/2\}$};
            \node at (6,0) {$\longrightarrow$};
            \begin{scope}[shift={(7.3,0.8)},scale=0.3]
                \wtrbox{0}{0}\btrbox{0}{0}\wtrbox{1}{0}\btrbox{1}{0}\wtrbox{0}{-1}\btrbox{0}{-1}\wtrbox{1}{-1}\btrbox{1}{-1}
                \wtrbox{0}{-5}\btrbox{0}{-5}
                \wtrbox{0}{-6}\wtrbox{1}{-5}
                \node at (0.5,-2.5) {$\vdots$};
                \wtrbox{5}{0}\btrbox{5}{0}
                \wtrbox{6}{0}\wtrbox{5}{-1}
                \node at (3,-0.5) {$\ldots$};
                \draw[dashed,thick] (-0.5,0.5) -- (6.5,0.5);
                \draw[dashed,thick] (-0.5,0.5) -- (-0.5,-6.5);
                \draw[dashed,thick] (6.5,0.5) -- (-0.5,-6.5);
                \draw [decorate,decoration = {brace},thick] (-0.5,1) --  (6.5,1);
                \node at (3,2) {\footnotesize $k_1-k_0$};
                \node at (6.5,0.5) {\small ${\color{red} \bullet}$};
                \node at (-0.5,-6.5) {\small ${\color{blue} \bullet}$};
            \end{scope}
        \end{scope}
        \begin{scope}[shift={(3,-2.5)}]
            \node at (-2,0) {$\lambda^+$};
            \node at (-1,0) {$\longrightarrow$};
            \begin{scope}[shift={(0,0.3)},scale=0.4]
                \sqbox{0}{0}\sqbox{1}{0}\sqbox{2}{0}\sqbox{3}{0}
                \sqbox{0}{-1}\sqbox{1}{-1}
                \sqbox{0}{-2}
            \end{scope}
            \node at (2.2,0) {$\longrightarrow$};
            \begin{scope}[shift={(3.8,0.3)},scale=0.4]
                \btrbox{-2}{-2}\wtrbox{-1}{-2}\btrbox{-1}{-1}\wtrbox{0}{-1}\btrbox{0}{0}\wtrbox{1}{0}
                \btrbox{0}{-1}\wtrbox{1}{-1}\btrbox{1}{0}\wtrbox{2}{0}\btrbox{2}{0}\wtrbox{3}{0}\btrbox{3}{0}\wtrbox{4}{0}
                \node at (0.5,0.5) {\small ${\color{red} \bullet}$};
            \end{scope}
        \end{scope}
        \begin{scope}[shift={(3,-5)}]
            \node at (-2,0) {$\lambda^-$};
            \node at (-1,0) {$\longrightarrow$};
            \begin{scope}[shift={(0,0.3)},scale=0.4]
                \sqbox{0}{0}\sqbox{1}{0}\sqbox{2}{0}
                \sqbox{0}{-1}\sqbox{1}{-1}
                \sqbox{0}{-2}\sqbox{1}{-2}
            \end{scope}
            \node at (2.2,0) {$\longrightarrow$};
            \begin{scope}[shift={(3.5,-0.1)},scale=0.4]
                \wtrbox{0}{-2}\btrbox{0}{-1}\wtrbox{0}{-1}\btrbox{0}{0}\wtrbox{0}{0}\btrbox{0}{1}
                \wtrbox{1}{-1}\btrbox{1}{0}\wtrbox{1}{0}\btrbox{1}{1}\wtrbox{1}{1}\btrbox{1}{2}
                \wtrbox{2}{2}\btrbox{2}{3}
                \node at (-0.5,0.5) {\small ${\color{blue} \bullet}$};
            \end{scope}
        \end{scope}
    \end{tikzpicture}
    \caption{The three pieces $(k_1-k_0,\lambda^+,\lambda^-)$ of a separable super-partition where the original super-partition is obtained by gluing together the red dots and the blue dots.}
    \label{fig:SPtotriplet}
\end{figure}

\paragraph{}
The blow-up chamber is defined as the chamber infinitesimally close and above the line $\zeta_0+\zeta_1=0$ with $\zeta_0 <0$ and $\zeta_1>0$. As discussed in Section~\ref{subsec:chambers}, the blow-up chamber is obtained as the limit of the process of crossing the walls in the domain $\zeta_1 >-\zeta_0>0$ towards the accumulation line $\zeta_0+\zeta_1=0$. At the level of the bipartite trees, as we cross the sequence of walls a subset of the trees originally appearing in the $\mathcal{SP}$-chamber destabilizes, the resulting configurations are a subset of super-partitions which are {\it separable}. We say a super-partition is separable if it can be formed by gluing together three pieces: a super-partition of the form $\{k_1-k_0+1/2,k_1-k_0-1/2,\ldots,3/2,1/2\}$, a horizontally sheared partition $\lambda^+$ and a vertically sheared partition $\lambda^-$ as depicted by~\figref{fig:SPtotriplet}. We write a separable super-partition as a triplet $(k_1-k_0,\lambda^+,\lambda^-)\in \mathbb{N}\times \mathcal{P}^2$.
Importantly, the central triangle super-partition is fixed by $k_1-k_0$ which corresponds physically to the flux through the exceptional $\mathbb{P}^1$ at the origin of the blow-up, i.e. the first Chern class on $\mathbb{P}^1$. In the gluing process the $\lambda^+$ and $\lambda^-$ partitions cannot overlap and must therefore satisfy $\lambda^-_1 + (\lambda^+)^\vee_1 \leq k_1-k_0$ with $^\vee$ denoting the transpose partition in this context. The first contributing separable super-partitions are given by Table~\ref{tab:buchamber}.
\begin{table}[tbp]
    \centering
    \begin{tabular}{c|c|c|c}
        $(k_0,k_1)$ &  & $(k_0,k_1)$ & \\
        \hline \hline
        $(0,1)$ & \begin{tikzpicture}[scale=0.4,baseline=(current bounding box.center)]
            \wtrbox{0}{0}
        \end{tikzpicture} & $(2,3)$ & \begin{tikzpicture}[scale=0.4,baseline=(current bounding box.center)]
            \node at (0,0.5) {$ $};
            \wtrbox{0}{0}\btrbox{0}{0} \wtrbox{0}{-2} \wtrbox{0}{-1}\btrbox{0}{-1}
            \node at (0,-2.5) {$ $};
        \end{tikzpicture}\quad\begin{tikzpicture}[scale=0.4,baseline=(current bounding box.center)]
            \wtrbox{0}{0}\btrbox{0}{0} \wtrbox{1}{0} \wtrbox{2}{0}\btrbox{1}{0}
        \end{tikzpicture} \\
        \hline
        $(1,2)$ & \begin{tikzpicture}[scale=0.4,baseline=(current bounding box.center)]
            \wtrbox{0}{0}\btrbox{0}{0} \wtrbox{1}{0} 
        \end{tikzpicture}\quad\begin{tikzpicture}[scale=0.4,baseline=(current bounding box.center)]
            \wtrbox{0}{0}\btrbox{0}{0} \wtrbox{0}{-1}
        \end{tikzpicture} & $(2,4)$ & \begin{tikzpicture}[scale=0.4,baseline=(current bounding box.center)]
            \node at (0,0.5) {$ $};
            \wtrbox{0}{0}\btrbox{0}{0} \wtrbox{1}{0} \wtrbox{0}{-1}\btrbox{0}{-1} \wtrbox{0}{-2}
            \node at (0,-2.5) {$ $};
        \end{tikzpicture}\quad\begin{tikzpicture}[scale=0.4,baseline=(current bounding box.center)]
            \node at (0,0.5) {$ $};
            \wtrbox{0}{0}\btrbox{0}{0} \wtrbox{1}{0} \btrbox{1}{0} \wtrbox{2}{0}\wtrbox{0}{-1}
        \end{tikzpicture} \\
        \hline
        $(1,3)$ & \begin{tikzpicture}[scale=0.4,baseline=(current bounding box.center)]
            \node at (0,0.5) {$ $};
            \wtrbox{0}{0}\btrbox{0}{0} \wtrbox{1}{0} \wtrbox{0}{-1}
        \end{tikzpicture} & $(3,4)$ & \begin{tikzpicture}[scale=0.4,baseline=(current bounding box.center)]
            \node at (0,0.5) {$ $};
            \wtrbox{0}{0}\btrbox{0}{0} \wtrbox{1}{0} \btrbox{1}{0} \wtrbox{2}{0} \btrbox{2}{0} \wtrbox{3}{0}
        \end{tikzpicture} + tr. \\
    \end{tabular}
    \caption{Configurations of super-partitions in the blow-up chamber with the + tr. indicating additional transposed partitions, all other cases for $k_{0,1}\leq4$ admit no configurations.}
    \label{tab:buchamber}
\end{table}

\paragraph{}
Similarly to the super-partition chamber, we can construct a partition function refined with the monopole charge. The physical instanton number, as described in the counting of Section~\ref{subsec:spchamber}, corresponds to the numbers of boxes in the partitions $\lambda^+$ and $\lambda^-$: $|\Lambda|-\binom{|\partial\Lambda|}{2}=|\lambda^+|+|\lambda^-|$. Denoting $\mathcal{SP}_\infty \subset \mathbb{N} \times \mathcal{P}^2$ the set of separable super-partitions, we obtain:
\begin{align}
    Z_{\mathcal{SP}_\infty}(\Qt,z):=\sum_{(p,\lambda^+,\lambda^-) \in \mathcal{SP}_\infty} \Qt^{|\lambda^+|+|\lambda^-|}z^p = (z;\Qt)_\infty^{-1} (z\Qt;\Qt)_\infty^{-1}\,,
\end{align}
using the constraint $\lambda^-_1 + (\lambda^+)^\vee_1 \leq p$ together with the identity:
\begin{align}
    \sum_{\lambda \in \mathcal{P}}  \Qt^{|\lambda|} z^{\lambda_1} = \frac{1}{(z\Qt;\Qt)_\infty}\,.
\end{align}

\subsubsection{$n$-chamber}\label{subsec:nchamber}

\paragraph{}
A simple way to describe the configuration appearing in each chamber is to describe which super-partitions becomes unstable as we cross the $n$-th wall. We call such configurations {\it n-stable}. Such that in the $n$-th chamber only $(n+k)$-stable super-partitions for $k \in \mathbb{N}$ appear. In order to understand the combinatorial properties of such object we examine some $n$-stable configurations for $n=0,1$ before detailing the general case.

\paragraph{$0$-stable super-partitions} From the stability condition~\eqref{eq:stabcond}, we can deduce that a $0$-stable object with dimension $(k_0,k_1)$ must admit a sub-super-partition with dimension $(k_0-1,k_1)$, graphically this corresponds to a removable black triangle \begin{tikzpicture}[scale=0.4]\btrbox{0}{0}\end{tikzpicture}, or more directly a removable box $\Xi_0:=$ \begin{tikzpicture}[scale=0.4]\wtrbox{0}{0}\btrbox{0}{0}\end{tikzpicture}. Some examples of $0$-stable objects are given by:
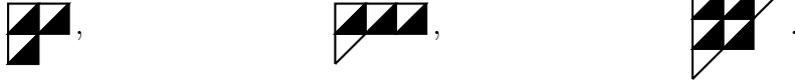
\begin{align*}
    \begin{tikzpicture}[scale=0.4,baseline=(current bounding box.center)]
        \wtrbox{0}{0}\btrbox{0}{0} \wtrbox{1}{0}\btrbox{1}{0}
        \wtrbox{0}{-1}\btrbox{0}{-1}
    \end{tikzpicture}\,,
    &&
    \begin{tikzpicture}[scale=0.4,baseline=(current bounding box.center)]
         \wtrbox{0}{0}\btrbox{0}{0} \wtrbox{1}{0}\btrbox{1}{0}\wtrbox{2}{0}\btrbox{2}{0}
        \wtrbox{0}{-1}
    \end{tikzpicture}\,,
    &&
    \begin{tikzpicture}[scale=0.4,baseline=(current bounding box.center)]
        \wtrbox{0}{0}\btrbox{0}{0} \wtrbox{1}{0}\btrbox{1}{0}\wtrbox{1}{-1}\btrbox{1}{-1}
        \wtrbox{0}{-1}\btrbox{0}{-1}
        \wtrbox{0}{-2}\wtrbox{2}{0}
    \end{tikzpicture}\,.
\end{align*}
All such configurations become unstable as we cross the wall corresponding to $\zeta_0=0$ and $\zeta_1>0$. In particular, we remark that all partitions are $0$-stable.

\paragraph{$1$-stable super-partitions} Similarly, we can read from the stability condition that a $1$-stable object of dimension $(k_0,k_1)$ admits a sub-super-partition of dimension $(k_0-2,k_1-1)$ but no sub-super-partition of dimension $(k_0-1,k_1)$. In practice, it corresponds to super-partitions to which one can remove a white triangle \begin{tikzpicture}[scale=0.4]\wtrbox{0}{0}\end{tikzpicture} uncovering a pair of removable black triangles \begin{tikzpicture}[scale=0.4]\btrbox{0}{0}\end{tikzpicture}. More concretely it corresponds to a super-partition with a removable block of the form:
\begin{align}
    \Xi_1:=\begin{tikzpicture}[scale=0.5,baseline=(current bounding box.center)]\wtrbox{0}{0}\btrbox{0}{0}\wtrbox{1}{0}\btrbox{1}{0}\btrbox{0}{-1}\wtrbox{0}{-1}\wtrbox{1}{-1}
    \node at (0,-1.7) {$ $};
    \end{tikzpicture}\,.
\end{align}
Some examples of $1$-stable super-partitions are given by:
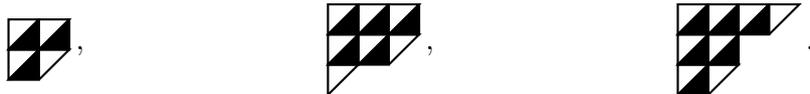
\begin{align*}
    \begin{tikzpicture}[scale=0.4,baseline=(current bounding box.center)]
        \wtrbox{0}{0}\btrbox{0}{0} \wtrbox{1}{0}\btrbox{1}{0}\wtrbox{1}{-1}
        \wtrbox{0}{-1}\btrbox{0}{-1}
    \end{tikzpicture}\,,&&
    \begin{tikzpicture}[scale=0.4,baseline=(current bounding box.center)]
        \wtrbox{0}{0}\btrbox{0}{0} \wtrbox{1}{0}\btrbox{1}{0}\wtrbox{1}{-1}
        \wtrbox{0}{-1}\btrbox{0}{-1}\wtrbox{-1}{0}\btrbox{-1}{0}\wtrbox{-1}{-1}\btrbox{-1}{-1}\wtrbox{-1}{-2}
    \end{tikzpicture}\,,&&
    \begin{tikzpicture}[scale=0.4,baseline=(current bounding box.center)]
        \wtrbox{0}{1}\btrbox{0}{1}\btrbox{1}{1}\wtrbox{1}{1}\wtrbox{2}{1}\btrbox{2}{1}\wtrbox{3}{1}
        \wtrbox{0}{0}\btrbox{0}{0} \wtrbox{1}{0}\btrbox{1}{0}\wtrbox{1}{-1}
        \wtrbox{0}{-1}\btrbox{0}{-1}
    \end{tikzpicture}\,.
\end{align*}

\paragraph{$n$-stable super-partitions} Recursively, the stability condition~\eqref{eq:stabcond} for a $n$-stable object of dimension $(k_0,k_1)$ implies the existence of a sub-object of dimension $(k_0-n,k_0-(n-1))$ but no sub-objects of dimension $(k_0-(n-1),k_0-(n-2)),\ldots,(k_0-1,k_1)$. Such super-partition therefore admits a removable block $\Xi_n$ corresponding to:
\begin{align}
    \Xi_n := \begin{tikzpicture}[scale=0.3,baseline=(current bounding box.center)]
                \wtrbox{0}{1}\btrbox{0}{1}\wtrbox{1}{1}\btrbox{1}{1}\wtrbox{5}{1}\btrbox{5}{1}\wtrbox{6}{1}\btrbox{6}{1}
                \wtrbox{-1}{1}\btrbox{-1}{1}\wtrbox{-1}{0}\btrbox{-1}{0}\wtrbox{-1}{-1}\btrbox{-1}{-1}\wtrbox{-1}{-5}\btrbox{-1}{-5}\wtrbox{-1}{-6}\btrbox{-1}{-6}
                \wtrbox{0}{0}\btrbox{0}{0}\wtrbox{1}{0}\btrbox{1}{0}\wtrbox{0}{-1}\btrbox{0}{-1}\wtrbox{1}{-1}\btrbox{1}{-1}
                \wtrbox{0}{-5}\btrbox{0}{-5}
                \wtrbox{0}{-6}\wtrbox{1}{-5}
                \node at (0,-2.5) {$\vdots$};
                \wtrbox{5}{0}\btrbox{5}{0}
                \wtrbox{6}{0}\wtrbox{5}{-1}
                \node at (3,0) {$\ldots$};
                \draw[dashed,thick] (-0.5,1.5) -- (6.5,1.5);
                \draw[dashed,thick] (-1.5,0.5) -- (-1.5,-6.5);
                \draw[dashed,thick] (6.5,0.5) -- (-0.5,-6.5);
                \draw [decorate,decoration = {brace},thick] (-1.5,2) --  (6.5,2);
                \node at (2.5,3) {\footnotesize $n+1$};
    \end{tikzpicture}\,,
\end{align}
but no removable blocks of the form $\Xi_{n-1},\ldots ,\Xi_0$.

\paragraph{}
We obtain a complete description of the content of each chamber: let $\mathcal{SP}_n$ be the set of super-partitions which are stable in the $n$-th chamber as defined in Section~\ref{subsec:chambers} the $\mathcal{SP}_n$ is given by:
\begin{align}
    \mathcal{SP}_n := \left\{\Lambda\in \mathcal{SP} \,|\, \Xi_i \text{ is not a removable block of } \Lambda\,, i\in \{0,\ldots,n-1 \} \right\}.
\end{align}
The separable super-partitions of the blow-up chamber defined in Subsection~\ref{subsec:buchamber} are given by the set $\mathcal{SP}_\infty$ obtained through the limit $n \to \infty$. This can be understood as follows: First, given $\Lambda\in \mathcal{SP}_\infty$, there exists no $\Box \in \Lambda$ such that both arm and leg associated to this box also end on a full box $\Box$. Then, we distinguish three types of boxes in $\Lambda\in \mathcal{SP}_\infty$:
\begin{itemize}
    \item the boxes with both arm and leg ending on a triangle (among which we count the triangles by convention),
    \item the boxes with only leg ending on a triangle,
    \item the boxes with only arm ending on a triangle.
\end{itemize}
These three sets are in bijection with respectively the central triangle, horizontally sheared partition and vertically sheared partitions of separable super-partitions as depicted by~\figref{fig:SPtotriplet}.

\paragraph{}
Interestingly, some of the sets $\mathcal{SP}_n$ can also be associated with simple generating functions. For instance in the $1$-chamber we obtain:
\begin{align}
    Z^\prime_{\mathcal{SP}_1}(\Qt,z):= \sum_{\Lambda \in \mathcal{SP}_1} \Qt^{|\Lambda|} z^{|\partial \Lambda|} = \frac{(\Qt(1-z);\Qt)_\infty}{(\Qt;\Qt)_\infty}\,,
\end{align}
by numerical check of the series expansion of the r.h.s. up to order $12$ in $\Qt$ and order $5$ in $z$. This final expression in terms of Pochhammer symbols is surprisingly the generating function for the number of partitions of $n$ in $k$ part sizes~\cite{MacMahon1921}.
In this chamber, it seems that the physical partition function $Z_{\mathcal{SP}_1}$ does not admit a simple expression. We leave the exploration of the generating functions for other chambers and their mutual relations for future work.

\section{Instanton partition function and the blow-up formula}\label{sec:pfandblowup}

\paragraph{}
In this Section, we discuss the 4d $\mathcal{N}=2$ instanton partition function on the blow-up, we start by focusing on the general case $G=U(N)$, we then show how the blow-up partition function factorizes through the blow-up formula in the limit chamber. Finally, we discuss the particularities of the reduction to $G=SU(N)$, $G=PSU(N)$ and $G=SU(N)/\mathbb{Z}_l$ for $l | N$.

\subsection{4d $\mathcal{N}=2$ instanton partition function}
\paragraph{}
As discussed in Section~\ref{subsec:coutourintegral}, the 4d $\mathcal{N}=2$ instanton partition function on the blow-up can be formulated as a series of contour integral which we used to characterize the fixed points in the different chambers. For explicit computation it is however simpler to formulate the partition function using character formulas. To do so, we start by writing the tangent space to the instanton moduli space $T\widehat{\mathcal{M}}_{N,k_0,k_1}$ as the middle cohomology of the following complex~\cite{Nakajima:2008eq}:
\begin{align}
    \begin{matrix}
        {\rm Hom}(\mathfrak{K}_0,\mathfrak{K}_0)\\
        \oplus\\
        {\rm Hom}(\mathfrak{K}_1,\mathfrak{K}_1)
    \end{matrix} \quad\xrightarrow[]{\,\,\,\, dg\,\,\,\,}\quad
    \begin{matrix}
        {\rm Hom}(\mathfrak{K}_1,\mathfrak{K}_0)\\
        \oplus \\
       \mathfrak{Q}\otimes{\rm Hom}(\mathfrak{K}_0,\mathfrak{K}_1)\\
        \oplus\\
        {\rm Hom}(\mathfrak{N}, \Lambda^2 \mathfrak{Q} \otimes \mathfrak{K}_0)\\
        \oplus\\
        {\rm Hom}(\mathfrak{K}_1,\mathfrak{N})
    \end{matrix}\quad
    \xrightarrow[]{\,\, d\mu_{\mathbb{C}}\,\,} \quad 
    {\rm Hom}(\mathfrak{K}_0,\mathfrak{K}_1)\otimes \Lambda^2 \mathfrak{Q}\,,
\end{align}
where the maps $dg$ and $d\mu_\mathbb{C}$ are respectively the differential of the group action and the differential of the complex moment map which are given explicitly in~\cite{Nakajima:2003pg}.
$\mathfrak{Q}$ is defined as the cotangent to the blow-down $\Omega$-background:
\begin{align}
    \mathfrak{Q} = T_o^\vee(\mathbb{C}^2_{\varepsilon_1,\varepsilon_2})=\mathfrak{Q}_1 \oplus \mathfrak{Q}_2\,,
\end{align}
with ${\rm ch}\, \mathfrak{Q}_1=q_1^{-1}$, ${\rm ch}\, \mathfrak{Q}_2 = q_2^{-1}$, and also $\operatorname{ch} \Lambda^2 \mathfrak{Q} = q_{12}^{-1}$.
The character of the tangent space to the instanton moduli space can therefore be expressed in terms of the characters of the instanton and framing vector spaces:
\begin{align}\label{eq:characterN2}
    {\rm ch} \, T \widehat{\cal M}_{N,k_0,k_1} = - {\bf K}_1 {\bf N}^\vee - q_{12}^{-1} {\bf N} {\bf K}_0^\vee + {\bf K}_0 {\bf K}_0^\vee +{\bf K}_1 {\bf K}_1^\vee - (q_1^{-1}+q_2^{-1}-q_{12}^{-1}){\bf K}_1 {\bf K}_0^\vee - {\bf K}_0{\bf K}_1^\vee\,,
\end{align}
where ${\bf K}_{0,1}$ are defined by~\eqref{eq:defk01} and ${\bf N}:= {\rm ch}\, \mathfrak{N}$, in this context ${\bf K}^\vee$ denotes the character of the dual vector space. This character can be recast in a simpler form using the $\kappa$-matrix~\eqref{eq:kappamatrix} of the free-field realization~\ref{subsec:freefield}:
\begin{align}
    {\rm ch} \, T \widehat{\cal M}_{N,k_0,k_1} = - {\bf K}_1 {\bf N}^\vee - q_{12}^{-1} {\bf N} {\bf K}_0^\vee + \underline{\bf K} \cdot \kappa\cdot \underline{\bf K}^\vee\,, && \kappa = \begin{bmatrix}
        1 & -q_1^{-1}-q_2^{-1}-q_{12}^{-1}\\ -1 & 1
    \end{bmatrix}\,,
\end{align}
with $\underline{\bf K} = ({\bf K}_0,{\bf K}_1)$.

\paragraph{$\mathcal{P}$-chamber}
As discussed in Section~\ref{subsec:pchamber}, in the $\mathcal{P}$-chamber the vector spaces $\mathfrak{K}_0$ and $\mathfrak{K}_1$ are identified and $k_0=k_1=k$. The character~\eqref{eq:characterN2} reduces to the character of the tangent space to the instanton moduli space on $\mathbb{C}^2$:
\begin{align}
    {\rm ch}\, T \mathcal{M}_{N,k} = - {\bf KN}^\vee - q^{-1}_{12} {\bf NK}^\vee + (1-q_1^{-1})(1-q_2^{-1}) {\bf KK}^\vee\,,
\end{align}
the fixed points are characterised by standard partitions and the partition function corresponds to the formula~\cite{Nekrasov:2002qd,Nekrasov:2003rj}\footnote{One may incorporate hypermultiplets contributions to the instanton partition function~\eqref{eq:instpf}. Adding a hypermultiplet in the adjoint representation and taking the massless limit for $N=1$, one recovers the generating function~\eqref{eq:ZP}.}:
\begin{align}\label{eq:instpf}
    \mathcal{Z}^{\rm inst.}_N = \sum_{k\geq 0} \Qt^k \,\mathbb{I}_{\rm 4d}[{\rm ch}\, T \mathcal{M}_{N,k}]=\sum_{\underline{\lambda} \in \mathcal{P}^N} \Qt^{|\underline{\lambda}|} \prod_{\alpha,\beta=1}^N \frac{1}{\mathcal{N}_{\lambda_\alpha \lambda_\beta}({\sf a}_\beta-{\sf a}_\alpha|\varepsilon_{1,2})}\,,
\end{align}
where we define the 4d index $\mathbb{I}_{\rm 4d}$ as follows:
\begin{align}\label{eq:blowdownPF}
    \mathbb{I}_{\rm 4d}\Big[\sum_k n_k e^{2i\pi x_k}\Big] = \prod_k [x_k]^{n_k}\,,
\end{align}
and the Nekrasov subfunction $\mathcal{N}_{\lambda_\alpha\lambda_\beta}$ defined as~\cite{Nakajima:1999,Flume:2002az}:  
\begin{align}
    \mathcal{N}_{\lambda_\alpha \lambda_\beta}({\sf a}|\varepsilon_{1,2}) &= \prod_{\Box \in \lambda_\alpha } ({\sf a}-l_{\lambda_\beta}(\Box)\varepsilon_1+(a_{\lambda_\alpha}(\Box)+1)\varepsilon_2)\nonumber\\ & \hspace{3cm}\times \prod_{\Box \in \lambda_\beta} ({\sf a}+ (l_{\lambda_\alpha}(\Box)+1)\varepsilon_1 - a_{\lambda_\beta}(\Box)\varepsilon_2)\,,
\end{align}
with $a_\lambda(\Box)$ and $l_\lambda(\Box)$ repsectively the arm-length and leg-length of a box $\Box$ in $\lambda$ defined as:
\begin{align}
    a_{\lambda}(\Box) := \lambda_r -s\,, && l_{\lambda}(\Box) :=\lambda^\vee_s - r\,,
\end{align}
for a box $\Box$ with coordinates $(r,s)$ and $\lambda^\vee$ denoting the transpose of the partition $\lambda$. In this context, the instanton counting parameter corresponds to $\Qt = \exp(2\pi i \tau)$ with $\tau$ the complexified coupling constant with $\operatorname{Im}\tau > 0$ and hence $|\Qt|<1$.

\paragraph{$\mathcal{SP}$/$\mathcal{SP}_n$-chambers}

In the $\mathcal{SP}$- and $\mathcal{SP}_n$- and blow-up chambers, the situation is modified and the fixed points arrange as super-partitions as detailed in Sections~\ref{subsec:spchamber} and~\ref{subsec:buchamber}. More precisely, the fixed points are given by an $N$-tuple of super-partitions $\underline{\Lambda}=(\Lambda_1,\ldots,\Lambda_N)$ for $U(N)$ gauge theory where in the context of this discussion we will represent a super-partition $\Lambda_\alpha$ by a pair of partitions $(\lambda_\alpha^{(0)},\lambda_\alpha^{(1)})$. The characters of instanton vector spaces at a fixed point labelled by $\underline{\Lambda}$ are given by:
\begin{align}
    \mathbf{K}_0\big|_{\underline{\Lambda}} = \sum_{\alpha=1}^N e^{2i\pi {\sf a}_\alpha} c_{\lambda_\alpha^{(0)}}(q_1,q_2)\,, &&\mathbf{K}_1\big|_{\underline{\Lambda}} = \sum_{\alpha=1}^N e^{2i\pi {\sf a}_\alpha} c_{\lambda_\alpha^{(1)}}(q_1,q_2)
\end{align}
where $c_\lambda(q_1,q_2)$ which we will now abbreviate as $c_\lambda$ is the content sum of the partition $\lambda$ defined as:
\begin{align}
    c_{\lambda}(q_1,q_2):= \sum_{(r,s)\in \lambda} q_1^{r-1} q_2^{s-1}\,,
\end{align}
where the sum runs on the coordinates of the boxes in the partition $\lambda$. Using these notations, we now show that the partition function can be written in a form which resembles the partition function on $\mathbb{C}^2$.
The character of the tangent space to the moduli space near the fixed point labelled by $\underline{\Lambda}$ can be decomposed as:
\begin{align}
    {\rm ch} \, T_{\underline{\Lambda}} \widehat{\cal M}_{N,k_0,k_1} = -\sum_{\alpha,\beta=1}^N e^{2i\pi({\sf a}_\beta - {\sf a}_\alpha)} {\cal V}_{\Lambda_\alpha \Lambda_\beta}(q_1,q_2)\,,
\end{align}
with ${\cal V}_{\Lambda_\alpha \Lambda_\beta}$ given by:
\begin{align}
    {\cal V}_{\Lambda_\alpha \Lambda_\beta} :=  c_{\lambda^{(0)}_\alpha} + q_{12}^{-1} c^\vee_{\lambda_\beta^{(1)}} - c_{\lambda_\alpha^{(0)}} c^\vee_{\lambda_\beta^{(0)}} - c_{\lambda_\alpha^{(1)}} c^\vee_{\lambda_\beta^{(1)}} + (q_1^{-1} + q_2^{-1} - q_{12}^{-1}) c_{\lambda_\alpha^{(1)}} c^\vee_{\lambda_\beta^{(0)}} + c_{\lambda_\alpha^{(0)}} c^\vee_{\lambda_\beta^{(1)}}\,.
\end{align}
This can be further rearranged as:
\begin{align}
    {\cal V}_{\Lambda_\alpha \Lambda_\beta}(q_1,q_2) = f_{\lambda_\alpha^{(0)}\lambda_\beta^{(1)}}(q_1,q_2) - c_{\lambda_\alpha^{(1)}\setminus\lambda_\alpha^{(0)}} c^\vee_{\lambda_\beta^{(1)}\setminus\lambda_\beta^{(0)}}\,,
\end{align}
with $f_{\mu\nu}(q_1,q_2)$ denoting the usual vector multiplet character~\cite{Nakajima:1999,Flume:2002az}:
\begin{align}
    f_{\mu\nu}(q_1,q_2):= \sum_{\Box \in \mu} q_1^{-l_{\nu}(\Box)}q_2^{a_\mu(\Box)+1} + \sum_{\Box \in \nu} q_1^{l_\mu(\Box)+1}q_2^{-a_\nu(\Box)}\,.
\end{align}
Based on \cite[Proposition 5.5]{Nakajima:2008ss}, $\mathcal{V}_{\Lambda_\alpha\Lambda_\beta}$ can be interpreted as a $f_{\lambda_\alpha^{(0)}\lambda_\beta^{(1)}}$ with a summation restricted \modif{to {\it pair-relevant}\footnote{This should not be confused with the notion of relevant boxes for a single super-partition which we shall introduce later on.} boxes as defined in \cite{Nakajima:2008ss}.
Given a pair of super-partitions $(\Lambda_\alpha,\Lambda_\beta)$ the for each pair of triangle boxes in $(\partial\Lambda_\alpha,\partial\Lambda_\beta)$ with coordinates $(r_\alpha,s_\alpha)$ and $(r_\beta,s_\beta)$ then the box with coordinates $(r_\alpha,s_\beta)\in\Lambda_\alpha\setminus\partial\Lambda_\alpha$ is pair-irrelevant if $s_\alpha>s_\beta$, otherwise (i.e. if $s_\alpha\leq s_\beta$) the box with coordinates $(r_\beta,s_\alpha)\in\Lambda_\beta$ is pair-irrelevant. This condition defines $|\partial\Lambda_\alpha||\partial\Lambda_\beta|$ pair-irrelevant boxes. A box is then said to be pair-relevant if it is not pair-irrelevant.
The natural vector multiplet character for super-partitions is then~\cite{Nakajima:2008ss,Kanno:2025ifd}:
\begin{align}
    \mathcal{V}_{\Lambda_\alpha\Lambda_\beta}(q_1,q_2)=\sum^\prime_{\Box \in \lambda_\alpha^{(0)}} q_1^{-l_{\lambda_\beta^{(1)}}(\Box)}q_2^{a_{\lambda_\alpha^{(0)}}(\Box)+1} + \sum^\prime_{\Box \in \lambda_\beta^{(1)}} q_1^{l_{\lambda_\alpha^{(0)}}(\Box)+1}q_2^{-a_{\lambda_\beta^{(1)}}(\Box)}\,,
\end{align}
with $\prime$ denoting the restriction to pair-relevant boxes with respect to the pair $(\Lambda,\Sigma)$.}

\paragraph{}
 The instanton partition function on the blow-up is therefore given by:
 \begin{align}\label{eq:ZSPn}
     \widehat{\mathcal Z}^{\rm inst.}_{N,n} :=& \sum_{k\geq 0}\sum_{p\in \mathbb{Z}} \Qt^k z^p \mathbb{I}_{\rm 4d} [{\rm ch} T \widehat{\mathcal M}_{N,k_0,k_1}]\nonumber\\
     =&\sum_{\underline{\Lambda} \in \mathcal{SP}_n^N} \Qt^{-\int_{\widehat{\mathbb{C}}^2}{\rm ch}_2(F_{\underline{\Lambda}})} z^{\int_{\mathcal{C}} c_1(F_{\underline{\Lambda}})} 
     \prod_{\alpha,\beta=1}^N \frac{1}{\widehat{\mathcal{N}}_{\Lambda_\alpha \Lambda_\beta}({\sf a}_\beta - {\sf a}_\alpha|\varepsilon_{1,2})}\,,
 \end{align}
with combinatorial formulas for the integrated Chern characters given by~\eqref{eq:ch2} and~\eqref{eq:c1}, and $\widehat{\cal N}_{\Lambda\Sigma}({\sf a}|\varepsilon_{1,2})$ the super-partition Nekrasov subfunction~\cite{Nakajima:2008ss,Kanno:2025ifd} defined as:
\begin{align}
    \widehat{\cal N}_{\Lambda\Sigma}({\sf a}|\varepsilon_{1,2}):= &\prod_{\Box \in \lambda^{(0)}}^\prime ({\sf a}-l_{\sigma^{(1)}}(\Box)\varepsilon_1 +(a_{\lambda^{(0)}}(\Box)+1)\varepsilon_2)\nonumber\\
    &\hspace{2cm} \times\prod_{\Box \in \sigma^{(1)}}^{\prime} ({\sf a}+(l_{\lambda^{(0)}(\Box)}+1)\varepsilon_1 - a_{\sigma^{(1)}}(\Box) \varepsilon_2)\,,
\end{align}
with $\prime$ denoting the \modif{restriction of the product to pair-relevant boxes in the pair of super-partitions $(\Lambda,\Sigma)$}. The formulation~\eqref{eq:ZSPn} allows to compute the instanton partition function in any $\mathcal{SP}_n$ chamber. The summation is indexed by a pair of integers corresponding to the Chern character at the second order evaluated on the blow-up $\widehat{\mathbb{C}}^2$ and the first Chern class evaluated on the exceptional divisor $\mathcal{C}$. To each, we associate a counting parameter, naturally the second order Chern character corresponds to the exponentiated complex gauge coupling $\Qt$, while the first Chern class is associated with $z$ which for now plays the role of a refinement parameter. This pair of integers can be nicely related to the number of boxes and boundary boxes in the $N$-tuple of super-partitions $\underline{\Lambda}$. The second order Chern character on $\widehat{\mathbb C}^2$ is given by:
\begin{align}\label{eq:ch2}
    -\int_{\widehat{\mathbb{C}}^2} {\rm ch}_2(F_{\underline{\Lambda}}) = |\underline{\Lambda}| - \frac{1}{2}\sum_{\alpha=1}^N |\partial \Lambda_\alpha|(|\partial \Lambda_\alpha|+1) + \frac{1}{2N}\sum_{\alpha<\beta} (|\partial\Lambda_\alpha|-|\partial\Lambda_\beta|)^2\,.
\end{align}
In general, this integral is a rational number, however for certain choices of co-character (see Section~\ref{subsec:reductiontosun}) it can be integer-valued. The first Chern class on $\mathcal{C}$ is then given by:
\begin{align}\label{eq:c1}
    \int_\mathcal{C} c_1(F_{\underline{\Lambda}}) = \sum_{\alpha=1}^N |\partial \Lambda_\alpha|\,,
\end{align}
with $F_{\underline{\Lambda}}$ the field-strength associated with the instanton configuration labelled by $\underline{\Lambda}$. The first Chern class on $\mathcal{C}$ corresponds to $\pi_1(U(N))=\mathbb{Z}$, which could be perceived as a contradiction since $\sum_{\alpha=1}^N|\partial \Lambda_\alpha|\in \mathbb N$ however we should stress that the integers $|\partial \Lambda_\alpha|$ can be freely shifted to be $\mathbb{Z}$-valued since only the relative differences $|\partial \Lambda_\alpha|-|\partial \Lambda_\beta|$  for $\alpha,\beta\in \{1,\ldots ,N\}$ matter in the counting problem\footnote{This is manifest in the blow-up chamber as we will show in Section~\ref{subsec:blowupformula}. It is however not clear if this observation can be generalised to all other chambers.}. In order to avoid confusion, we will now call $p_\alpha$ the relative integers and $|\partial\Lambda_\alpha|$ the positive number of boundary boxes. The redefined first Chern class on $\mathcal{C}$ therefore becomes:
\begin{align}
    \int_{\mathcal{C}}c_1(F_{\underline \Lambda}) = \sum_{\alpha=1}^N p_\alpha \in \mathbb{Z}\,.
\end{align}

\subsection{Blow-up formula}\label{subsec:blowupformula}

\begin{figure}[tbp]
    \centering
        \begin{tikzpicture}[scale=0.5]
		\rsqbox{0}{0}\rsqbox{1}{0}\gsqbox{2}{0}\rsqbox{3}{0}\gsqbox{4}{0}\rsqbox{5}{0}\gsqbox{6}{0}\trbox{7}{0}
        \rsqbox{0}{-1}\rsqbox{1}{-1}\gsqbox{2}{-1}\rsqbox{3}{-1}\gsqbox{4}{-1}\trbox{5}{-1}
        \gsqbox{0}{-2}\gsqbox{1}{-2}\gsqbox{2}{-2}\gsqbox{3}{-2}
        \rsqbox{0}{-3}\rsqbox{1}{-3}\gsqbox{2}{-3}\trbox{3}{-3}
        \gsqbox{0}{-4}\gsqbox{1}{-4}
        \rsqbox{0}{-5}\trbox{1}{-5}
        \gsqbox{0}{-6}
        \gsqbox{0}{-7}
        \trbox{0}{-8}
		\end{tikzpicture}
    \caption{A $1$-stable super-partition where red boxes are irrelevant and green boxes are relevant.}
    \label{fig:irrelevantboxes}
\end{figure}
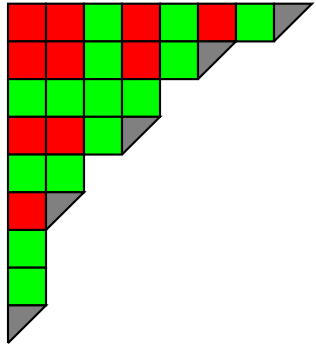

\paragraph{}
As observed in Section~\ref{subsec:buchamber}, in the blow-up chamber the super-partitions can be described by a triple in $\mathbb{N}\times \mathcal{P}^2$ by constructing the super-partition through the gluing of two sheared partitions to a triangle super-partition of fixed size. \modif{For separable super-partitions, the appropriate notion of {\it relevant} and {\it irrelevant} boxes from the character perspective is no longer pairwise. We should now define an irrelevant box in a given super-partition as a box with both arm and leg ending on a triangle box. A relevant box is then a box which is not irrelevant. \figref{fig:irrelevantboxes} provides an example of a super-partitions and its relevant and irrelevant boxes.} It is possible to understand separable configurations as the limiting case of the process of removing $n$-stable sub-blocks by observing that the irrelevant boxes can be rearranged to form the triangle super-partition and we are left with two disjoint sets of relevant boxes. This observation is only valid for $\infty$-stable super-partition since $n$-stable objects cannot admit two disjoint sets of relevant boxes. This is explicit for instance in~\figref{fig:irrelevantboxes} where the $1$-stable sub-block $\Xi_1$ forces the two disjoint sets to interact. This combinatorial principle allows to formulate the partition function on the blow-up in the blow-up chamber as a bilinear combination of the partition function in the $\mathcal{P}$-chamber.

\paragraph{Blow-up formula from characters}
We now proceed to demonstrate the bilinear formula explicitly. Let $\Lambda_\alpha,\Lambda_\beta \in \mathcal{SP}_\infty$ be a pair of separable super-partitions (as defined in Section~\ref{subsec:buchamber}) such that $\Lambda_\alpha=(p_\alpha,\lambda^+_\alpha,\lambda^-_\alpha)$ and $\Lambda_\beta = (p_\beta,\lambda_\beta^+,\lambda_\beta^-)$.
\modif{
First, the character $\mathcal{V}_{\Lambda_\alpha\Lambda_\beta}$ splits into:
\begin{align}
    \mathcal{V}_{\Lambda_\alpha \Lambda_\beta}(q_1,q_2) = \mathcal{F}_{\Lambda_\alpha \Lambda_\beta}(q_1,q_2) + \mathcal{L}_{\alpha \beta}(q_1,q_2)\,,
\end{align}
where $\mathcal{F}_{\Lambda_\alpha\Lambda_\beta}$ is the super-partition analogue of the vector multiplet character given by
\begin{align}\label{eq:charV}
    \mathcal{F}_{\Lambda\Sigma}(q_1,q_2) = \sum_{\Box \in \lambda^{(0)}}^\sim q_1^{-l_{\sigma^{(1)}}(\Box)}q_2^{a_{\lambda^{(0)}}(\Box)+1} + \sum^\sim_{\Box \in \sigma^{(1)}} q_1^{l_{\lambda^{(0)}}(\Box)+1}q_2^{-a_{\sigma^{(1)}}(\Box)}\,,
\end{align}
where the $\sim$ symbol denotes that the summation is restricted to the relevant boxes of each super-partition and $\Lambda=(\lambda^{(0)},\lambda^{(1)})$, $\Sigma=(\sigma^{(0)},\sigma^{(1)})$. While $\mathcal{L}_{\alpha\beta}$ only depends on the number of boundary boxes $|\partial \Lambda_\alpha|$ and $|\partial \Lambda_\beta|$ and is given by:
\begin{align}\label{eq:monofac}
    \mathcal{L}_{\alpha\beta}(q_1,q_2):= \begin{cases}
        c_{\nu_{|\partial\Lambda_\alpha|-|\partial\Lambda_\beta|}}^\vee(q_1,q_2) & \text{if} \quad |\partial \Lambda_\alpha|-|\partial \Lambda_\beta|>0\,,\\
        q_{12}c_{\nu_{|\partial\Lambda_\beta|-|\partial \Lambda_\alpha|-1}}(q_1,q_2) & \text{if} \quad |\partial\Lambda_\beta|-|\partial \Lambda_\alpha|-1 > 0\,, \\
        0 & \text{otherwise}\,,
    \end{cases}
\end{align}
with $\nu_n$ denoting the partition $\{n,n-1,\ldots,2,1\}$ for $n\geq1$.
Physically, $\mathcal{F}_{\Lambda_\alpha \Lambda_\beta}$ arises from the bound states formed by the instanton-monopole configurations of the Cartan subgroup while $\mathcal{L}_{\alpha\beta}$ purely depends on the monopole configurations. We should also stress that the virtual dimensions of the associated vector spaces are given by:
\begin{align}
    &\mathcal{F}_{\Lambda_\alpha\Lambda_\beta}(1,1)= |\Lambda_\alpha|+|\Lambda_\beta| - \frac{|\partial\Lambda_\alpha|(|\partial \Lambda_\alpha|+1)}{2}-\frac{|\partial\Lambda_\beta|(|\partial \Lambda_\beta|+1)}{2}\,,\nonumber\\
    &\mathcal{L}_{\alpha\beta}(1,1)=\frac{1}{2}(|\partial \Lambda_\alpha|-|\partial\Lambda_\beta|)^2-\frac{1}{2}(|\partial \Lambda_\alpha|-|\partial\Lambda_\beta|)\,.
\end{align}}
Then, the instanton-monopole counting character~\eqref{eq:charV} satisfies the property:
\begin{align}\label{eq:chblowup}
    \mathcal{F}_{\Lambda_\alpha \Lambda_\beta}(q_1,q_2) = q_1^{p_\alpha-p_\beta} f_{\lambda_\alpha^- \lambda_\beta^-}(q_1,q_2/q_1) + q_2^{p_\alpha - p_\beta} f_{\lambda_\alpha^+\lambda_\beta^+}(q_1/q_2,q_2)\,.
\end{align}
This property can be proven by analysing the arm-length and leg-length in the respective partitions. Let $\Box \in \Lambda_\alpha$ be a relevant box in the lower side of $\Lambda_\alpha$, then $\Box$ corresponds to $\Box^\prime\in \lambda^-_\alpha$. The relevant arm-length and leg-length satisfy:
\begin{align}\label{eq:armbla}
    a_{\Lambda_\alpha}(\Box) = a_{\lambda_\alpha^-}(\Box^\prime)\,, && l_{\Lambda_\beta}(\Box) = l_{\lambda_\beta^-}(\Box^\prime) + a_{\lambda_\alpha^-}(\Box^\prime) + p_\alpha - p_\beta +1\,.
\end{align}
The arm-length relation follows since there is no irrelevant boxes in the arm of $\Box$, the leg-length relation is obtained by noticing that $\lambda_\beta^-$ and $\lambda_\alpha^-$ most top-left box are separated by $p_\alpha - p_\beta$ boxes vertically and that there is $a_{\lambda_\alpha^-}(\Box^\prime)$ irrelevant boxes below $\Box$ in $\Lambda_\alpha$.
Upon exchanging $\Lambda_\beta$ and $\Lambda_\alpha$ we obtain the same relation for boxes in the lower side of $\Lambda_\beta$. Similarly for a relevant box $\Box \in \Lambda_\alpha$ in the top-right side of $\Lambda_\alpha$, it admits a corresponding box $\Box^\prime\in \lambda_\alpha^+$ such that:
\begin{align}\label{eq:armblb}
    a_{\Lambda_\beta}(\Box) = a_{\lambda_\beta^+}(\Box^\prime) + l_{\lambda_\alpha^+}(\Box^\prime) + p_\alpha - p_\beta +1\,, && l_{\Lambda_\alpha}(\Box) = l_{\lambda_\alpha^+}(\Box^\prime)\,, 
\end{align}
with a symmetric relation for relevant boxes in the top-right side of $\Lambda_\beta$. The relations~\eqref{eq:armbla} and~\eqref{eq:armblb} and their symmetric counter-part under the exchange of $\Lambda_\alpha$ and $\Lambda_\beta$ imply~\eqref{eq:chblowup}. 

\paragraph{Perturbative contribution}
The perturbative contribution is unchanged on the blow-up $\widehat{\mathbb{C}}^2$ since it only depends on the asymptotic behaviour far from the blow-up point. However as noticed in~\cite{Nakajima:2003uh}, \modif{the monopole factors arising from~\eqref{eq:monofac} and defined as:
\begin{align}
    L_{\alpha\beta}({\sf a}_\beta - {\sf a}_\alpha |\varepsilon_{1,2}):= \mathbb{I}_{\rm 4d}[e^{2i\pi({\sf a}_\beta - {\sf a}_\alpha)}\mathcal{L}_{\alpha\beta}(q_1,q_2)]
\end{align}
satisfy the relation:}
\begin{align}\label{eq:pertblowup}
    \prod_{\alpha,\beta=1}^N L_{\alpha\beta}({\sf a}_\beta - {\sf a}_\alpha |\varepsilon_{1,2}) = \frac{{\cal Z}^{\text{1-loop}}_N(\underline{\sf a},\varepsilon_1,\varepsilon_2)}{{\cal Z}^{\text{1-loop}}_N(\underline{\sf a}+\underline{p}\varepsilon_1,\varepsilon_1,\varepsilon_2-\varepsilon_1){\cal Z}^{\text{1-loop}}_N(\underline{\sf a}+\underline{p}\varepsilon_2,\varepsilon_1-\varepsilon_2,\varepsilon_2)}\,,
\end{align}
with the one-loop factor given by
\begin{align}
    {\cal Z}^{\text{1-loop}}_N(\underline{\sf a},\varepsilon_1,\varepsilon_2) := \prod_{\alpha,\beta=1}^N \Gamma_2({\sf a}_\beta-{\sf a}_\alpha;\varepsilon_1,\varepsilon_2)\,,
\end{align}
where we define the double Gamma function $\Gamma_2$ as~\cite{Nekrasov:2002qd}:
\begin{align}
    \Gamma_2(a;\varepsilon_1,\varepsilon_2) := \exp \Bigg( \frac{\partial}{\partial s}\Bigg|_{s=0} \sum_{n_1,n_2\geq 0} \frac{1}{(a+n_1 \varepsilon_1 +n_2 \varepsilon_2)^s}\Bigg)\,.
\end{align}

\paragraph{}
Combining~\eqref{eq:chblowup} and~\eqref{eq:pertblowup} we obtain the blow-up formula~\cite{Nakajima:2003pg}: 
\begin{align}\label{eq:buformula}
    \mathcal{Z}^{\text{1-loop}}_N \mathcal{Z}^{\rm inst.}_{N,\infty} = \sum_{\underline{p} \in \mathbb{Z}^N} \Qt^{\frac{1}{2N}\sum_{\alpha<\beta} (p_\alpha-p_\beta)^2} z^{\sum_{\alpha=1}^N p_\alpha} &{\cal Z}_{N}(\underline{\sf a}+\underline{p}\varepsilon_1,\varepsilon_1,\varepsilon_2-\varepsilon_1,\tau)\nonumber\\
    &\quad\quad\times{\cal Z}_{N}(\underline{\sf a}+\underline{p}\varepsilon_2,\varepsilon_1-\varepsilon_2,\varepsilon_2,\tau)\,,
\end{align}
with the full partition function $\mathcal{Z}_N$ defined as:
\begin{align}
    \mathcal{Z}_N(\underline{\sf a},\varepsilon_1,\varepsilon_2,\tau) := \mathcal{Z}^{\text{1-loop}}_N(\underline{\sf a},\varepsilon_1,\varepsilon_2) \mathcal{Z}^{\rm inst.}_N(\underline{\sf a},\varepsilon_1,\varepsilon_2,\tau)\,,
\end{align}
with $\mathcal{Z}^{\rm inst.}_N$ the blow-down instanton partition function defined by~\eqref{eq:blowdownPF}.

\subsection{Reduction to $SU(N)$, $PSU(N)$ and $SU(N)/{\mathbb{Z}_l}$}\label{subsec:reductiontosun}

\paragraph{}
Up to this point, we have only considered the case $G=U(N)$ which satisfies $\pi_1(U(N))=\mathbb{Z}$. The summation over the number of triangle boxes on the boundary is therefore free. However, for physical theories we are interested in asymptotically free theories, one should therefore consider the reduction of the partition function to theories with gauge algebra $\mathfrak{su}(N)$ which leaves the gauge groups $SU(N)$, $PSU(N) = SU(N)/\mathbb{Z}_N$ and all $SU(N)/\mathbb{Z}_l$ for $l|N$. In the usual $\mathbb{C}^2$ space-time, it is not trivial to see the factorisation of the so-called $U(1)$ factor~\cite{Alday:2009aq}. This is, in principle, not the case on the blow-up since the partition function becomes sensitive to the first homotopy group of the gauge group, critically changing the set of allowed fluxes through the exceptional curve.

\paragraph{}
For $G=U(N)$, the summation over all possible $p_\alpha$ appearing in~\eqref{eq:buformula} corresponds to a summation over all points on the co-character lattice of $U(N)$ defined as:
\begin{align}
    L_{\text{co-char.}}^{U(N)} := {\rm Hom}(U(1),\mathbb{T}) = \big\{p_\alpha \in \mathbb{Z}\,, \quad \alpha \in \{1,\ldots,N\}\big\}\,,
\end{align}
where $\mathbb{T}\subset U(N)$ is the maximal torus in $U(N)$. Upon reduction to $SU(N)$ the Cartan subgroup is reduced, leading to the relation ${\sf a}_N=-\sum_{\alpha=1}^{N-1}{\sf a}_\alpha$. In addition, the first homotopy group vanishes $\pi_1(SU(N))=0$ which reduces the co-character lattice to:
\begin{align}
    L_{\text{co-char.}}^{SU(N)}=\bigg\{p_\alpha \in \mathbb{Z}\,,\quad \alpha\in \{1,\ldots,N\}\,\, \bigg|\ \sum_{\alpha=1}^N p_\alpha =0\bigg\} \subset L^{U(N)}_{\text{co-char.}}\,.
\end{align}
The co-character lattice corresponds in that situation to the co-root lattice. Imposing these two conditions leads, up to a $U(1)$ prefactor, to the $SU(N)$ partition function of the blow-up.

\paragraph{}
As it can be noticed from the previous discussion, the computation on the partition function on the blow-up depends critically on the first homotopy group $\pi_1(G)$ as it fixes the nature of the allowed magnetic flux through the exceptional curve $\mathcal{C}$. In particular, it allows to distinguish between $SU(N)$ and $PSU(N)$ since $\pi_1(PSU(N))=\mathbb{Z}_N$. While the reduction of the Cartan subgroup still leads to ${\sf a}_N = - \sum_{\alpha=1}^{N-1} {\sf a}_\alpha$, the co-character lattice is now given by:
\begin{align}
    L_{\text{co-char.}}^{PSU(N)} = \bigg\{p_\alpha \in \frac{1}{N} \cdot \mathbb{Z}\,,\quad \alpha\in \{1,\ldots,N\}\,\, \bigg|\ \sum_{\alpha=1}^N p_\alpha =0\,, \quad p_\alpha-p_\beta\in \mathbb{Z}\bigg\} \,,
\end{align}
corresponding to the co-weight lattice, hence it is electro-magnetic dual to the $SU(N)$ theory. Similarly, one can define the co-character lattice of $SU(N)/\mathbb{Z}_l$ for $l\,|\,N$:
\begin{align}
    L_{\text{co-char.}}^{SU(N)/\mathbb{Z}_l} = \bigg\{p_\alpha \in \frac{1}{l} \cdot \mathbb{Z}\,,\quad \alpha\in \{1,\ldots,N\}\,\, \bigg|\ \sum_{\alpha=1}^N p_\alpha =0\,, \quad p_\alpha-p_\beta\in \mathbb{Z}\bigg\} \,.
\end{align}
these lattices satisfy the inclusion relation:
\begin{align}
    L_{\text{co-char.}}^{SU(N)} \subset L_{\text{co-char.}}^{SU(N)/\mathbb{Z}_l} \subset L_{\text{co-char.}}^{PSU(N)}\,.
\end{align}
As described in~\cite{Nakajima:2003pg}, such lattices can be realised starting from the $U(N)$ case by considering the sub-sector satisfying $\sum_{\alpha=1}^N p_\alpha =k$ for $k\in \{1,\ldots,l\}$ and by taking $\Tilde{p}_\alpha := p_\alpha + k/l$ to be the coefficient of the lattice vectors.

\section{Conclusion}\label{Sect:Conclusion}

\paragraph{}
In this work, we have analysed the problem of counting instantons on the blow-up of $\mathbb{C}^2$. On this modified space-time, background magnetic fluxes refine the counting problem. Using the description of the moduli space of instantons in terms of a quiver variety~\cite{Nakajima:2008eq,Nakajima:2008ss,Nakajima:2009qjc}, we show how the new configurations are counted by super-partitions instead of standard partitions. We interpret these objects as instanton-monopole configurations or equivalently in five dimensions as $D0$-$D2$ bound states. Contrary to the blow-down situation, the moduli space on the blow-up is not hyper-K\"ahler and admits non-trivial stability conditions. We analyse this stability by using the JK residue prescription and show how the stable instanton-monopole configurations vary across the walls. We make this manifest by describing the set of stable super-partitions in each chambers. Finally, we discuss the implication of the chamber dependence at the level of the instanton partition function and show how the wall-crossing process for $n\to\infty$ (\emph{i.e.} to the blow-up chamber) implies the blow-up formula of~\cite{Nakajima:2003pg}.

\paragraph{}
Our analysis provides a description of the stable configurations in each chamber separately, allowing to compute the complete partition function. However, in order to fully capture the wall-crossing phenomenon, it would be necessary to have a more microscopic understanding by describing how a given unstable object re-arranges into stable sub-objects as we cross a wall. Such information could be accessed through a more in depth analysis of the partition functions of Section~\ref{sec:classification} and their relations as we cross the walls. Furthermore, to a given quiver one can associate an affine Yangian~\cite{Li:2020rij} (and its uplift to the quantum toroidal algebra~\cite{Noshita:2021ldl}) governed by the gymnastic of adding and removing cells/boxes to the crystals/partitions. In~\cite{Galakhov:2024foa}, the correspondence with the affine Yangian has been explored for a similar problem to the one studied in our work. Contrary to~\cite{Galakhov:2024foa}, the set of stable super-partitions appearing in our work varies significantly across the walls, it would therefore be interesting to understand how this manifests itself at the level of the affine Yangian.

\paragraph{}
The blow-up introduces a collection of instanton partition functions associated with all the different stability chambers. It is however not clear what is a physical realisation of the partition function in the intermediary chambers such as the $\mathcal{SP}$-chamber. Notably, the blow-up renders the instanton counting problem dependent to the global structure of the gauge group~\cite{Aharony:2013hda}; the flux through the exceptional curve being indexed by $\pi_1(G)$. While in the blow-up chamber this does not seem to lead to non-trivial modifications, it could be interesting to see how the partition function depends on the choice of gauge group in the intermediary chambers. 

\paragraph{}
Finally, the approach to blow-up through stability can be generalised in many directions. The first is the extension to five and six dimensional theories on $\widehat{\mathbb{C}}^2\times\mathbb{S}^1$ and $\widehat{\mathbb{C}}^2\times\mathbb{T}^2$ as well as $\mathcal{N}=2^*$ theories and little string theories. In particular, for the latter, the blow-up approach could allow to regularize the thermodynamic limit~\cite{Grekov:2024ayn} and access the doubly elliptic Seiberg-Witten curve~\cite{Kanazawa:2016tnt,Braden:2003gv,Filoche:2024knd,Filoche:2024vne}.  Another is the inclusion of co-dimension 2 and 4 defects as discussed in~\cite{Jeong:2020uxz}. In the presence of such a defect, it was already observed that the change of stability condition leads to non-trivial modifications of the defect partition function~\cite{Lee:2020hfu}. \modif{In those cases, an important subtlety is introduced by the addition of matter in some representation of the gauge group. In the blow-up chamber we expect that it would significantly change the discussion of the monopole contributions similarly to the type of modification observed for ALE spaces in~\cite{Ito:2013kpa}.}
Two natural generalisations are given by the blow-up of $\mathbb{C}^3$ given by ${\rm Tot}(\mathcal{O}(-1)\to \mathbb{P}^2)$ and the blow-up of $\mathbb{C}^4$, ${\rm Tot}(\mathcal{O}(-1)\to \mathbb{P}^3)$. It would be interesting to understand what is the equivalent object to super-partitions for plane and solid partitions. We expect that some limit chamber should be associated with a notion of blow-up formula which could be related to the topological vertex~\cite{Aganagic:2003db,Iqbal:2007ii} for $\mathbb{C}^3$ and 4G networks~\cite{Nekrasov:2023nai} for $\mathbb{C}^4$.

\section*{Acknowledgements}
We would like to thank Timothé Alezraa for inspiring exchanges \modif{as well as Hiroaki Kanno, Hiraku Nakajima and Go Noshita for very useful comments on a previous version of this work}.
BF would also like to thank Kimyeong Lee, Nicolò Piazzalunga, Yuuji Tanaka and Xia Gu for invaluable discussions. The work of TK was supported by EIPHI Graduate School (No.~ANR-17-EURE-0002) and the Bourgogne-Franche-Comté region.

\printbibliography

\end{document}